\documentclass[journal,10pt]{IEEEtran}
\usepackage{indentfirst}
\usepackage{inputenc}
\usepackage[switch,mathlines]{lineno}
\usepackage{soul}
\usepackage{graphicx}
\usepackage{cite}
\usepackage{CJK}
\usepackage{algorithmic}
\usepackage{array}
\usepackage{bm}
\usepackage{amssymb}
\usepackage{color}
\usepackage{nomencl}
\usepackage{pdfpages}
\usepackage{booktabs} 
\usepackage{multicol}
\usepackage{multirow}
\usepackage{threeparttable}
\usepackage[ruled, linesnumbered]{algorithm2e}
\usepackage[hyphens]{url}
\usepackage[colorlinks, linkcolor=blue, citecolor=blue, urlcolor=blue]{hyperref}
\usepackage{supertabular}
\usepackage{longtable}
\usepackage{amsmath}
\usepackage{mathtools}
\allowdisplaybreaks
\usepackage{adjustbox}
\usepackage{tabularx}
\usepackage{ltablex}
\usepackage{pifont}
\keepXColumns

\ifCLASSOPTIONcompsoc
 \usepackage[caption=false,font=normalsize,labelfont=sf,textfont=sf]{subfig}
\else
 \usepackage[caption=false,font=footnotesize]{subfig}
\fi
\usepackage{url}
\begin{document}

\title{A Review of Safe Reinforcement Learning Methods for Modern Power Systems}

\author{{Tong Su, \IEEEmembership{Graduate Student Member, IEEE}, Tong Wu, \IEEEmembership{Member, IEEE}, Junbo Zhao, \IEEEmembership{Senior Member,~IEEE},\\ Anna Scaglione, \IEEEmembership{Fellow,~IEEE}, Le Xie, \IEEEmembership{Fellow,~IEEE}}

\thanks{This work is supported by the U.S. Department of Energy Solar Energy Technologies Office under award 10422 (Corresponding author: Junbo Zhao).
\par Tong Su and Junbo Zhao are with the Department of Electrical and Computer Engineering, University of Connecticut, Storrs, CT 06269, USA. Junbo Zhao is also with Dartmouth College, Hanover, NH 03755, USA (e-mail: tongsu@uconn.edu; junbo@uconn.edu).
\par Tong Wu is with the Department of Electrical and Computer Engineering, University of Central Florida, Orlando, FL 32816, USA (e-mail: tong.wu@ucf.edu).
\par Anna Scaglione is with the Department of Electrical and Computer Engineering, Cornell Tech, Cornell University, New York City, NY 10044, USA (e-mail: as337@cornell.edu).
\par Le Xie is with the Harvard John A. Paulson School of Engineering and Applied Sciences, Allston, MA 02134, USA (e-mail: xie@seas.harvard.edu).
}}
\markboth{Published in the Proceedings of the IEEE, 2025}%
{Shell \MakeLowercase{\textit{et al.}}: Bare Demo of IEEEtran.cls for Journals}

\maketitle
\begin{abstract}


Given the availability of more comprehensive measurement data in modern power systems, reinforcement learning (RL) has gained significant interest in operation and control. Conventional RL relies on trial-and-error interactions with the environment and reward feedback, which often leads to exploring unsafe operating regions and executing unsafe actions, especially when deployed in real-world power systems. To address these challenges, safe RL has been proposed to optimize operational objectives while ensuring safety constraints are met, keeping actions and states within safe regions throughout both training and deployment. Rather than relying solely on manually designed penalty terms for unsafe actions, as is common in conventional RL, safe RL methods reviewed here primarily leverage advanced and proactive mechanisms. These include techniques such as Lagrangian relaxation, safety layers, and theoretical guarantees like Lyapunov functions to rigorously enforce safety boundaries. This paper provides a comprehensive review of safe RL methods and their applications across various power system operations and control domains, including security control, real-time operation, operational planning, and emerging areas. It summarizes existing safe RL techniques, evaluates their performance, analyzes suitable deployment scenarios, and examines algorithm benchmarks and application environments. The paper also highlights real-world implementation cases and identifies critical challenges such as scalability in large-scale systems and robustness under uncertainty, providing potential solutions and outlining future directions to advance the reliable integration and deployment of safe RL in modern power systems.
\end{abstract}

\begin{IEEEkeywords}
Safe reinforcement learning, machine learning, power system operation, security control, energy management, real-time operation, operational planning, real-world deployment and roadmap.
\end{IEEEkeywords}

\section*{Nomenclature}
\subsection*{Notations}
\addcontentsline{toc}{section}{Nomenclature}
\begin{IEEEdescription}[\IEEEusemathlabelsep\IEEEsetlabelwidth{$i, j, x, y, zz$}]
\item[$\gamma$] Discount factor $\gamma \in [0, 1)$
\item[$\varepsilon$] Safety constraint bound
\item[$\zeta$] Safety probability
\item[$\lambda$] Penalty coefficient or Lagrange multiplier
\item[$\Pi_{S}, \pi_\theta$] Policy set, policy with parameters $\theta$
\item[$\rho_0$] Starting state distribution $\rho_0: \mathcal{S} \rightarrow [0,1]$
\item[$\tau$] Trajectory $\tau = (s_0, a_0, s_1, \ldots)$
\item[$\mathcal{A}, \bm a$] Action set, action
\item[$\mathcal{B}/\mathcal{G}/\mathcal{N}/\mathcal{R}$] BESS/SG/node/RES set
\item[$\mathcal{C}, C$] Constraint set $\mathcal{C} = \{(C_i, \varepsilon_i)\}^m_{i=1}$, constraint cost function $C: \mathcal{S} \times \mathcal{A} \times \mathcal{S} \rightarrow \mathbb{R}$
\item[$\text{ch}/\text{dis}$] Subscript for charging/discharging of devices
\item[$\mathbb{E}, E$] Expectation function, energy of devices
\item[$f, g/h$] State transition dynamics or the model of the environment, equality/inequality constraints with a total number of $m/n$
\item[$\mathcal{J}_R^{\pi_\theta}$, $\mathcal{J}_{h_i}^{\pi_\theta}$] Reward performance, constraint cost performance of inequality constraints
\item[$\mathcal{L}$] Lagrangian (Lag)
\item[$\mathcal{M}$, $\mathcal{M}_C$] MDP $\mathcal{M} = (\mathcal{S}, \mathcal{A}, \mathcal{P}, r, \rho_0, \gamma)$, CMDP $\mathcal{M}_C = (\mathcal{S}, \mathcal{A}, \mathcal{P}, R, \rho_0, \gamma, \mathcal{C})$
\item[$\mathbb{P}, \mathcal{P}$] Probability function, $\mathcal{P}: \mathcal{S} \times \mathcal{A} \times \mathcal{S} \rightarrow [0, 1]$ is the transition matrix, where $\mathcal{P}(s_{t+1}|s_t, a_t)$ denotes the probability of state transition from $s_t$ to $s_{t+1}$ after taking action $a_t$
\item[$\bm p/\bm q$] Active/reactive power generation/load vector
\item[$R$] Reward function $R: \mathcal{S} \times \mathcal{A} \times \mathcal{S} \rightarrow \mathbb{R}$
\item[$\mathcal{S}, \bm s$] State set, state
\item[$\mathcal{T}, t$] Time step set of trajectory $\tau$, time instant
\item[$\bm v$] Voltage phasor 
\item[$\overline{\ }/\underline{\ }$] Maximum/minimum values of the variables
\end{IEEEdescription}
\subsection*{Abbreviations}
\addcontentsline{toc}{section}{Nomenclature}
\begin{IEEEdescription}[\IEEEusemathlabelsep\IEEEsetlabelwidth{$i, j, x, y, zz$}]
\item[(B/T)ESS] (Battery/thermal) energy storage system
\item[(C)MDP] (Constrained) Markov decision process
\item[DER] Distributed energy resource
\item[DG] Distributed generation
\item[(D/H/R)RL] (Deep/hierarchical/robust) reinforcement learning
\item[(D/IC)NN] (Deep/input convex) neural network
\item[EV, V2G] Electric vehicle, vehicle-to-grid
\item[G(C/N)N] Graph (convolution/neural) network
\item[GPT] Generative pre-trained transformer
\item[HVAC] Heating, ventilation, and air-conditioning
\item[IPO] Interior-point policy optimization
\item[LLM] Large language model
\item[MA] Multi-agent
\item[MI(N)LP] Mixed-integer (non-)linear programming
\item[MPC] Model predictive control
\item[PDO] Primal-dual optimization
\item[PPO] Proximal policy optimization
\item[(P/R)CPO] (Projection-based/Reward) constrained policy optimization
\item[RES, SG] Renewable energy source, synchronous generator
\item[SAC] Soft actor-critic
\item[(SC)(O)PF] (Security constrained) (optimal) power flow
\item[SoC] State of change
\item[TR(PO/M)] Trust region (policy optimization/method)
\end{IEEEdescription}

\section{Introduction}\label{sec1}
\IEEEPARstart{W}{ith} the extensive integration of RESs, ESSs, and advanced power electronic devices, modern power systems face increased uncertainty and complexity, resulting in a significantly higher computational burden to model stochastic, nonlinear control and decision-making \cite{aien2016comprehensive}. Additionally, ensuring system stability, managing renewable variability, and maintaining safe operations under dynamic conditions remain persistent issues \cite{roald2023power}. However, thanks to the widespread deployment of smart sensors, such as PMUs and AMIs, along with advanced communication technologies, a vast amount of power system data can be measured and utilized for state estimation and control \cite{cheng2023survey, li2024artificial}. As a result, data-driven approaches like RL have emerged as the key candidates for the numerical optimization of power systems decision and/or control policies\cite{chen2022reinforcement, li2023deep}, which would be otherwise intractable to derive. RL training is based on trial-and-error interactions with the environment and reward feedback, updating policy parameters to maximize expected cumulative rewards. Recently, DRL, which embeds NNs as the policy function, has proven expressive enough to solve complicated control tasks \cite{nguyen2020deep}. The NN is used to reduce computation costs for online implementation. Once the NNs are trained, they approximate closed-form solutions and produce results quickly \cite{li2017deep}.
However, conventional RL lacks effective constraint handling mechanisms, which can lead agents to explore unsafe regions during training or perform unsafe actions in deployment, creating an unacceptable risk in safety‐critical energy systems. These limitations highlight the urgent need to move beyond conventional RL for real-world power system applications \cite{gu2024review}.

In 2015, safe RL was first defined as {\it ``the process of learning policies that maximize the expectation of the reward in problems, where it is crucial to ensure reasonable system performance and/or respect safety constraints during the learning and/or deployment processes"} \cite{garcia2015comprehensive}. Concurrently, the safe RL literature has garnered increasing attention, offering mechanisms to integrate safety directly into the learning process, particularly in dynamic and high-stakes applications like power systems, where stability, reliability, and operational constraints must be strictly upheld. Safe RL methods can be broadly categorized into three groups. The first group focuses on incorporating safety factors into the reward function to penalize violations \cite{zhao2022deep}. While this approach is straightforward, it often struggles to enforce the physics-hard constraints of power systems effectively \cite{zhang2023data, yan2023multi}. The other two groups, which have gained significant attention in recent years, involve either structural adjustments to the RL framework or modifications to the learning process. These methods leverage advanced safety mechanisms to ensure safety-compliant policies, which is the primary focus of this review \cite{garcia2015comprehensive}. Based on these latter two categories, numerous safe RL methods have been proposed and many have been applied and tailored for solving power systems decision and control problems, such as energy management, economic dispatch, EV charging, voltage control, and stability control.

There exist many review papers on RL in general, such as \cite{li2017deep, arulkumaran2017deep, nguyen2020deep, wells2021explainable, prudencio2023survey}. Additionally, many reviews have focused on the application of RL in power systems, such as \cite{zhang2019deep, glavic2019deep, cao2020reinforcement, chen2022reinforcement, li2023deep}. However, these works primarily address broad aspects of RL and provide little to no discussion on safe RL. In addition, there are several review papers on safe RL in general domains, which provide a comprehensive analysis of safe RL algorithms, historical background, and development trends, such as \cite{garcia2015comprehensive, zhao2023state, wang2023safe, gu2024review}. Before this submission, \cite{li2023research} was the only paper reviewing safe RL in power systems, while \cite{chen2022reinforcement} covered RL applications generally, only briefly noting safety as future work. Our review fills this gap by comprehensively linking RL methods to safety requirements in power systems. After our preprint \cite{su2024review}, \cite{yu2024safe} and \cite{bui2025critical} (also on arXiv as \cite{bui2024critical}) appeared, but our paper provides a more comprehensive overview of safe RL applications in power systems, including a wide range of application domains, practical implementation guidance, and the challenges and potential solutions. We also maintain a GitHub repository to keep the field’s developments up to date \cite{su2025github}. The main contributions are as follows:
\begin{enumerate}
\item{This paper offers a comprehensive review of safe reinforcement learning by presenting its core concepts and definitions, categorizing constraints and environments, comparing RL/DRL with model‐based analytical optimization approaches, surveying existing safe RL methods and benchmarks, and providing a detailed analysis of each method’s distinctive features, limitations, convergence, and optimality. Through this rigorous analytical approach, the paper lays a solid foundation for addressing complex power system challenges and delivers reliable, tailored solutions.}
\item{This review formulates safety requirements in power systems as concrete mathematical constraints grounded in physical principles. By mapping nearly all existing work to specific application domains, our review shifts safety analysis from qualitative descriptions to quantitative, physics-driven constraints. This enables more precise, actionable insights than general safe AI surveys.}
\item{We identify critical challenges such as scalability, distributed implementations, uncertainty, topology changes, user-centric design, real-world deployment, hybrid/fused methods, and LLM-in-the-loop integration, and we propose physics-based solutions tailored to power system complexities. These insights offer a clear roadmap for advancing safe RL in energy applications.}
\end{enumerate}

\begin{figure*}[htb]
    \centering
    \includegraphics[width=18.0cm]{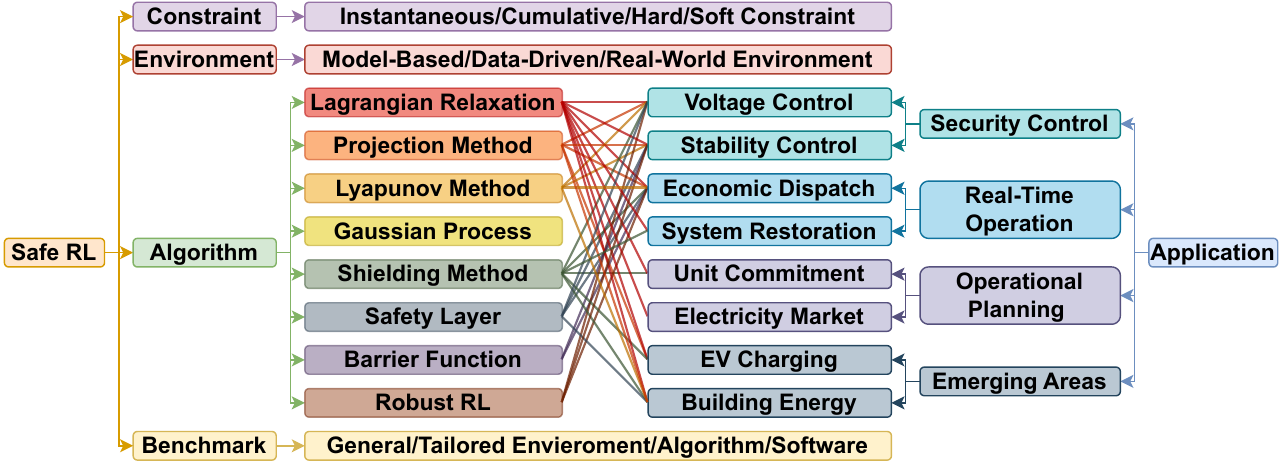}
    \caption{The framework of safe RL in power system applications. Safe RL methods are developed based on standard RL and include learning process modifications, such as Lagrangian relaxation, the Lyapunov method, the GP method, the barrier function method, and RRL, as well as RL structure adjustments, including the projection method, the shielding method, and the safety layer method.}
\label{Fig_framework}
\end{figure*}

The framework of this paper is shown in Fig. \ref{Fig_framework}. The rest of the paper is organized as follows. Section \ref{sec2} introduces the CMDP, constraints, environments, safety, and motivations. Section \ref{sec3} introduces and classifies safe RL methods, with algorithm comparisons and benchmarks. Section \ref{sec4} reviews and analyzes safe RL applications across power system domains. Section \ref{sec5} summarizes existing real-world deployment cases and outlines a roadmap. Section \ref{sec6} discusses challenges and future directions, and Section \ref{sec7} concludes the paper.

\section{Preliminaries of Safe RL in Power Systems}\label{sec2}
\subsection{Constrained Markov Decision Process}\label{sec2Problem}
MDPs are defined by $\mathcal{M} = (\mathcal{S}, \mathcal{A}, \mathcal{P}, R, \rho_0, \gamma)$ which are, respectively, the state space $\mathcal{S}$, action space $\mathcal{A}$, probability distribution $\mathcal{P}$, reward function $R$, initial state $\rho_0 \in {\mathcal{S}}$, and discount factor $\gamma$. When the decision problem fits in an MDP, the objective is to determine the policy $\pi$ that maximizes the expected discounted reward $\mathcal{J}_R^{\pi_\theta}$, i.e. \cite{gu2024review, zhao2023state, sutton2018reinforcement}:
\begin{equation}
\label{Eq_J}
\mathcal{J}_R^{\pi_\theta} = \mathbb{E}_{\tau \sim \pi} \left[ \sum_{t=0}^{\infty} \gamma^t R(\bm s_t, \bm a_t, \bm s_{t+1}) \right]
\end{equation}
where $\tau \sim \pi$ indicates that the distribution over trajectories depends on the policy $\pi$; similarly $\bm s_0 \sim \rho_0$, $\bm a_t \sim \pi(\cdot|\bm s_t)$, $\bm s_{t+1} \sim \mathcal{P}(\cdot|\bm s_t, \bm a_t)$.
Even if the transition probabilities and reward function are fully known, this task is often intractable. However, the approach taken normally is to learn the policy, using some parametrization. 

The CMDP $\mathcal{M}_C = (\mathcal{S}, \mathcal{A}_t, \mathcal{P}, R, \rho_0, \gamma, \mathcal{C})$ extends a standard MDP to handle a common variation where the action space $\mathcal{A}_t$ depends on the state space $\mathcal{S}$, i.e., $\bm s_t \mapsto \mathcal{A}_t$. This accounts for environmental changes that affect which actions are safe or feasible, or for actions with state-dependent costs that must stay below a specified threshold. This occurs in physical systems in which the boundary conditions, the state and the laws of physics limit what is feasible, what would lead to operations that are unsafe and how expensive is a certain agent action. In a nutshell, what differentiates the various instances of CMDP from a conventional MDP is the class of constraints that characterize the action space as a function of the system dynamics and the specific engineering problem and context that define the constraints. 
When feasible actions represent constraint satisfaction, a CMDP can be defined as:
\begin{subequations}
\label{Eq_CMDP}
\begin{gather}
    \max _{\pi_\theta \in \Pi_{S}} \mathcal{J}_{R}^{\pi_\theta} \\
    \text {s.t.} ~~ \bm a_t \text { is feasible }
\end{gather}
\end{subequations}
where “$\bm a_t$ is feasible” means not only that actions respect their upper and lower limits (e.g., SG/RES/ESS outputs or HVAC setpoints) but also that the resulting state $\bm s_t$ lies within safe sets (e.g., voltage, line flow, temperature bounds and stability constraints on voltage, frequency, and rotor angles). Safe RL must therefore generate actions that guarantee both action and state safety, relying on an accurate environment model and a reliable safety evaluation mechanism \cite{krasowski2023provably}. In power systems, enforcing action bounds is straightforward by restricting the RL action space, but ensuring $\bm s_{t+1}$ remains feasible is challenging due to the system’s nonlinear, nonconvex dynamics. This difficulty in finding actions that keep states safe is the primary challenge in training safe RL for power systems.

\subsection{Constraints in the Safe RL}\label{sec2Constraint}
In safe RL, constraints are classified as instantaneous or cumulative based on the time horizon over which the constraints are enforced \cite{liu2021policy, wachi2024survey}. We draw on the definitions of objective functions and constraints from power system optimization and control to provide a detailed introduction.

\subsubsection{Instantaneous Constraints}
Instantaneous constraints require that states or actions meet specific safety conditions at every time step. In power systems, constraints include real‐time power flow limits, BESS restrictions, voltage magnitude bounds, generation capacity limits, stability requirements, EV charging demands, and building energy constraints. In general, these constrained power system optimization problems can be formulated as follows:
\begin{subequations}
\label{Eq_Constraint}
\begin{gather}
    \max _{\pi_\theta \in \Pi_{S}} \mathcal{J}_{R}^{\pi_\theta} \\
    \text {s.t.} ~~ g_j(\bm{s}_t, \bm{a}_t, \bm{s}_{t+1})= 0, ~~ j = 1, \cdots, m \\
    ~ h_k(\bm{s}_t, \bm{a}_t, \bm{s}_{t+1})\le 0, ~~ k = 1, \cdots, n
\end{gather}
\end{subequations}
where the control action must fulfill both the $m$ equality and $n$ inequality constraints. We incorporate the terms $\bm{s}_t$ and $\bm{s}_{t+1}$ within these constraints to represent the time-varying bounds of $\bm{a}_t$. Additionally, the dynamic constraints are also integrated into the aforementioned constraints.

\subsubsection{Cumulative Constraints}
Cumulative constraints require that the sum or average of a specific constraint cost remains within prescribed limits over time. Examples include total revenue and network throughput. Common in robotics \cite{achiam2017constrained}, they can be viewed as flexible alternatives to instantaneous constraints in power systems. For example, \cite{li2022learning} relaxes instantaneous voltage, SoC, and power quality bounds into a discounted cumulative form for distribution network management. Similarly, \cite{zhang2020multi, ye2023safe} apply cumulative formulations. However, these constraints may not fully capture all safety requirements, although they provide some improvement in safety measures and are significantly better than having no constraints. To make the review more self-contained, three cumulative constraints are reviewed. The discounted cumulative constraint is of the form:
\begin{equation}
\label{Eq_discounted_cumulative}
    \mathcal{J}_{h_i}^{\pi_\theta} = \mathbb{E}_{\tau \sim \pi} \left[ \sum_{t=0}^{\infty} \gamma^t h_i(\bm s_t, \bm a_t, \bm s_{t+1}) \right] \leq \varepsilon_i
\end{equation}
where $\varepsilon_i$ is the limit for each cumulative constraint.

The mean valued constraint is of the form:
\begin{equation}
\label{Eq_mean_valued}
    \mathcal{J}_{h_i}^{\pi_\theta} = \mathbb{E}_{\tau \sim \pi} \left[ \frac{1}{t_\text{tot}} \sum_{t=0}^{t_\text{tot}-1} h_i(\bm s_t, \bm a_t, \bm s_{t+1}) \right] \leq \varepsilon_i
\end{equation}
where $t_\text{tot}$ is the total number of time steps in each trajectory.

The third category is probabilistic constraints, which ensure that the probability of cumulative costs meeting a specified threshold $\varepsilon$ remains above a given probability $\zeta$ \cite{liu2021policy}:
\begin{equation}
\label{Eq_probabilistic}
    \mathcal{J}_{h_i}^{\pi_\theta} = \mathbb{P} \left[ \sum_t h_i(\bm s_t, \bm a_t, \bm s_{t+1}) \leq \varepsilon_i \right] \ge \zeta
\end{equation}
where $\zeta_i \in (0, 1) $ is the probability limit.

Some studies use cumulative constraints because they simplify strict instantaneous limits, focus on long‐term safety, and avoid myopic decisions. This allows constrained RL methods to be applied to the power system planning, storage optimization, and load scheduling, where brief local deviations are acceptable if long‐term averages remain safe. Similarly, for certain voltage or line capacity limits, temporary exceedances may be permitted and can be modeled as cumulative constraints. However, in power systems, the majority of constraints must be satisfied at every instant, thus, they are commonly implemented as instantaneous constraints. For example, \cite{yi2023model} utilizes the expected discounted reward, whereas constraints related to branch power flow and security operations are treated as instantaneous constraints.

\subsubsection{Constraints in Power Systems}\label{sec2ConstraintPS}
\begin{figure}[htb]
    \centering
    \includegraphics[width=6.0cm]{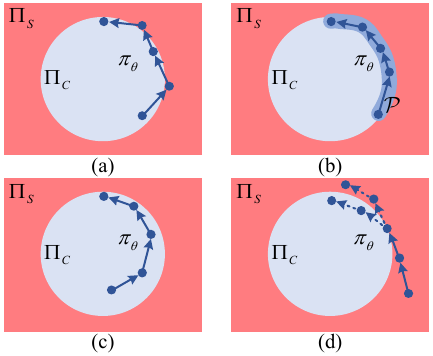}
    \caption{Illustrations of different constraints of safe RL. \textbf{(a)}: Cumulative constraints \eqref{Eq_discounted_cumulative}-\eqref{Eq_mean_valued}. \textbf{(b)}: Probabilistic constraints \eqref{Eq_probabilistic}. \textbf{(c)}: Instantaneous constraints and hard constraints. \textbf{(d)}: Soft constraint, where the final $\pi_\theta$ may be either safe or unsafe.}
\label{Fig_constraint}
\end{figure}

In power systems, constraints are classified as instantaneous or cumulative and as hard or soft, depending on the time horizon, strictness, and the selected safe RL method. Typically, bus balance equations, equipment limits, ESS capacities, certain voltage amplitudes, and some stability constraints are considered hard constraints. Safe RL algorithms that guarantee hard‐constraint satisfaction include projection \eqref{sec3b}, Lyapunov \eqref{sec3c}, shielding \eqref{sec3e}, and safety layer \eqref{sec3f} methods. \cite{zhang2023data} embeds safe policy projection in RL to prevent any physical‐constraint violations. 
Due to discrepancies between simulation models and real-world systems, various uncertainties of RESs and loads, and algorithmic shortcomings, even if constraints are theoretically satisfied, they may not be guaranteed in real-world deployment. To address this, GP \eqref{sec3d} and RRL \eqref{sec3h} methods have been proposed using the probabilistic/chance constraint \eqref{Eq_probabilistic}. However, their application in power systems remains underexplored. A more common approach is to use RRL to enhance adaptability under uncertainty \cite{yi2023model, liu2020two}.
Furthermore, by design, some safe RL methods can only encourage but not guarantee constraint satisfaction. Such methods include Lagrangian relaxation \eqref{sec3a}, barrier function \eqref{sec3g}, and penalty functions. For example, \cite{yan2023multi} uses the voltage constraint metric $\mathcal{J}_{h_i}^{\pi_\theta} = \sum_{i \in \mathcal{N}} \max \left\{ |\bm v_{i,t} - 1| - 0.05|, 0 \right\}$ and Lagrangian relaxation for voltage control, which cannot guarantee absolute adherence to voltage constraints, thus classifying it as a soft constraint. 
For some constraints, like user satisfaction with EV charging and voltage at certain nodes, the goal is to approach standard values as closely as possible, making them inherently soft constraints. The illustrations of different constraints of safe RL are shown in Fig. \ref{Fig_constraint}.

\subsection{Environments in the Safe RL}\label{sec2Environment}

In safe RL, the environment represents the power system tailored to a specific problem, including its state information, dynamics, and constraints. It simulates state transitions in response to the agent's actions and provides feedback through rewards and constraint costs.

\subsubsection{Classification of Environments}
The environment is generally categorized into three types: real-world environments \cite{lin2023reinforcement}, model-based simulation environments \cite{vu2023multi}, and data-driven simulation environments \cite{zhao2022deep}. The adoption of the real-world environment is relatively rare. Examples are mainly found in low-risk scenarios, such as building energy management systems (Section \ref{sec5}). In these cases, actions can be implemented on real buildings with manageable safety and minimal potential risks \cite{lin2023reinforcement}. In contrast, applications targeting power grids predominantly use the other two types of environments, as real-world deployment faces significant challenges, especially in terms of safety. Additionally, some safe RL methods struggle to enforce constraints during the early stages of training, necessitating pre-training in model-based or data-driven environments.

Model-based simulation environments rely heavily on the model fidelity, as inaccuracies in system dynamics can lead to unsafe actions \cite{yan2022hybrid}. However, safe RL methods, especially those incorporating robustness, can to some extent ensure safety and handle uncertainty and inaccuracy \cite{yi2023model}.

Data-driven simulation environments can be broadly divided into two categories: (1) Offline RL \cite{prudencio2023survey}, which learns policies directly from datasets, and (2) RL that interacts with surrogate models, such as well-trained NN based on data \cite{cao2022model, su2025safe}. The datasets for these methods may originate from direct measurements of real systems or synthetic data generated by simulation models \cite{yi2023model}.

\subsubsection{Discussions on Environments}
The difference between conventional RL and safe RL in terms of the environment lies in how constraints and safety considerations are incorporated during the learning process. In conventional RL, the environment is typically used to explore a wide range of state-action pairs with minimal restrictions, even if it means encountering unsafe states. In contrast, safe RL explicitly incorporates physics-based safety constraints within the environment, avoiding unsafe actions and states during training and execution through enforcement or penalty mechanisms. As a result, while conventional RL prioritizes exploration and reward maximization, safe RL focuses on constraint satisfaction and risk mitigation, which is essential in high-stakes domains like power systems where safety failures can have severe consequences.

\subsection{Power System Security and Safe RL}\label{sec2Safety}


In power systems, security mainly encompasses steady-state security and dynamic stability. Steady-state security ensures that the system operates within all physical and operational constraints, focusing on factors such as transmission capacity, voltage margins, power flow distribution, and component limitations. Modeling and solution methods often rely on steady-state power flow calculations and constraint optimization \cite{kundur1994power}. Thus, the security of safe RL in this context can be defined as ensuring compliance with operational constraints during steady-state operations.

Dynamic stability focuses on the system's ability to return to a stable state following large disturbances, such as faults, line outages, or generator disconnections. It includes electromagnetic transients, small disturbance stability, and large disturbance stability, all of which are tied to the system's dynamic behavior and controller performance. Modeling and solution methods are based on time-domain simulations or energy function analysis to evaluate system dynamics \cite{kundur1994power}. In the context of safe RL, dynamic stability can be defined as ensuring that the agent's actions do not compromise the system's stability under disturbances, addressing aspects such as frequency stability, voltage stability, and rotor angle stability.

In addition, grid security can be extended to include robustness under worst-case scenarios, contingencies and probabilistic extremes, ensuring the system remains safe under adverse conditions or rare high-impact events. Robustness in worst-case scenarios focuses on keeping the system operational under the most unfavorable disturbances or uncertainties. This is often achieved through robust optimization or adversarial training, such as using RRL in safe RL. Probabilistic robustness focuses on minimizing the likelihood of extreme violations by integrating stochastic modeling or risk-based penalties into decision-making.

\subsection{Motivations for Safe RL: A Comparative Perspective}\label{sec2Comparison}
Model-based analytical optimization methods rely on physical modeling and mathematical equations (such as differential and algebraic equations) to describe system dynamics and perform computations, including steady-state analysis (e.g., PF calculations), dynamic analysis (e.g., time-domain simulation), and OPF \cite{kundur2004definition, kundur1994power}. The advantages of these methods include high reliability, strong interpretability, and applicability to known systems. However, as system complexity increases, for example due to the integration of new devices such as inverters, greater uncertainty, and rapidly fluctuating RESs, modeling becomes more challenging and the model may become unreliable. Additionally, computational challenges arise when addressing complex problems in large-scale power systems \cite{prostejovsky2016distribution}. Furthermore, some problems may lack explicit models, such as bidding behaviors in an electricity market, making data-driven approaches particularly crucial \cite{tang2022multi}.

Conventional RL primarily focuses on maximizing rewards, often without explicitly addressing constraints. Some studies incorporate constraints by introducing penalty terms into the reward function, forming a reward-based optimization problem \cite{zhao2022deep}. However, this approach has limitations. If the penalty weight $\lambda$ is set too low, constraints may be ignored. Conversely, if the penalty is too large, RL may become overly conservative, avoiding exploration. In such cases, the RL policy may oscillate near constraint boundaries, occasionally violating them, as it only optimizes the overall reward and constraint cost rather than enforcing strict constraint satisfaction. To address this issue, safe RL has been proposed, which handles the objective function and constraints jointly but explicitly. During the agent's exploration, the action space is restricted using physical models, expert knowledge, or constraint rules, ensuring that exploration remains within or eventually returns to the safe feasible region \cite{gu2024review, garcia2015comprehensive}.

\begin{figure}[htb]
    \centering
    \includegraphics[width=7.0cm]{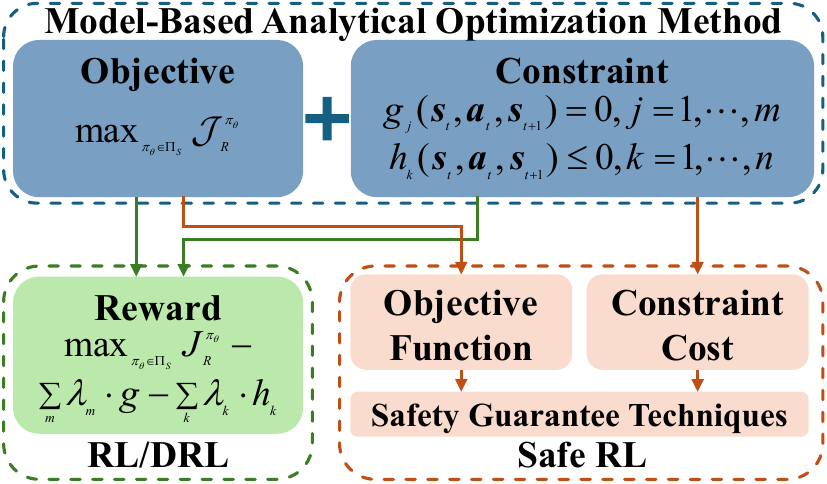}
    \caption{Comparison between model-based analytical optimization methods, RL/DRL, and safe RL. Model-based analytical optimization methods can accurately model and solve objective functions and physical constraints, but face challenges related to parameter inaccuracies and high computational costs. Conventional RL optimizes the sum of the objective function and constraints, but may fail to satisfy constraints. In contrast, safe RL can handle constraints separately, promoting or ensuring their satisfaction.}
\label{Fig_Comparison_with_Model}
\end{figure}

Conventional RL and safe RL, unlike model-based analytical optimization methods, do not require pre-built physical models but instead learn optimal policies directly from data, making them highly adaptable to unknown or changing environments. In scenarios with fluctuating RESs or variable loads, where uncertainty distributions may be unknown or nonstationary, RL can use trial-and-error and real-time feedback to adjust its policy and maintain performance \cite{lockwood2022review}. The comparison of how model-based analytical optimization methods, RL/DRL, and safe RL handle objective functions and constraints is illustrated in Fig. \ref{Fig_Comparison_with_Model}, while the feature comparison of model-based methods and safe RL is summarized in Table \ref{Table_Comparison_with_Model} \cite{kundur1994power, li2023deep, gu2024review, glanois2024survey, zanon2020safe}.

\renewcommand{\arraystretch}{1.1}
\begin{table*}[htb]
\caption{Comparison Between Model-Based Analytical Optimization Methods and Safe RL}
\label{Table_Comparison_with_Model}
\centering
\begin{tabular}{|>{\centering\arraybackslash}m{1.6cm} >{\raggedright\arraybackslash}m{8.0cm} >{\raggedright\arraybackslash}m{7.2cm}|}
\toprule
Dimension & Model-Based Analytical Optimization Methods & Safe RL \\
\midrule
Dependency & Physical models + precise parameter estimation & Large and high-quality data \\
\hline
Efficiency & Heavy online computation (dynamic analysis + large optimization) & Intensive offline training; fast online inference \\
\hline
Safety & Theoretical constraint guarantees after convergence & Safety guarantee depends on the specific algorithm \\
\hline
Interpretability & Strong (physical + math foundations $\rightarrow$ simple debugging) & Black-box; some interpretable/provably convergent variants \\
\hline
Robustness & Sensitive to model/parameter errors; requires uncertainty assumptions & Adaptable to uncertainty; some methods are robust\\
\hline
Challenges & Accurate models; significant compute resources; uncertainty and randomness; struggles with non-analytic problems & Data quality/availability issues; topology change; deployment safety/interpretability issues\\
\bottomrule
\end{tabular}
\end{table*}

\section{Safe Reinforcement Learning Methods}\label{sec3}
Safe RL is often formulated as a CMDP problem, where the objective is to maximize the reward of agents while ensuring that the agents satisfy safety constraints \cite{altman1999constrained, gu2024review}. Based on the definitions of $\mathcal{J}_{R}^{\pi_\theta}$ and $\mathcal{J}_{h_i}^{\pi_\theta}$ in Section \ref{sec2}, the unified CMDP formulation can be written as:
\begin{equation}
\label{Eq_CMDP_Standard}
    \max _{\pi_\theta \in \Pi_{S}} \mathcal{J}_{R}^{\pi_\theta}, ~~ \text {s.t.} ~~ \mathcal{J}_{h_i}^{\pi_\theta}, ~~ i = 1, \cdots, n
\end{equation}
The safe RL techniques introduced in this section are all based on \eqref{Eq_CMDP_Standard}. The primary difference between the various safe RL methods lies in how they handle constraints.

\begin{figure}[htb]
    \centering
    \includegraphics[width=8.8cm]{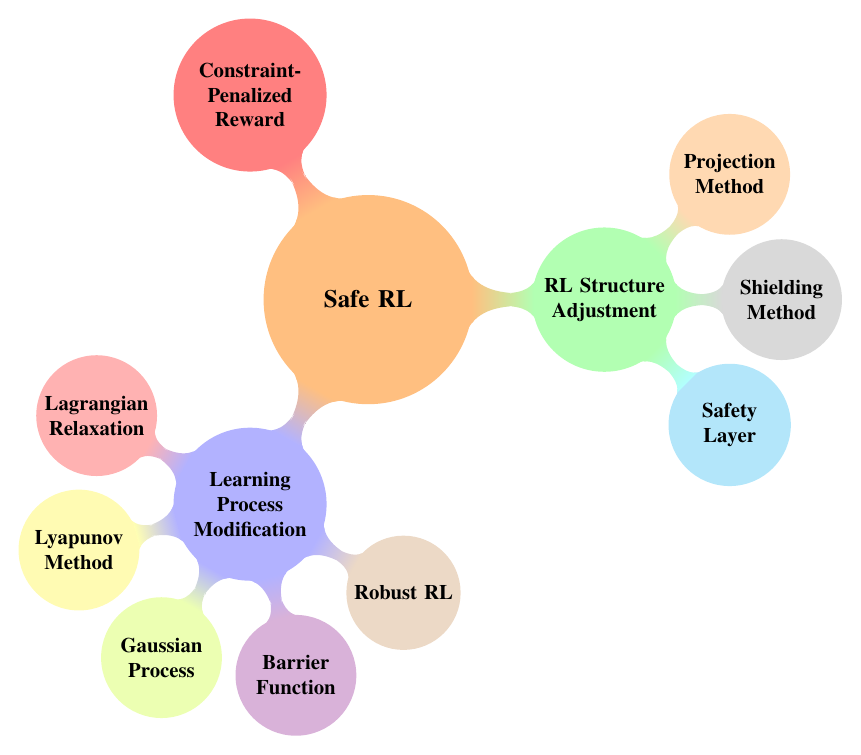}
    \caption{Classification of safe RL techniques. Safe RL methods can be broadly categorized into three groups based on their mechanisms: (1) Constraint-penalized reward, where constraints are incorporated into the reward function; (2) Learning process modification, where safety constraints or metrics are directly integrated into the policy iteration or gradient update process; and (3) RL structure adjustment, which explicitly introduces structural constraints or modules into the policy architecture to ensure safety during execution or updates.}
\label{Fig_Safe_RL_Classification}
\end{figure}

This section categorizes safe RL methods based on the techniques used to ensure constraint satisfaction, and provides detailed introductions to their fundamentals, characteristics, and benchmarks. The specific classification is shown in Fig. \ref{Fig_Safe_RL_Classification}.
In \ref{Fig_Safe_RL_Classification}, the techniques are categorized into three groups. The first group focuses on incorporating safety factors into the reward function to penalize violations. While this approach is straightforward, it often struggles to effectively enforce the physics-hard constraints of power systems. The second group involves learning process modifications, where safety constraints or metrics are directly integrated into the policy iteration or gradient update process. This ensures that safety considerations are embedded in the policy itself and includes methods such as Lagrangian relaxation \eqref{sec3a}, the Lyapunov method \eqref{sec3c}, the GP method \eqref{sec3d}, the barrier function method \eqref{sec3g}, and RRL \eqref{sec3h}. The third group focuses on RL structure adjustments, explicitly introducing structural constraints or modules into the policy framework to ensure safety during execution or updates. Examples include the projection method \eqref{sec3b}, the shielding method \eqref{sec3e}, and the safety layer method \eqref{sec3f}.

While extensive theoretical research exists on safe RL algorithms, this paper focuses on their practical application in power systems. It highlights the core concepts and representative algorithms of each approach instead of reiterating the broader theoretical developments. For a general introduction to safe RL, please refer to references \cite{gu2024review, garcia2015comprehensive, zhao2023state, wang2023safe}.

\subsection{Lagrangian Relaxation / Primal-Dual Method}\label{sec3a}
Lagrangian relaxation, also known as the primal-dual method, is the most common technique in safe RL. The key idea of this method is to transform the CMDP problem into an unconstrained dual problem. This is achieved by employing adaptive Lagrange multipliers to penalize constraints \cite{chow2018risk}:
\begin{subequations}
\label{Eq_Lagrangian_relaxation}
\begin{alignat}{2}
&\textbf{Instantaneous:} \notag \\
&\min_{\lambda_i \geq 0}\max_{\theta} \mathcal{L}(\lambda_i, \theta) =\min_{\lambda_i \geq 0}\max_{\theta} \left[J_{R}^{\pi_{\theta}}-\sum_i\lambda_i \cdot h_i \right]\\
&\textbf{Cumulative:} \notag \\
&\min_{\lambda_i \geq 0}\max_{\theta} \mathcal{L}(\lambda_i, \theta) =\min_{\lambda_i \geq 0}\max_{\theta} \left[J_{R}^{\pi_{\theta}}-\sum_i\lambda_i \cdot\left(J_{h_i}^{\pi_{\theta}}-\varepsilon_i\right)\right]
\end{alignat}
\end{subequations}

\begin{figure}[htb]
    \centering
    \includegraphics[width=6.2cm]{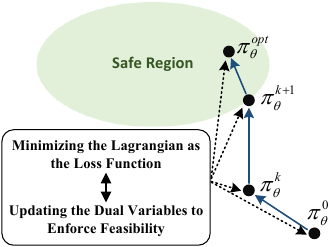}
    \caption{Policy update based on the Lagrangian relaxation method. The agent starts from an initial policy $\pi_\theta^0$ and iteratively updates the policy by minimizing the Lagrangian, while dual variables are updated to gradually enforce constraint satisfaction. Although early iterations (e.g., $\pi_\theta^k$) may fall outside the safe region, the method aims to converge to an approximately feasible policy $\pi_\theta^{\text{opt}}$ within the safe region.}
\label{Fig_Lagrangian}
\end{figure}

The solution of \eqref{Eq_Lagrangian_relaxation} relies on Danskin's theorem and convex analysis \cite{bertsekas2015convex}. 
Due to its straightforward implementation and compatibility with both on-policy and off-policy methods, Lagrangian relaxation has been widely adopted in RL. It has been integrated with various algorithms, leading to many variants such as DDPG-Lag, PPO-Lag, TRPO-Lag, TD3-Lag, SAC-Lag, MAPPO, RCPO, PDO, TRPO-PID, CPPO-PID, DDPG-PID, TD3-PID, SAC-PID \cite{ray2019benchmarking, gu2021multi, chow2018risk, stooke2020responsive}. The policy updates based on the Lagrangian relaxation method are shown in Fig. \ref{Fig_Lagrangian}.

Lagrangian relaxation is the most commonly used approach in power systems because it easily handles a variety of constraints and can be applied across diverse domains. Based on instantaneous or hard constraints, \cite{wu2023constrained} utilizes a primal-dual approach to optimize power generation and BESS charging and discharging actions in a multi-stage real-time stochastic dynamic OPF. Additionally, \cite{wang2019safe} applies constrained SAC to the Volt-VAR control problem by synergistically combining the merits of the maximum-entropy framework, the method of multipliers, a device-decoupled NN structure, and an ordinal encoding scheme. Furthermore, \cite{wu2023network} employs constrained RL for the predictive control of OPF, paired with EV charging control. 
On the other hand, based on cumulative or soft constraints, \cite{yan2022hybrid} approximates the actor gradients by solving the Karush-Kuhn-Tucker conditions of the Lagrangian, instead of constructing reward critic networks and cost critic networks through interactions with the environment. Then, the interior point method is incorporated to derive the parameter updating rule for the DRL agent. Similarly, \cite{sayed2023online} develops a soft-constraint enforcement method to adaptively encourage the control policy in the safety direction with nonconservative control actions and find decisions with near-zero degrees of constraint violations.
However, the Lagrangian relaxation method cannot guarantee strict constraint satisfaction, requires fine-tuning of Lagrange multipliers, and may oscillate near the constraints' boundaries.

\subsection{Projection Method / Trust Region Method}\label{sec3b}
\begin{figure}[htb]
    \centering
    \includegraphics[width=5.5cm]{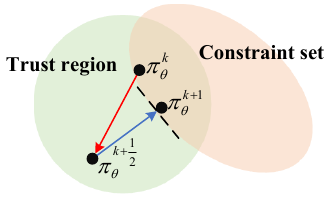}
    \caption{Update procedures for PCPO. In step one (red arrow), PCPO follows the reward improvement direction in the trust region (light green). In step two (blue arrow), PCPO projects the policy onto the constraint set (light orange).}
\label{Fig_PCPO}
\end{figure}

The projection method ensures constraint satisfaction at every step and enhances performance by updating the trust region policy gradient and projecting the policy into a safe feasible set during each iteration \cite{ding2020natural}. TRPO enforces a KL-divergence trust‐region constraint on policy updates. CPO is developed based on TRPO, and both belong to the category of TRM \cite{achiam2017constrained}. A series of projection-based methods have subsequently been developed from this foundation. Typical projection methods include PCPO \cite{yang2020projection}, FOCOPS \cite{zhang2020first}, CUP \cite{yang2022constrained}, and MACPO\cite{gu2021multi}. Among these, PCPO follows a two-step process: it first performs a local reward update and then projects the policy onto the constraint set to correct any constraint violations, as depicted in Fig. \ref{Fig_PCPO}.

In the power system domain, projection methods have also seen widespread application. For instance, \cite{zhang2023data} introduced a projection-embedded MA-DRL algorithm that smoothly and effectively restricts the DRL agent action space to prevent any violations of physical constraints, thereby achieving decentralized optimal control of distribution grids with a guaranteed 100\% safety rate. Additionally, in the area of EV charging problems, \cite{jiang2021data} utilizes a penalty function to penalize the NN output if it exceeds the action space and uses a projection operator to avoid incurring a negative reward when no EV is occupying the charging bay. In addition, \cite{wang2019volt} employs CPO for Volt-VAR control to minimize the total operation costs while satisfying the physical operation constraints.
However, TRMs, primarily based on TRPO or PPO, are not easily integrated with other RL types and are computationally intensive in high dimensions, limiting their suitability for large-scale safe RL problems \cite{wang2023safe}. Similarly, projection methods guarantee strict constraint satisfaction at each step but require accurate feasible‐region estimation and a suitable projection operator. In addition, in power system applications, many projection methods are implemented using projection operations derived from system physical rules.

\subsection{Lyapunov Method}\label{sec3c}
\begin{figure}[htb]
    \centering
    \includegraphics[width=7.2cm]{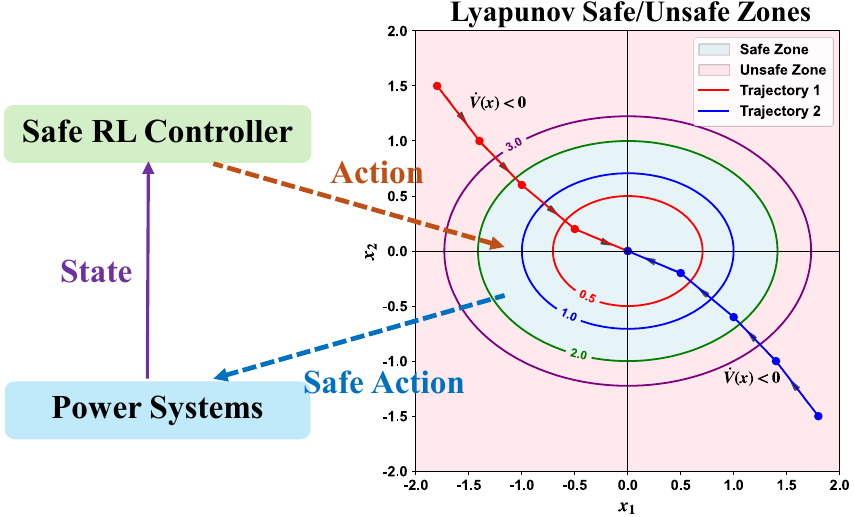}
    \caption{Lyapunov-based safe RL control. Contours represent level sets of $V(x)$, where $\dot{V}(x) < 0$ indicates decreasing system energy and movement toward a stable equilibrium. The green region denotes the safe zone where the Lyapunov condition is satisfied, while the pink region represents unsafe states with $\dot{V}(x) \geq 0$. The controller receives the system state from the power system and selects actions accordingly. Safe actions keep the state trajectory within or steer it back into the safe zone, ensuring stability over time.}
\label{Fig_Lyapunov}
\end{figure}

Lyapunov functions, widely used in control engineering for controller design \cite{sepulchre2012constructive}, were first applied to safe RL in \cite{perkins2002lyapunov}. They are used to constrain the action space, ensuring the safety of all policies while maintaining agent performance. Additionally, a set of control laws is constructed under the assumption that the Lyapunov domain knowledge is known beforehand \cite{gu2024review}. The application of the Lyapunov method in power systems is limited because it requires prior knowledge of a Lyapunov function and is difficult to handle multiple complex constraints. If the model of environmental dynamics is unknown, identifying a suitable Lyapunov function can be challenging. For example, \cite{cui2022reinforcement} integrates a Lyapunov function into the structural properties of primary frequency controllers, guaranteeing local asymptotic stability over a large set of states. Additionally, \cite{cui2022decentralized} utilizes Lyapunov theory to design the controller that satisfies specific Lipschitz constraints for decentralized inverter-based voltage control. In addition, \cite{shi2022stability} utilizes a stability-constrained RL method for real-time voltage control in distribution grids, providing a formal voltage stability guarantee using the Lyapunov function. A visualization of a Lyapunov-based safe RL control is shown in Fig. \ref{Fig_Lyapunov}.

\subsection{Gaussian Process Method}\label{sec3d}
\begin{figure}[htb]
    \centering
    \includegraphics[width=6.8cm]{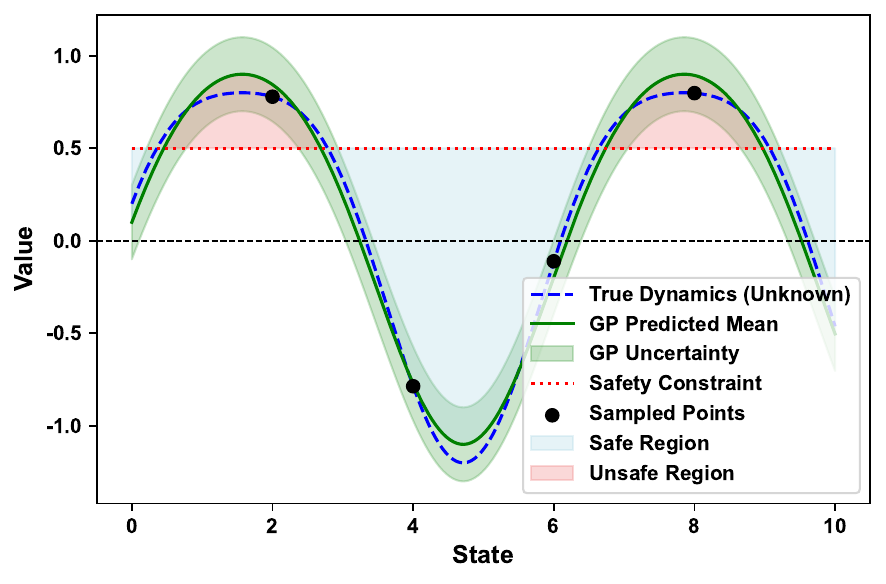}
    \caption{GP-based safe RL with uncertainty-aware safety assessment. The true dynamics (dashed blue) are approximated by the GP predicted mean (solid green), with shaded areas representing uncertainty bounds. A red dotted line marks the safety constraint threshold. The pink shaded region highlights the unsafe area where the upper confidence bound exceeds the safety constraint, while the blue region indicates safe estimates. Sampled data points (black dots) are used to update the GP model. This approach allows the agent to avoid unsafe actions with high probability, enabling probabilistic safety guarantees in safe RL.}
\label{Fig_GP}
\end{figure}

GP \cite{williams2006gaussian} is widely utilized in numerous approaches to estimate uncertainty and identify unsafe areas. Consequently, assessments based on GP can be incorporated into the learning process to enhance agent safety \cite{akametalu2014reachability}. The GP method ensures that the rewards of decisions during exploration always meet the predefined safety threshold. GP-based safe RL algorithms include SafeOpt \cite{sui2015safe}, SafeMDP \cite{turchetta2016safe}, PILCO \cite{cowen2022samba, deisenroth2011pilco}, etc. For example, SafeOpt uses a GP to model the unknown objective function, leveraging the posterior mean for prediction and confidence intervals to quantify uncertainty. It exploits Lipschitz continuity to expand the safe set, enabling efficient exploration and optimization while adhering to safety constraints \cite{sui2015safe}. The application of the GP method-based safe RL in power systems is limited, meriting further research to adequately address the various uncertainties inherent in power systems. The potential disadvantage of GP methods is their high computational complexity and limited scalability as problem dimensionality grows, along with sensitivity to kernel selection and hyperparameter tuning \cite{wang2023safe}. A visualization of GP-based safe RL with uncertainty assessment is shown in Fig. \ref{Fig_GP}.

\subsection{Shielding Method}\label{sec3e}
\begin{figure}[htb]
    \centering
    \includegraphics[width=6.8cm]{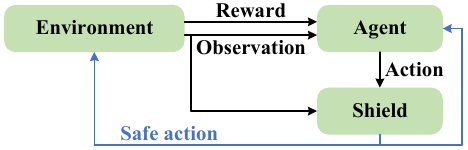}
    \caption{The framework of shielded RL. The shield monitors the actions selected by the learning agent and corrects them if and only if the chosen action is unsafe. The correctness of the system's execution against a given specification is assured during both the learning and controller execution phases, regardless of the convergence speed of the learning process.}
\label{Fig_Shielding}
\end{figure}

In \cite{alshiekh2018safe}, the shield is introduced for the first time in RL. Shielding methods explicitly enforce safety by pre-defining rules to prevent unsafe actions, ensuring strict constraint satisfaction and excellent real-time applicability. This shield is explicitly computed in advance, based on the safety component of the system specification and an abstraction of the dynamics of the agent's environment. It guarantees safety with minimal interference, implying that the shield restricts the agent's actions only as much as necessary, prohibiting actions that could jeopardize the safe behavior of the system. The shielded RL is shown in Fig. \ref{Fig_Shielding}.

Shielding is a method that enforces constraint satisfaction, making it highly suitable for power system problems with hard constraints. For instance, in \cite{chen2022physics}, actions that would lead to dangerous states, such as the SoC of BESSs being fully charged or depleted, are substituted by the shielding mechanism with safe actions to maintain system stability. Additionally, \cite{zhang2022residual} combines a correction model adapted from gradient descent with the prediction model as a post-posed shielding mechanism to enforce safe actions in computer room air conditioning unit control problems. In addition, in unit commitment scheduling, \cite{ajagekar2022deep} utilizes action space clipping to ensure that uncertainty estimates are reasonable and within appropriate bounds obtained from historical data. A potential drawback of shielding methods is the challenge of identifying safe, feasible actions from infeasible ones, as this requires detailed knowledge of the system dynamics and constraints. As a result, these methods can be especially challenging to apply in complex or uncertain systems or specific control scenarios, significantly limiting scalability and flexibility in practical applications \cite{wang2023safe}.

\subsection{Safety Layer Method}\label{sec3f}
\begin{figure}[htb]
    \centering
    \includegraphics[width=6.5cm]{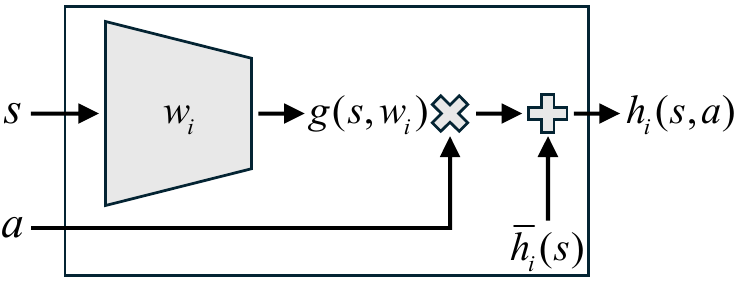}
    \caption{Safety layer. Each safety signal $h_i(s,a)$ is approximated with a linear model with respect to $a$, whose coefficients are features of $s$, extracted with a NN.}
\label{Fig_Safetylayer}
\end{figure}

Both the safety layer and the shielding method restrict actions within a safe region. However, the essential distinction lies in their approach to ensuring safety: the shielding method computes the shield rules prior to training, based on system safety specifications and an abstracted model of the environment dynamics. As a result, before the RL agent selects an action, the shield proactively filters out potentially unsafe actions. In contrast, the safety layer allows the RL agent to first generate an action, and then adjust it to a safe region through a safety layer. In other words, it is a reactive safety mechanism, requiring the solution of an optimization problem at each step during training or execution to ensure the action satisfies safety constraints. The safety layer method, first proposed in \cite{dalal2018safe} for continuous action spaces in RL, emphasizes maintaining zero-constraint violations throughout the learning process. It expresses safety constraints as linear functions of action through a first-order approximation. Assuming that at most one constraint is violated at any time, an analytical solution to the safety layer optimization problem can be directly obtained. The linearization transition equation and visualization of the safety layer are shown in \eqref{Eq_Safetylayer} and Fig. \ref{Fig_Safetylayer}, respectively.
\begin{equation}
\label{Eq_Safetylayer}
\overline{h}_i(s_{t+1}) \triangleq h_i(s_t, a_t) \approx \overline{h}_i(s_t) + g(s_t; w_i)^\top a_t
\end{equation}
where $w_i$ are weights of NN; $g(s_t; w_i)$ denotes first-order approximation to $h_i(s_t, a_t)$ with respect to $a_t$.

Safety-layer methods ensure real-time safety and offer modular integration independent of specific RL algorithms, leading to their widespread application in power systems. For example, in economic dispatch, \cite{yi2023real} proposes a hybrid knowledge-data-driven safety layer to convert unsafe actions into the safety region, which is accelerated by a security-constrained linear projection model. Additionally, in Volt-VAR control, \cite{gao2022model} adds a safety layer to the policy NN to enhance operational constraint satisfaction during both the initial exploration phase and the convergence phase. In addition, \cite{du2022deep} uses action clipping, reward shaping, and expert demonstrations to ensure safe exploration and accelerate the training process during the online training stage for the assist service restoration problem.
However, the linear approximation in the safety layer might not accurately capture the complex dynamics of highly non-linear systems, and iterating at every time step could introduce a significant computational burden. Moreover, assuming only one constraint at a time may not be valid in complex environments where multiple safety constraints are concurrently active. In addition, methods based on complex optimization can further increase computation per step.

\subsection{Barrier Function Method}\label{sec3g}
\begin{figure}[htb]
    \centering
    \includegraphics[width=6.5cm]{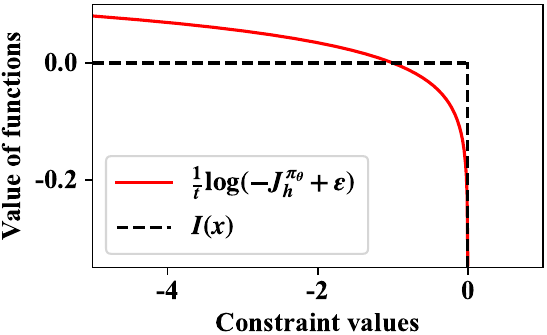}
    \caption{Barrier function. The solid red line represents the logarithm barrier function $\log(-J_{h}^{\pi_\theta} + \varepsilon)/t$, which is a differentiable approximation of the indicator function $I(x)$.}
\label{Fig_Barrier}
\end{figure}

The barrier function method involves adding a barrier function penalty term to the original objective function. When the system state approaches the safety boundary, the value of the constructed barrier function tends to infinity, thereby ensuring that the state remains within the safe boundary \cite{wang2023enforcing}. The most typical barrier function method is IPO, which augments the objective with logarithmic barrier functions, drawing inspiration from the interior-point method \cite{liu2020ipo}:
\begin{subequations}
\begin{alignat}{2}
\textbf{Instantaneous:}&~ \max_\theta J_R^{\pi_\theta} + \sum_{i} \frac{1}{t_i} \log(-h_i)\\
\textbf{Cumulative:}&~ \max_\theta J_R^{\pi_\theta} + \sum_{i} \frac{1}{t_i} \log(-J_{h_i}^{\pi_\theta} + \varepsilon_i)
\end{alignat}
\end{subequations}
where $t_i$ is a hyperparameter for $h_i$. The illustration of IPO is shown in Fig. \ref{Fig_Barrier}.

Barrier function method and IPO have been widely applied in power systems to ensure the safety of constraints. For example, \cite{ye2023safe} utilizes IPO to ensure the fulfillment of distribution network constraints without the need for designated penalty terms and the associated tuning of penalty factors, or repeatedly solving optimization problems for action rectification. Additionally, \cite{cui2023online} uses IPO to facilitate desirable learning behavior towards constraint satisfaction and policy improvement simultaneously during online preventive control for transmission overload relief. In addition, \cite{vu2021barrier} proposes a safe RL method for emergency load shedding in power systems, where the reward function includes a barrier function that approaches negative infinity as the system state approaches safety bounds. However, the accurate formulation and tuning of barrier functions necessitate knowledge of system dynamics, which can be challenging in complex environments. Additionally, the barrier function method tends to be overly conservative in optimization problems, making it suitable for scenarios with high safety requirements. Moreover, when environmental uncertainty is low and constraint boundaries are clearly defined and structurally simple, it is possible to construct barrier functions that closely follow the constraint boundaries, significantly reducing conservativeness.

\subsection{Robust Reinforcement Learning}\label{sec3h}
One of the challenges in RL is generalization under uncertainties not seen during training. To address this, RRL frameworks have been developed, focusing on enhancing the reliability and robustness of RL agents for the worst-case scenarios \cite{zanon2020safe, li2021safe}. Two notable approaches in this context are chance-constrained RRL and constrained game-theoretic RL. It is important to note that RRL is not universally recognized as a safe RL algorithm in other fields. However, due to the significant uncertainties in power systems, RRL is employed to enhance control robustness and is reviewed here.

\subsubsection{Chance-Constrained RRL}
Chance-constrained RRL, in particular, focuses on ensuring that policies perform well under uncertain conditions by incorporating probabilistic constraints into the learning process \cite{kordabad2022safe}. In this framework, the goal is not just to maximize expected rewards but to do so while ensuring that the probability of undesirable outcomes (e.g., safety violations) remains below a specified threshold \cite{pfrommer2022safe}. This is particularly important in scenarios where safety and reliability are critical, such as autonomous driving or robotics \cite{coulson2021distributionally}.
The general form can be expressed as:
\begin{subequations}
\begin{gather}
\max_{\pi} \mathcal{J}_R^{\pi_\theta} \\ 
\text {s.t.} ~~ \mathbb{P} \left[\max_{1\leq i\leq n} h_i(\bm{s}_t, \bm{a}_t, \bm{s}_{t+1}) \leq \varepsilon_i \right] \ge \zeta, \forall t \in \mathcal{T}
\end{gather}
\end{subequations}

\subsubsection{Constrained Game-Theoretic RL}
Constrained game-theoretic RL is a framework that models the interaction between the RL agent and its environment as a game, specifically focusing on scenarios where there are constraints that the agent must respect during the learning and decision-making processes \cite{yu2021robust}. The objective is to maximize the agent's rewards while minimizing the possible losses or costs, considering the worst-case scenarios posed by adversaries' actions or environmental uncertainties $a_t^{\text{adv}}$ \cite{rajeswaran2020game}. Here's a more accurate representation using a minimax optimization framework \cite{yu2021robust}:
\begin{subequations}
\begin{gather}
\max_{\pi_\theta} \min_{\pi_\theta^{\text{adv}}} \mathbb{E}_{\tau \sim \pi} \left[\sum_{t=0}^{\infty} \gamma^t R(s_t, a_t, a_t^{\text{adv}}, s_{t+1})\right] \\
\text {s.t.}~~ h_i(s_t, a_t, a_t^{\text{adv}}, s_{t+1}) \leq 0, \forall t \in \mathcal{T} \label{Eq_adv_h}
\end{gather}
\end{subequations}
where $\pi_\theta^{\text{adv}}$ denotes the policy of adversary; \eqref{Eq_adv_h} represents the game-theoretic or environmental constraints, incorporating both the agent’s and the adversary’s policies.

One of the key benefits of constrained game-theoretic RL is its ability to manage both competitive and cooperative interactions in complex environments. This makes it well-suited for applications such as strategic games, mobile edge computing \cite{asheralieva2019hierarchical}, and coordination in robotic teams \cite{tessler2019action}.

RRL is applied in power systems to ensure control strategies remain effective under uncertainties. For example, \cite{yi2023model} employs adversarial safe RL to address model inaccuracies in virtual power plants without relying on precise environmental models. In sequential OPF problems, \cite{yi2023real} utilizes a bi-level robust optimization approach to improve the Q-network’s robustness against uncertainties. Similarly, \cite{liu2020two} develops an adversarial RL algorithm for inverter-based Volt-VAR control, training an offline agent capable of handling model mismatches.
Game-theoretic RL has also been explored for multistage games, optimizing attack-defense strategies and internal trading price dynamics \cite{ni2019multistage, guo2021reinforcement, bui2022dynamic, surani2024competitive}. Meanwhile, chance-constrained RRL methods \cite{coulson2021distributionally, pfrommer2022safe, peng2022model} and robust optimization techniques \cite{hassan2018optimal, ciftci2019data, liang2024joint} have demonstrated potential in power flow control. However, significant opportunities remain for applying these approaches to power system control and optimization. 


\subsection{Constraint Satisfaction Levels: Soft, Hard, Probabilistic}\label{sec3constraint}
In this paper, we consider the Lagrangian relaxation method, the barrier function method, and the GP method as capable of only satisfying soft constraints. Specifically, the Lagrangian relaxation method incorporates penalty terms to guide constraint satisfaction, but it cannot guarantee strict adherence, and minor violations often occur in practice. The barrier function method gradually approaches the constraint boundary but typically does not strictly prohibit all violations; the degree of constraint satisfaction depends on parameter tuning. The GP method provides probabilistic uncertainty estimation, clearly categorizing it as a soft or probabilistic constraint method. However, with specialized safety designs or when combined with other methods, these safe RL methods can still potentially enforce hard constraints.

In contrast, the projection method, Lyapunov method, shielding method, and safety layer method are considered capable of satisfying hard constraints. The projection method explicitly projects actions into the safe region, ensuring constraint satisfaction at every step. The Lyapunov method can theoretically ensure asymptotic stability and long-term safety under deterministic settings. The shielding method precomputes shield rules and explicitly excludes unsafe actions at each step. The safety layer method can enforce hard constraints if strict projection or adjustment is applied at every step; however, if approximate projection (e.g., linear approximation) is used, strict constraint satisfaction may not be guaranteed. Similarly, if there are significant model uncertainties, mismatches between execution and design, or only numerical approximations are applied, methods such as the projection method, Lyapunov method, and shielding method may also face risks of brief or localized constraint violations during real-world deployment.

RRL improves robustness against worst-case scenarios and reduces constraint violation risks, but unless it explicitly incorporates hard constraint formulations, it should be regarded as offering probabilistic or soft constraint enforcement.

In real-world power system deployments, the selection of safe RL methods should be based on specific factors such as problem complexity, the strictness of constraints, computational efficiency, and the level of uncertainty. In addition, these methods should be complemented with external safety verification, fallback mechanisms, or conservative operational margins to ensure system reliability during real-world operation.


\begin{table*}[htb]
\caption{Comparison of Advantages, Disadvantages, and Computational Complexities of Different Safe RL Methods}
\label{Table_Comparison_SafeRL_Methods}
\centering
\begin{tabular}{|>{\centering\arraybackslash}m{1.5cm} >{\raggedright\arraybackslash}m{4.9cm} >{\raggedright\arraybackslash}m{5.0cm} >{\raggedright\arraybackslash}m{5.0cm}|}
\toprule
Methods & Advantages & Disadvantages & Computational Complexities\\
\midrule
Lagrangian relaxation method & Simple implementation; easily integrable with multiple RL algorithms; efficiently handles various types of constraints; High scalability to large-scale problems \cite{ray2019benchmarking}. & Require fine-tuning of Lagrange multipliers; does not strictly guarantee constraint satisfaction; risk of oscillation near constraint boundaries \cite{yu2019convergent}. & Depends on the convergence rate of the multipliers and the complexity of the underlying RL algorithm; does not significantly increase the complexity \cite{yu2019convergent}.\\
\hline
Projection method & Strict constraint satisfaction at every step; projection can be efficiently performed using traditional optimization methods if the feasible region is convex \cite{achiam2017constrained}. & Limited scalability to large-scale systems; requires an accurate estimation of the feasible region; needs selecting an appropriate projection method \cite{wang2023safe}. & Computationally expensive for high-dimensional and non-linear systems; scalability issues may limit real-time applications \cite{yi2023model}.\\
\hline
Lyapunov method & Offers rigorous theoretical stability guarantees; suitable for voltage and frequency stability control problems in power system control \cite{cui2022reinforcement, shi2022stability}. & Requires prior knowledge of Lyapunov functions; challenging for complex or unknown dynamics; difficult to handle multiple complex constraints; poor scalability \cite{chow2018lyapunov}. & Requires the computation or learning of Lyapunov functions, which can be resource-intensive; high training overhead in large-scale systems \cite{gu2024review}.\\
\hline
GP method & Effectively enhances safety under uncertainty; well-suited for managing stochastic system dynamics \cite{sui2015safe}. & Challenging to apply to large-scale systems; sensitive to kernel functions and hyperparameter selection \cite{wang2023safe}. & High computational complexity; scalability issues with increasing problem dimensions \cite{wang2023safe}.\\
\hline
Shielding method & Guarantees safety at every step with minimal intervention; ensures adherence to hard constraints \cite{berkenkamp2017safe}. & Requires detailed prior system knowledge to identify feasible actions; less effective in complex or uncertain environments \cite{politowicz2024safety}. & Scalability depends on the complexity of the specific algorithm; high computational costs arise with complex reachability analysis or online optimization \cite{politowicz2024safety}.\\
\hline
Safety layer method & Ensures real-time safety; features modular integration independent of specific RL algorithms; adaptable to continuous high-dimensional action spaces \cite{dalal2018safe}. & Linear approximations for non-linear systems may inadequately capture complex system dynamics \cite{wang2023safe}. & Solving optimization at each policy step causes significant computational overhead in high-dimensional multi-constraint scenarios; scalability depends on specific safety layer design \cite{gu2024review}.\\
\hline
Barrier function method & Ensures safety near boundaries; particularly effective in systems with explicitly defined constraint sets \cite{wang2023enforcing}. & Requires accurate system dynamics and safe set; challenging to deal with complex or multi-constraint problems; tends to prioritize safety over optimality, limiting exploration and rewards \cite{wang2023safe}. & Depends on the form of the barrier function and the constraint problem; computational efficiency may decrease as the number of constraints increases and the system scales up \cite{liu2020ipo}.\\
\hline
RRL & Capable of handling worst-case scenarios; strong adaptability to uncertain and adversarial environments; improves control robustness \cite{li2021safe}. & Difficult to define uncertainty sets or adversarial models; overly conservative, sacrificing average performance; challenging to design algorithms \cite{moos2022robust}. & Introduces additional overhead for worst-case policy learning than standard RL, potentially significantly increasing computational burden; scalability depends on the specific design \cite{moos2022robust}.\\
\bottomrule
\end{tabular}
\end{table*}

\begin{table*}[htb]
\caption{Comparison of Convergence and Optimality of Different Safe RL Methods}
\label{Table_Comparison_SafeRL_Convergence_Optimality}
\centering
\begin{tabular}{|>{\centering\arraybackslash}m{1.6cm} >{\raggedright\arraybackslash}m{7.6cm} >{\raggedright\arraybackslash}m{7.6cm}|}
\toprule
Methods & Convergence & Optimality\\
\midrule
Lagrangian relaxation method & Convergence is theoretically guaranteed for convex problems via duality theory; however, for non-convex scenarios, oscillations or convergence to local optima may arise \cite{chow2018risk, tessler2018reward}. & In general non-convex settings, only local optimality or convergence near saddle points can typically be guaranteed \cite{paternain2019constrained}. \\
\hline
Projection method & Projection affects the convergence speed, stability, and the final feasible solution; choice of trust region and specific projection algorithm directly affects convergence performance \cite{yang2020projection}. & In nonconvex and high-dimensional scenarios, global optimality is generally not guaranteed, and only local or approximate optima are typically achieved \cite{gros2020safe}. \\
\hline
Lyapunov method & Convergence strongly depends on the selected Lyapunov function; rigorous theoretical guarantees are typically achievable if the system dynamics are known or accurately estimated \cite{perkins2002lyapunov}. & Restricts the feasible policy search space to ensure stability, which may limit optimality and result in suboptimal performance \cite{perkins2002lyapunov}. \\
\hline
GP method & GPs are theoretically capable of ensuring asymptotic consistency, but the performance of GP-based RL methods depends on the specific algorithm used \cite{berkenkamp2017safe}. & Effective for robustness, but may yield overly conservative solutions in complex or uncertain environments, depending on confidence levels \cite{sui2015safe}. \\
\hline
Shielding method & Frequent shield interventions may disrupt smooth and continuous policy updates, negatively impacting convergence speed \cite{alshiekh2018safe}. & Shielding often results in conservative policies, compromising global optimality; the degree of conservatism strongly depends on the specific shielding mechanism design \cite{alshiekh2018safe}. \\
\hline
Safety layer method & Action clipping and first-order linearization adjustments used in safety layers may adversely affect convergence, particularly in complex, nonlinear systems \cite{dalal2018safe}. & When using first-order linear approximations, the policy may be confined to a narrower or even incorrect feasible region, potentially converging to suboptimal solutions \cite{dalal2018safe}. \\
\hline
Barrier function method & Excessively steep or heavily weighted barrier functions can introduce substantial gradient variations, potentially reducing training stability and slowing convergence \cite{wang2023safe}. & Excessively steep or heavily weighted barrier functions frequently lead to overly conservative policies, significantly restricting exploration and performance \cite{wang2023safe}.\\
\hline
RRL & Training convergence under uncertainty and adversarial conditions depends heavily on accurate environment modeling and well-designed adversarial strategies \cite{li2021safe}. & Excessive focus on worst-case scenarios often sacrifices average performance, typically leading to lower optimality compared to standard RL methods \cite{pinto2017robust}. \\
\bottomrule
\end{tabular}
\end{table*}

\subsection{Performance Comparison of Different Safe RL Methods}\label{sec3performance}
Different safe RL algorithms have varying advantages, disadvantages, and computational complexities. These characteristics are summarized in Table \ref{Table_Comparison_SafeRL_Methods}. According to Table \ref{Table_Comparison_SafeRL_Methods}, different safe RL methods are suited for specific applications. Lagrangian relaxation is well-suited for low-risk scenarios such as economic dispatch and energy management. The projection method is ideal for cases with detailed system knowledge to guide action projection, enabling efficient enforcement of strict feasibility. Lyapunov methods excel in stability control, particularly for voltage and frequency regulation, while GP methods are effective in handling uncertainty, such as renewable forecasting or stochastic load variations. Shielding methods are preferred in applications requiring hard constraint enforcement, such as BESS charging and discharging. Safety layer methods are most suitable for scenarios where the system's state provides clear guidance on how actions should be adjusted, such as in voltage control. Barrier function methods are designed for fields with strict safety requirements, such as frequency stability control and OPF. Finally, RRL is tailored for worst-case scenarios, including control under extreme weather or environmental conditions.
In addition, the sample complexity and safety violation analysis of specific model-based and model-free algorithms are summarized in \cite{gu2024review}. Table \ref{Table_Comparison_SafeRL_Methods} provides a general comparison of the 8 categories of safe RL methods. Each category includes specific algorithms with varying applicability to different problems. In addition, for methods such as the shielding method and safety layer method, the computational complexity and scalability largely depend on the design of the specific shield or safety layer. Therefore, practical implementation and performance need to be analyzed on a case-by-case basis.

A comparison of their convergence and optimality is also provided in Table \ref{Table_Comparison_SafeRL_Convergence_Optimality}. It is evident that most methods can ensure a certain level of convergence and optimality for simple convex problems, but achieving the same for nonlinear and high-dimensional scenarios remains challenging. Moreover, their convergence efficiency and speed, as well as their tendency to be overly conservative, are influenced by the specific methods and their accuracy.

\subsection{Benchmark Environments, Algorithms, and Software}\label{sec3benchmark}
Benchmarks consist of environments, algorithms, and software, all essential for developing and evaluating safe RL in power systems. Environments are power system models that accept agent actions and return dynamic responses for training and testing. Algorithms provide standardized implementations of safe RL methods to ensure reproducibility, enable improvements, and allow fair comparison under various safety constraints. Software tools offer interfaces to integrate detailed power system models into RL frameworks, supporting seamless data exchange for accurate simulation, analysis, and validation.

\subsubsection{General Benchmark Environments and Algorithms}\mbox{}

General benchmarks refer to universal environments or algorithms designed specifically for safe RL, offering comprehensive components, scalability, active maintenance, and broad applicability. \cite{gu2024review} maintains a GitHub repository with safe RL baselines, benchmarks, and recent algorithms in the general field \cite{gu2024saferlbaselines}.

Safety Gym, developed by OpenAI, is the first widely recognized safe benchmark environment, featuring an environment builder and several pre-configured tasks \cite{ray2019benchmarking, ray2019gym}. Correspondingly, Safety Starter Agents is a benchmark algorithm library built on Safety Gym that supports PPO, PPO-Lag, TRPO, TRPO-Lag, SAC, SAC-Lag, and CPO \cite{ray2019algorithm}.

\begin{figure}[htbp]
    \centering
    \includegraphics[width=8.4cm]{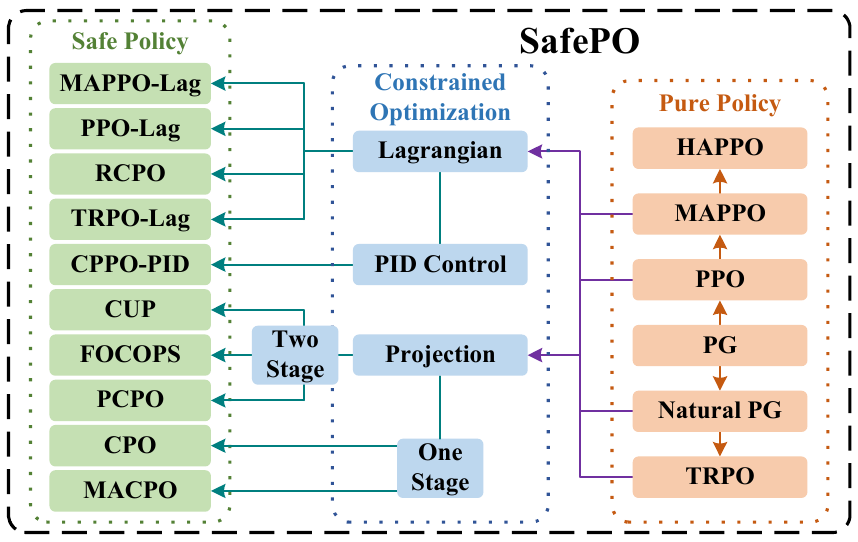}
    \caption{Supported safe RL algorithms of SafePO.}
\label{Fig_SafePO}
\end{figure}

Safety Gymnasium extends Safety Gym and has become the current mainstream platform \cite{ji2023paper, ji2023gymnasium}. Its corresponding algorithm benchmark repository, SafePO, provides implementations of safe RL algorithms \cite{ji2023algorithm}, as illustrated in Fig. \ref{Fig_SafePO}.

OmniSafe is a unified learning framework for safe RL, offering a modular structure with a comprehensive set of algorithms tailored to various domains. Its abstracted algorithm design and well-defined API enable seamless component integration, making extension and customization straightforward. Additionally, OmniSafe accelerates learning through process-level and agent-level parallelism \cite{ji2023omnisafe, ji2023omnisafealgorithm}. The supported safe RL algorithms of OmniSafe are listed in Table \ref{Table_OmniSafe}.
\begin{table}[htbp]
\caption{Supported Safe RL Algorithms of OmniSafe}
\label{Table_OmniSafe}
\centering
\begin{tabular}{|@{}lll@{}|}
\toprule
Domains & Types & Algorithms Registry \\
\midrule
On Policy & Primal-Dual & PPO/TRPO-Lag \cite{ray2019benchmarking}; RCPO \cite{tessler2018reward}; \\
        &             & PDO \cite{chow2018risk}; TRPO/CPPO-PID \cite{stooke2020responsive} \\
        \cmidrule{2-3}
        & Convex Optimization & CPO \cite{achiam2017constrained}; PCPO \cite{yang2020projection}; CUP \cite{yang2022constrained}; \\
        &             & FOCOPS \cite{zhang2020first}\\
        \cmidrule{2-3}
        & Penalty Function & IPO \cite{liu2020ipo}; P3O \cite{zhang2022penalized} \\
        \cmidrule{2-3}
        & Primal & CRPO \cite{xu2021crpo} \\
\hline
Off Policy & Primal-Dual & DDPG/TD3/SAC-Lag \cite{ray2019benchmarking}; \\
        &             & DDPG/TD3/SAC-PID \cite{stooke2020responsive} \\
\hline
Model-based & Online Plan & SafeLOOP \cite{sikchi2022learning}; CCE-PETS \cite{wen2018constrained};\\
        &             & RCE-PETS \cite{liu2020constrained} \\
        \cmidrule{2-3}
        & Pessimistic Estimate & CAP-PETS \cite{ma2022conservative} \\
\hline
Offline & Q-Learning-Based & BCQ-Lag \cite{fujimoto2019off}; C-CRR \cite{wang2020critic} \\
        \cmidrule{2-3}
        & DICE-Based & COptDICE \cite{lee2022coptidice}\\
\hline
Other MDP & EarlyTerminated-MDP & PPO/TRPO-EarlyTerminated \cite{sun2021safe} \\
        \cmidrule{2-3}
        & SauteRL & PPO/TRPOSaute \cite{sootla2022saute}\\
        \cmidrule{2-3}
        & SimmerRL & PPO/TRPOSimmer-PID \cite{sootla2022effects} \\
\bottomrule
\end{tabular}
\end{table}

Overall, Safety Gymnasium is the leading benchmark environment, and OmniSafe integrates it to ensure overall code compatibility. However, Safety Gymnasium was designed for gaming, robotics, and autonomous driving (e.g., point, car, dog, and ant agents) with tasks such as safe navigation and safe velocity, and it does not directly address power system formulations. Therefore, power‐system‐specific environments must be developed using Safety Gymnasium’s templates. In terms of benchmark algorithms, SafePO and OmniSafe offer the most comprehensive collections of safe RL algorithms; however, because most power‐system tools run on Windows, benchmark compatibility with Windows must be considered in advance.

\subsubsection{Power System Benchmark Software}\mbox{}

Current safe RL research for power system optimization and control relies on integrating power system simulators into RL environments. These simulation tools provide PF, continuation PF, OPF, small‐signal stability analysis, and time‐domain simulation, enabling reward calculation, enforcement of physical constraints, and validation of system safety to support various safe RL algorithm designs.

Power system simulation software can be categorized into two main types: commercial and open-source/free. Commercial software requires purchased licenses and offers stable performance, comprehensive models, and extensive libraries. It supports modeling and simulation of large‐scale power systems and accommodates nearly all static and dynamic simulations. Most commercial software provides interfaces with MATLAB, Python, or other programming languages, enabling seamless interaction with RL algorithms for real-time feedback. Examples include PSSE, PowerFactory, PowerWorld, EMTP, ETAP, RTDS, Simscape, and PSCAD \cite{bam2005power, vogt2018survey}. Open-source and free software provide unrestricted access to source code, enabling customization and transparency in modeling and simulation. These tools are widely used in academic research, enabling users to modify models, implement new algorithms, and conduct innovative studies. Many are developed in MATLAB, Python, or Julia, which facilitates integration with machine learning and RL frameworks. Notable examples include OpenDSS \cite{dugan2011open}, GridLAB-D \cite{chassin2014gridlab}, MATPOWER \cite{zimmerman2010matpower}, Pandapower \cite{thurner2018pandapower}, PyPSA \cite{PyPSA}, PowerModels \cite{coffrin2018powermodels}, PST \cite{chow1992toolbox}, PSAT \cite{milano2005open}, PowerSimulations.jl \cite{lara2021powersystems}, PowerModelsDistribution.jl \cite{fobes2020powermodelsdistribution}, ANDES \cite{cui2020hybrid}, PowerSimulationsDynamics.jl \cite{lara2023powersimulationsdynamics}, Dyna$\omega$o \cite{guironnet2018towards}.

Support for grid-forming inverters is crucial for dynamic simulation in power systems with high RES penetration. Most commercial software already provide grid‐forming inverter modules, including PowerWorld, PSSE, EMTP, RTDS, Simscape, and PSCAD, or allow user‐defined grid‐forming inverter models. Among open-source and free software, ANDES, PowerSimulationsDynamics.jl, Dyna$\omega$o, OpenDSS, and GridLAB-D provide built-in grid-forming inverter models with the flexibility for user modification \cite{su2024survey}.

\subsubsection{Tailored Benchmarks for Power System}\mbox{}

In addition to general benchmarks, several specialized environments following the Safety Gym/Gymnasium format have been developed to support power system optimization and control. These benchmarks facilitate developing new models and testing novel DR and safe DR algorithms. These benchmarks include:
\paragraph{OMG} Built on Safety Gym, OMG simulates and optimizes microgrid control via power‐electronic converters. It offers a plug-and-play grid design within OpenModelica and a Python interface for intuitive RL integration \cite{heid2020omg}.
\paragraph{RLGC} Using the InterPSS simulator, RLGC provides a Safety Gym–compatible environment for power grid dynamic simulation, enabling development, testing, and benchmarking of RL algorithms for grid‐level control tasks \cite{huang2019adaptive, huang2019RLGC}.
\paragraph{PowerGym} PowerGym is a Gym-like environment for Volt-Var control in power distribution systems, with networked constraints managed by the OpenDSS simulator \cite{fan2022powergym}.
\paragraph{OPF-Gym} Built on Safety Gymnasium and Pandapower, OPF-Gym offers five benchmark environments: economic dispatch, voltage control with reactive power, renewable feed-in maximization, reactive power market, and load shedding, enabling easy creation of custom OPF problems for RL research \cite{wolgast2024learning, wolgast2024opfgym}.
\paragraph{CommonPower} CommonPower applies safe RL to power system control by safeguarding decision-making and evaluating forecast quality’s impact. It uses an object-oriented, Pyomo‐based model to derive system equations and offers interfaces for single/multi‐agent RL \cite{eichelbeck2024commonpower, eichelbeck2024github}.

\section{Power System Applications of Safe RL}\label{sec4}
\begin{figure}[!htb]
    \centering
    \includegraphics[width=8.5cm]{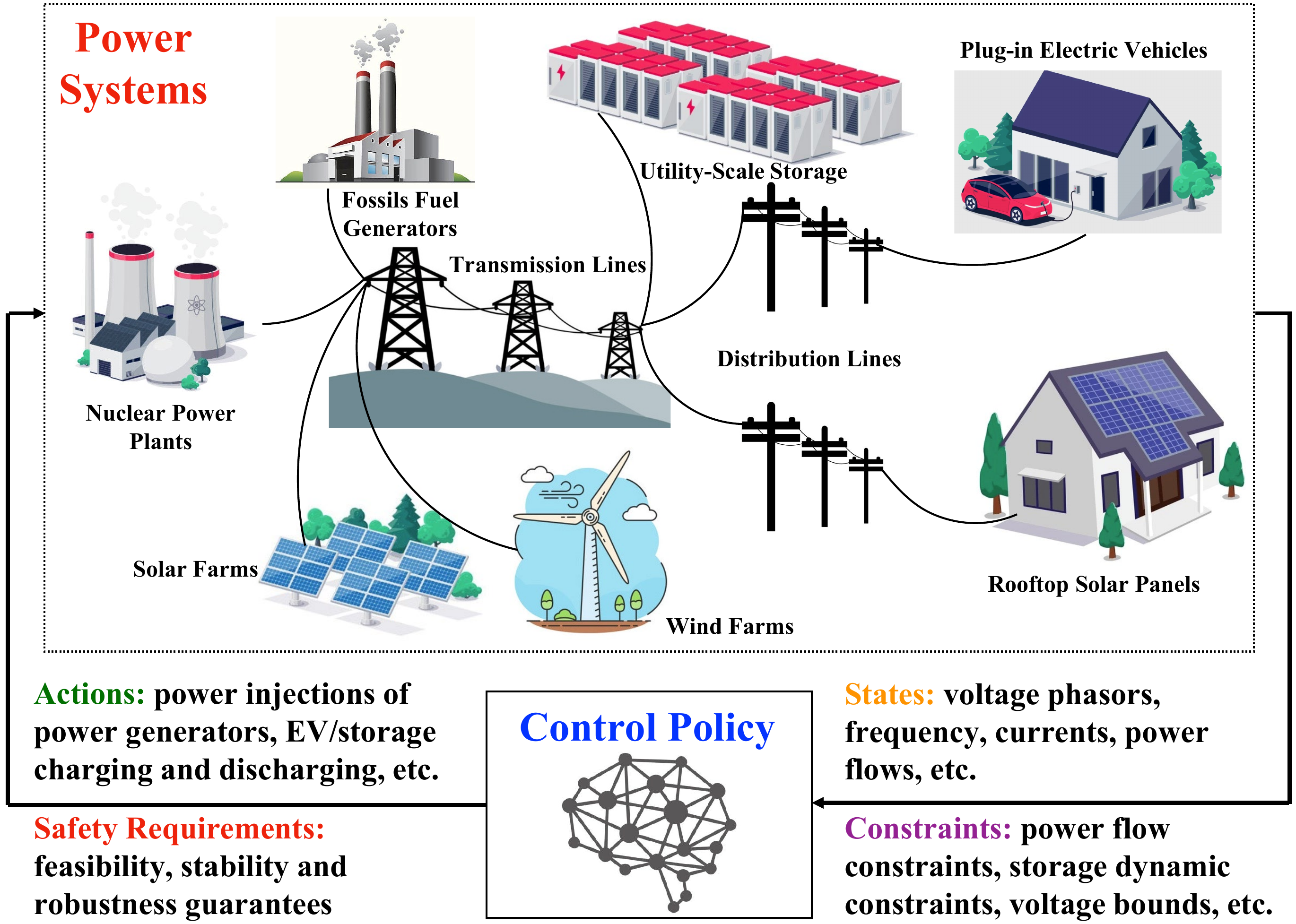}
    \caption{RL schemes for the safe control and decision-making in power systems. The RL agent observes system states that reflect current operating conditions and generates actions to influence control decisions. These actions must satisfy system constraints, which encode physical and operational limits. As a result, the RL policy must optimize performance while ensuring safety throughout both learning and deployment.}
\label{Fig_framework_x}
\end{figure}

This section synthesizes a broad collection of studies and applications of safe RL in power systems, spanning a wide range of domains, including security control, real-time operation, operational planning, and emerging areas. Specific examples reviewed within these domains include voltage control, stability control, economic dispatch, system restoration, unit commitment, electricity market, EV charging, and building energy management. Safe RL algorithms applied across various domains are presented in Fig. \ref{Fig_framework}. Fig. \ref{Fig_framework_x}, on the other hand, illustrates safe RL-based decision-making processes in power systems.
In these processes, agents gather power system measurements and incorporate system model knowledge into their policy training. They then execute actions to control power system devices, ensuring compliance with safety requirements such as feasibility, stability, and robustness.

\begin{figure}[!htb]
    \centering
    \includegraphics[width=6.5cm]{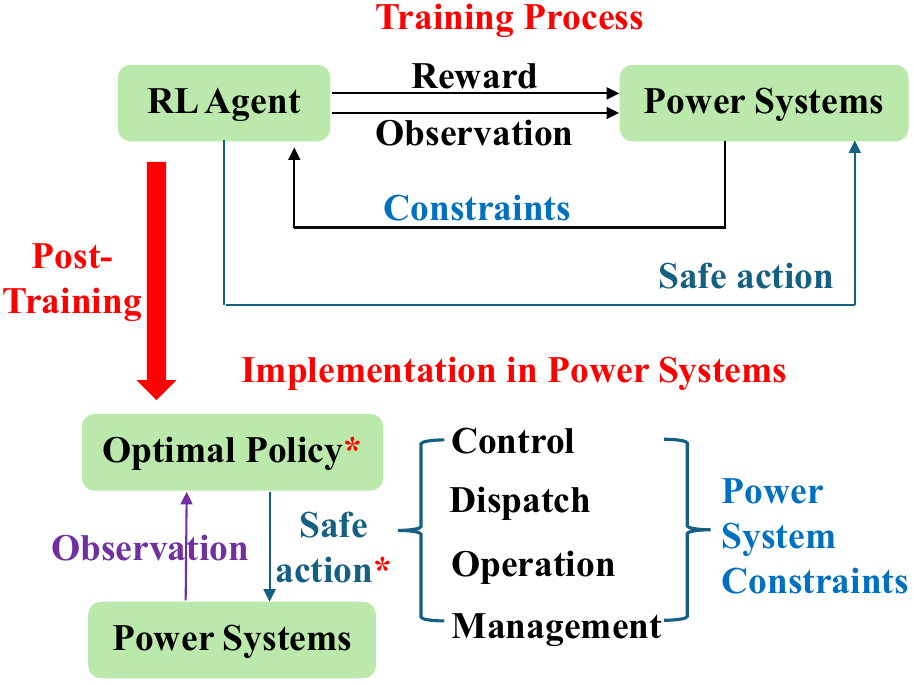}
    \caption{Training and implementation of safe RL for power system applications based on state, action, reward, and constraint. During the training process, the RL agent interacts with a simulated power system by observing states, receiving rewards, and generating actions, while accounting for system constraints to ensure safety. Once trained, the optimal policy is deployed in real-world applications, where it generates safe actions based on real-time observations to support control, dispatch, operation, and management decisions, while continuously satisfying power system constraints.}
\label{Fig_Implementation}
\end{figure}

For each application domain, this section summarizes its background, traditional methods, and the reason for applying safe RL. It then reviews existing work on objective functions, constraint formulations (cumulative vs. instantaneous, hard vs. soft), and applied safety techniques, enabling cross‐study comparisons. Additionally, it highlights the required modeling components for each application, including state, action, reward, and constraint. The training and deployment process of safe RL based on these four elements is illustrated in Fig. \ref{Fig_Implementation}.
It can be found that the integration of safe RL into power systems involves two key phases: training and implementation. During training, the agent interacts with a simulated power system, observes states, and takes actions under a safety mechanism that enforces operational limits. It receives rewards for optimal decisions that respect predefined safety constraints and iteratively learns a policy through this feedback loop. After training, the learned policy is deployed in real‐world power systems. During implementation, the policy uses real‐time system states to compute safe actions for tasks such as control, dispatch, operation, and planning that comply with system constraints. Continuous feedback between the deployed policy and the actual system ensures robust performance and adaptability, bridging simulation and real‐world application.

\subsection{Security Control}\label{sec4control}

Power system security control refers to the set of strategies and actions designed to maintain the stability and reliability of the power system under both normal and contingency conditions. It involves real-time monitoring, preventive measures, and corrective actions to ensure safe operation. These actions help keep the system within secure limits, such as voltage levels, frequency, and power flows, while preventing cascading failures or blackouts \cite{kundur1994power, balu1992line}.

\subsubsection{Voltage Control}\label{sec4voltage}\mbox{}

\begin{figure}[htb]
    \centering
    \includegraphics[width=8.5cm]{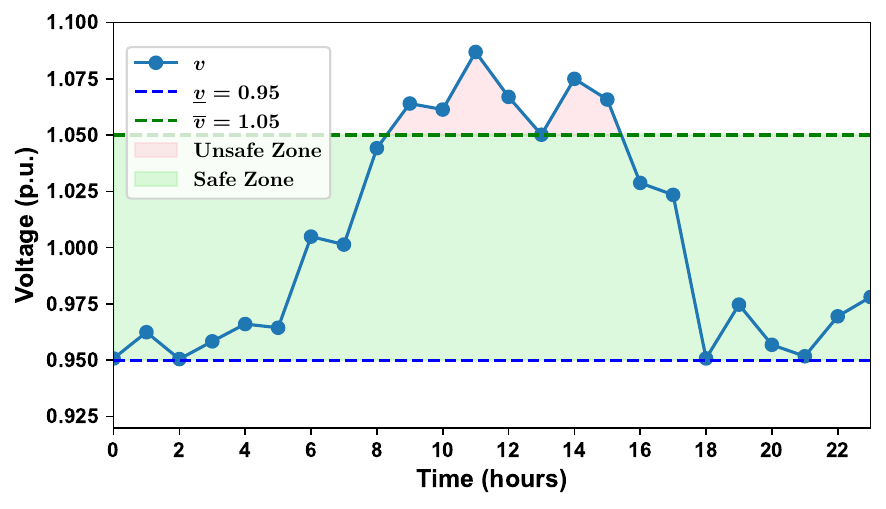}
    \caption{Voltage profile at the bus connected to PV. Strong sunlight around noon causes localized overvoltage, requiring safe RL for voltage regulation to bring it back to the safe range, i.e., $\underline{\bm v}$ to $\overline{\bm v}$.}
\label{Fig_Voltage}
\end{figure}

The increasing penetration of RESs, including wind, PVs, and EVs, has profoundly altered power system behavior. Distribution networks, which are often radial or meshed in structure and connect numerous intermittent and uncertain distributed RESs, now face heightened complexity in voltage management \cite{petinrin2016impact}. This complexity frequently results in voltage violations, where voltages fall below 0.95 p.u. or exceed 1.05 p.u. \cite{bastos2020machine}. For instance, Fig. \ref{Fig_Voltage} shows the voltage profile of nodes directly connected to PV systems, where strong sunlight around noon causes localized overvoltage and requires voltage regulation.

\begin{figure}[!htb]
    \centering
    \includegraphics[width=6.0cm]{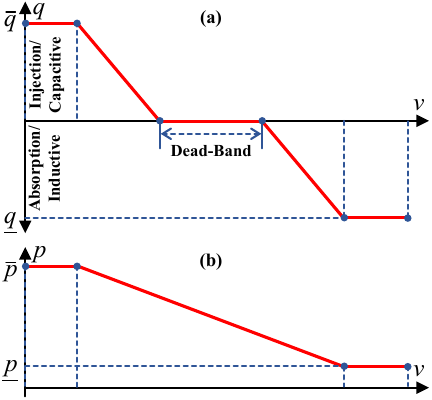}
    \caption{Volt-Var and Volt-Watt curves. (a) Volt-Var curve: In this mode, the inverter actively controls its reactive power output as a function of voltage; (b) Volt-Watt curve: In this mode, the inverter actively limits the maximum active power as a function of the voltage \cite{photovoltaics2018ieee}.}
\label{Fig_VV_VW}
\end{figure}

To address these challenges, voltage control aims to maintain voltage magnitudes across power networks within nominal or acceptable ranges, ensuring stable and reliable system operation \cite{sun2019review, murray2021voltage}. Traditional methods for voltage regulation often employ physical model-based optimization techniques. These methods leverage convex relaxation techniques, such as second-order cone programming, to simplify AC-PF constraints, enabling efficient resolution with standard solvers \cite{zhang2023data, wang2019safe, ruan2020distributed}. Additionally, instead of directly controlling the active and reactive power injections of smart inverters, some researchers have proposed resetting the Volt-Var and Volt-Watt curves to regulate voltage profiles \cite{wu2022reinforcement, roberts2021deep}. The Volt-Var and Volt-Watt curves for voltage control are illustrated in Fig. \ref{Fig_VV_VW} \cite{carreno2024voltage}.

Due to the integration of DERs, such as rooftop solar panels and EVs, distribution systems experience rapid and unpredictable fluctuations in generation and load profiles, posing significant challenges for real‐time voltage control in distribution grids using model‐based methods. As an alternative, RL has emerged as a promising approach for addressing model-free nonlinear control problems, driving interest in developing RL-based controllers to optimize voltage control performance. Moreover, the adoption of safe RL ensures adherence to voltage constraints, offering a robust solution for maintaining operational stability. A summary of existing papers applying safe RL to voltage control in power systems is detailed in Table \ref{Table_VoltageControl}.
\renewcommand{\arraystretch}{1.1}
\begin{table*}[htbp]
\caption{Safe RL Applications in Voltage Control}
\label{Table_VoltageControl}
\centering
\begin{adjustbox}{center, max width=\textwidth}
\begin{tabular}{|>{\centering\arraybackslash}m{0.7cm} >{\centering\arraybackslash}m{6.0cm} >{\centering\arraybackslash}m{3.2cm} >
{\centering\arraybackslash}m{2.2cm} >{\centering\arraybackslash}m{4.0cm} |}
    \toprule %
    Ref. & Problem/Objective & Constraint & Constraint Type & Safety Techniques\\
    \midrule %
    \cite{zhang2023data} & Transmission losses & Voltage & Ins/Hard & Projection layer \\ 
    \hline
    \cite{yan2023multi} & Total network energy loss & Voltage deviations & Cum/Soft & Primal-dual policy \\ 
    \hline
    \cite{liu2020two} & Voltage violations and network losses & Voltage bounds & Cum/Soft & Penalty function and RRL \\ 
    \hline
    \cite{wang2019safe} & Cost of losses and device switching & Voltage & Cum/Soft & Lagrangian relaxation \\ 
    \hline
    \cite{wang2019volt} & Total operation costs & Voltage & Cum/Soft & CPO \\ 
    \hline
    \cite{cui2022decentralized} & Operation cost & Voltage & Ins/Hard & Lyapunov stability \\ 
    \hline
    \cite{shi2022stability} & Voltage deviation and control cost & Voltage & Ins/Hard & Lyapunov function \\ 
    \hline
    \cite{chen2022physics} & Active voltage control & SoC of BESSs & Ins/Hard & Physics-based shielding \\ 
    \hline
    \cite{gao2022model} & Cost of network loss and device switching & Voltage and power flow & Ins/Hard & Safety layer \\ 
    \hline
    \cite{liu2021online} & Active power loss & Voltage violations & Cum/Soft & Lagrangian relaxation \\ 
    \hline
    \cite{chen2022saver} & Total control cost & Voltage & Ins/Hard & Safety projection layer \\ 
    \hline
    \cite{zhang2023dnn} & Transmission loss & Voltage and power flow & Ins/Hard & Finite iteration projection \\ 
    \hline
    \cite{kou2020safe} & Power losses and control efforts & Voltage and power grid & Ins/Hard & Safety layer \\ 
    \hline
    \cite{guo2023safe} & Network power loss & Nodal voltage & Ins/Hard & Safety projection \\ 
    \hline
    \cite{nguyen2022three} & Cost of electricity and BESSs maintenance & Voltage and network & Ins/Hard & SAC with safety module \\ 
    \hline
    \cite{deng2024safety} & Voltage control in flexible network topologies & Voltage & Cum/Hard & Lyapunov function \\ 
    \hline
    \cite{zhao2023explicit} & Inverter-based voltage regulation & System and voltage & Ins/Hard & Safety layer \\ 
    \bottomrule %
\end{tabular}
\end{adjustbox}
\begin{tablenotes}
    \item Cum: Cumulative; Ins: Instantaneous.
\end{tablenotes}
\end{table*}
Table \ref{Table_VoltageControl} highlights that most optimization objectives focus on minimizing voltage deviations, system losses, and control costs. The constraints typically involve voltage and other operational limits, which are represented in both instantaneous and cumulative forms. Due to the straightforward nature of voltage control problem formulations, they are well-suited for integration with various safe RL techniques, such as Lagrangian relaxation, projection, Lyapunov, shielding, and safety layer methods. Additionally, \cite{liu2020two} employs RRL to perform adversarial training, enhancing robustness against uncertainties. However, centralized approaches suffer from single‐point failures and significant communication overhead, making them impractical for large‐scale systems. Consequently, research is shifting toward distributed voltage regulation, which relies solely on local information exchange among neighboring units and has shown great promise \cite{yan2023multi}.

In the following, using DG and BESS smart inverters as key examples, we summarize the safe RL voltage control problem, focusing on Volt-Var control under AC‐PF or LinDistFlow constraints. The state, action, reward, and constraints are outlined as follows:

\paragraph{Safe RL for Volt-Var Control with AC-PF}
Volt-Var control maintains voltage within safe operating limits and optimizes reactive power flow in power systems. It is governed by nonlinear AC-PF constraints that relate to voltage magnitudes, phase angles, and reactive power.

\textbf{State:}
The state variables are PMU measurements from buses in $\mathcal{N}^\text{PMU}$, or AMI measurements from buses in $\mathcal{N}^\text{AMI}$. Thus, the state variable $\bm{s}$ is defined by:
\begin{subequations}
\label{Eq_PMU_AMI}
\begin{gather}
\bm{s^\text{PMU}} \triangleq \left( (v_i)_{i\in \mathcal{N}^\text{PMU}}, (i_i)_{i\in \mathcal{N}^\text{PMU}}\right)\\
\bm{s}^\text{AMI} \triangleq \left( ( {|v_i|}^2)_{i\in \mathcal{N}^\text{AMI}}, ({|i_i|}^2)_{i\in \mathcal{N}^\text{AMI}}, (s_i)_{i\in \mathcal{N}^\text{AMI}} \right)
\end{gather}
\end{subequations}
where $\bm v$, $\bm i$, $\bm s$ denote voltage, current, apparent power vectors, respectively. The system dynamics that depict the environment can be formulated as:
\begin{equation}
\bm{s}^\text{V}_{t+1} \triangleq \bm{f}(\bm{s}^\text{V}_t, \bm{a}^\text{V}_t)
\end{equation}

\textbf{Action:}
The control actions include regulating the DGs, BESSs, and other components.
\begin{equation}
   \bm{a}^\text{V}_t \triangleq \left( \bm{p}^\text{DG}_t, \bm{q}^\text{DG}_t, \bm{p}^\text{BESS}_t, \bm{p}^\text{other}_t \right)
\end{equation}

\textbf{Reward:}
The reward is to maintain the voltage magnitudes close to the nominal value $v_\text{ref}$ (typically 1.0 p.u.):
\begin{equation}
\label{Eq_Reward_V_ref}
R^\text{V}(\bm{s}, \bm{a}) = -\|{\bm{v}_t - v_{\text{ref}}}\|
\end{equation}

Another kind of reward design is to maintain the voltage as closely as possible within the safety range:
\begin{equation}
\label{Eq_Reward_V_soft}
R^\text{V}(\bm{s}, \bm{a}) = -\sum_{i \in \mathcal{N}}\big([ {v}_i - \overline{v}]_+ +  [ \underline{v} - {v}_i ]_+ \big)
\end{equation}

\textbf{Constraint:}
The AC-PF is shown in Section \ref{sec4economic}. The constraint for the active and reactive power injections of DGs is given by:
\begin{equation}
(\bm{p}^\text{DG})^2 + (\bm{q}^\text{DG})^2 \le (\overline{\bm{s}}^\text{DG})^2 \label{VC_action}
\end{equation}

However, \cite{carreno2024voltage} points out that the stability regions are more constrained than in \eqref{VC_action}. For simplicity, we omit the specific equations. Additionally, there are constraints that directly limit voltage $v$:
\begin{equation}
\label{Eq_Constraint_V}
\underline{\bm v} \leq \bm v \leq \overline{\bm v}
\end{equation}

\paragraph{Safe RL for Volt-Var Control with LinDistFlow} The LinDistFlow linearized branch flow model is applied within a tree-structured distribution network. The system consists of a set of nodes $\mathcal{N}_{+0} = \{0, 1, \cdots, N\}$ and an edge set $\mathcal{E}$. Node 0 is known as the substation, and $\mathcal{N} = \mathcal{N}_{+0}/\{0\} $ denotes the set of nodes excluding the substation node. Each node $i\in \mathcal{N}$ is associated with an active power injection $p_i$ and a reactive power injection $q_i$. Let $V_i$ be the squared voltage magnitude, and let $p, q$ and $V$ denote $\{p_i, q_i, V_i\}_{i\in \mathcal{N}}$ stacked into a vector.
The variables satisfy the following equations, $\forall i \in \mathcal{N}$, 
\begin{subequations}
\begin{gather}
p_i = -p_{ji} + \sum_{k: (i, k) \in \mathcal{E}} p_{ik}\\
q_i = -q_{ji} + \sum_{k: (i, k) \in \mathcal{E}} q_{ik}\\
V_i = V_j - 2(r_{ij}p_{ji} + x_{ji}q_{ji})\label{Eq_BranchFL}
\end{gather}
\end{subequations}
where $j$ is the parent node of $i$ in the distribution network. \eqref{Eq_BranchFL} can be written in the vector form: 
\begin{equation}
\bm{V} = \mathbf{R}\bm{p} + \mathbf{X}\bm{q} + V_0 \mathbf{1} = \mathbf{X} \bm{q} + \bm{V}_{\text{env}}
\end{equation}
where $\bm{V}_{\text{env}} = \mathbf{R} \bm{p} + V_0\mathbf{1}$ represents the non-controllable part; $\mathbf{R} = [2\mathrm{R}_{ij}]^{N\times N}$ and $\mathbf{X} = [2\mathrm{X}_{ij}]^{N\times N}$ are defined as $\mathrm{R}_{ij} \coloneqq 2\sum_{(h,k)\in \mathcal{P}_i \cap \mathcal{P}_j}r_{hk}$ and $\mathrm{X}_{ij} \coloneqq 2\sum_{(h,k)\in \mathcal{P}_i \cap \mathcal{P}_j}x_{hk}$, respectively; $\mathcal{P}_i$ is the set of lines on the unique path from bus $0$ to bus $i$; $V_0$ is the squared voltage magnitude at the substation bus; $\mathbf{R}$ and $\mathbf{X}$ are positive definite matrices, and all elements are positive \cite{feng2023bridging}.

\textbf{State:}
The state of LinDistFlow is also determined by PMU and AMI measurements, similar to the \eqref{Eq_PMU_AMI}.

\textbf{Action:}
The control actions is a mapping from the voltage to reactive power, which is defined by:
\begin{equation}
\bm{a}^\text{V}_t = \Delta \bm{q}_t \triangleq \bm{q}_{t} - \bm{q}_{t+1}
\end{equation}

The system dynamics can be given as
\begin{equation}
\bm{V}_{t+1} = \mathbf{R}\bm{p} + \mathbf{X}(\bm{q}_t - \bm{a}^\text{V}_t ) + V_0 \mathbf{1}
\end{equation}
where $\bm{p}$ lacks a time subscript because it pertains to a fast-response control mechanism, and $\bm{p}$ is assumed to be constant. 

\textbf{Reward:}
The reward is also designed to keep the voltage close to its nominal value \eqref{Eq_Reward_V_ref} or within its maximum and minimum limits \eqref{Eq_Reward_V_soft}.

\textbf{Constraint:}
The constraints include direct limitations on voltage \eqref{Eq_Constraint_V}, as well as action range and feasibility constraints:
\begin{subequations}
\begin{gather}
\underline{\bm{a}}^\text{V} \le \bm{a}^\text{V}_t \le \overline{\bm{a}}^\text{V}\\
\bm{a}^\text{V}_t ~\text{is feasible}
\end{gather}
\end{subequations}

\subsubsection{Stability Control}\label{sec4stability}\mbox{}

Power system stability control focuses on decision-making to prevent the system from entering undesired states, particularly to avert large-scale catastrophic faults \cite{machowski2020power, kundur1994power}. Based on the sequence of control actions and contingencies, stability control is generally categorized into two main categories: preventive control and emergency control. Preventive control aims to prepare the system while it is still in normal operation, ensuring it can satisfactorily handle future contingencies. In contrast, emergency control is initiated after contingencies have already occurred, with the objective of controlling the system's dynamics to minimize consequences \cite{wehenkel2004preventive}. Both types of control have stringent time requirements, with emergency control being particularly time-critical, often requiring actions to be executed within tens of milliseconds. From the perspective of key system variables that can indicate unstable behavior, traditional power system stability issues are classified into rotor angle stability, frequency stability, and voltage stability \cite{kundur2004definition}. With the increasing integration of power electronic devices, these categories have expanded to include resonance stability and converter-driven stability \cite{hatziargyriou2020definition}. Due to the complexity of stability issues and the rapidly changing system states, traditional analytical methods may struggle to find solutions and face computational efficiency limitations.

In this context, RL and safe RL have emerged as powerful tools to address these challenges, offering efficient and adaptive solutions. A summary of existing papers applying safe RL to stability control in power systems is detailed in Table \ref{Table_StabilityControl}.
\renewcommand{\arraystretch}{1.1}
\begin{table*}[htbp]
\caption{Safe RL Applications in Stability Control}
\label{Table_StabilityControl}
\centering
\begin{adjustbox}{center, max width=\textwidth}
\begin{tabular}{|>{\centering\arraybackslash}m{0.7cm} >{\centering\arraybackslash}m{6.0cm} >{\centering\arraybackslash}m{3.2cm} >
{\centering\arraybackslash}m{2.2cm} >{\centering\arraybackslash}m{4.0cm} |}
    \toprule %
    Ref. & Problem/Objective & Constraint & Constraint Type & Safety Techniques\\
    \midrule %
    \cite{su2025safe} & Emergency control for islanded microgrids & Rotor angle stability & Cum/Soft & RCPO \\
    \hline
    \cite{cui2022reinforcement} & Primary frequency control & Frequency stability & Ins/Hard & Lyapunov method \\ 
    \hline
    \cite{cui2023online} & Preventive control for transmission overload relief & Safety network & Cum/Soft & IPO \\ 
    \hline
    \cite{vu2021barrier} & Emergency control for under voltage load-shedding & Transient voltage stability & Cum/Soft & Barrier function \\ 
    \hline
    \cite{feng2023bridging} & Transient and steady-state voltage control & Reactive power capacity & Ins/Hard & Lagrangian, projection and barrier \\ 
    \hline
    \cite{zhang2023deep} & Emergency load-shedding control & Rated capacity and voltage & Cum/Soft & Lagrangian relaxation \\ 
    \hline
    \cite{xia2022safe} & Frequency control & Operation & Cum/Soft & Safety model \\ 
    \hline
    \cite{wan2023adapsafe} & Minimize the control cost & Frequency  & Cum/Soft & Barrier function \\ 
    \hline
    \cite{tabas2022computationally} & Primary frequency control & Frequency & Ins/Hard & Gauge map \\ 
    \hline
    \cite{zhou2022coordinated} & Frequency control & Operation & Cum/Soft & Lagrangian relaxation \\ 
    \hline
    \cite{gupta2021coordinated} & Wide-area damping control & System  & Ins/Hard & Bounded exploratory control \\ 
    \hline
    \cite{kwon2023risk} & Minimize large frequency oscillations & Mean-variance risk & Cum/Soft & Lagrangian relaxation \\ 
    \hline
    \cite{tarle2023safe} & FACTS setpoint control & System & Cum/Soft & Lagrangian relaxation \\ 
    \hline
    \cite{jin2020stability} & Power grid frequency regulation & Frequency stability & Ins/Hard & Lyapunov and RRL \\ %
    \hline
    \cite{gu2022recurrent} & Power grid frequency regulation & Frequency stability & Ins/Hard & Projection, Lyapunov and RRL \\ %
    \hline
    \cite{zhao2023barrier} & Transient stability of inverter-governed system & Transient stability & Ins/Hard & Barrier function method \\ 
    \bottomrule %
\end{tabular}
\end{adjustbox}
\end{table*}
From Table \ref{Table_StabilityControl}, it is evident that the current applications of safe RL in power systems span preventive and emergency control problems, as well as rotor angle stability control, frequency stability control, voltage stability control, damping control \cite{gupta2021coordinated}, flexible alternating current transmission system (FACTS) setpoint control \cite{tarle2023safe}, and transient stability control integrated with inverters \cite{zhao2023barrier}. However, the overall volume of research in this area remains limited, with only a few papers addressing each type of stability issue. Further research is needed to explore these stability domains more deeply, integrating their underlying mathematical dynamics.

In the following, we use frequency (F) control as a representative stability control example. The state, action, reward, and frequency‐dynamics constraints are outlined as follows:
\paragraph{Frequency Control by Safe RL}\mbox{}
Frequency control is a critical component of stability control in transmission power networks, ensuring a balance between power generation and demand to maintain system frequency \cite{obaid2019frequency, bevrani2021power, cui2022reinforcement}.

\textbf{State:} The state is the frequency $\omega$ and rotor angle $\delta$:
\begin{equation}
\bm s^\text{F} \triangleq \left(\bm \omega_t, \bm \delta_t \right)
\end{equation}

\textbf{Action:}
The control actions $\bm{a}_t$ are implemented through the control of active power injections:
\begin{equation}
\bm a^\text{F} \triangleq \left(\bm p^\text{SG}_t, \bm p^\text{RES}_t, \bm{p}^\text{Load}_t \right)
\end{equation}

\textbf{Reward:}
The reward is to minimize the frequency deviation $\Delta\omega$ and control action cost:
\begin{equation}
R^\text{F}(\bm{s}, \bm{a}) = - \sum_{i \in \mathcal{N}} \left(\|\Delta\omega_i\|_\infty + \lambda h_i(u_i)\right)
\end{equation}
where $\|\Delta\omega_i\|_\infty$ represents the maximum frequency deviation during the time horizon; the cost function $h_i(u_i)$ is a Lipschitz-continuous function; the cost coefficient $\lambda$ is used to balance the cost of actions relative to the frequency deviations.

\begin{figure}[htb]
    \centering
    \includegraphics[width=7.5cm]{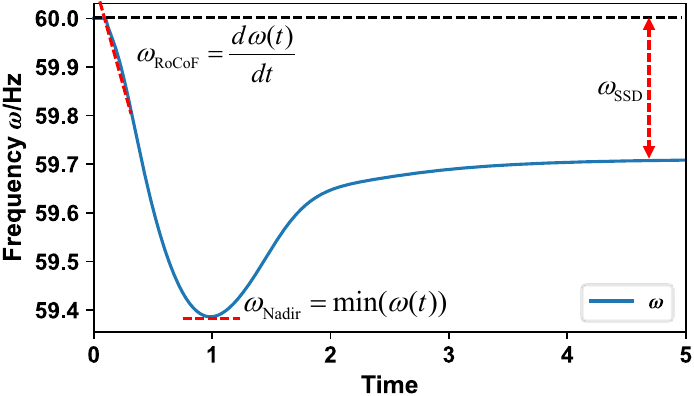}
    \caption{System frequency variations caused by sudden load increases or generator outages. $\omega_{\text{RoCoF}}$ represents the rate at which frequency changes, which is crucial in the initial stage of a disturbance and indicates the system's inertia response. $\omega_{\text{Nadir}}$ represents the lowest frequency reached after a disturbance, which is a critical metric in determining whether under-frequency load shedding will be triggered. $\omega_{\text{SSD}}$ represents the long-term frequency deviation from the nominal value, which depends on the frequency regulation strategy and reflects the system's steady-state frequency stability.}
\label{Fig_Frequency}
\end{figure}

\textbf{Constraint:}
The system frequency dynamics is given by the swing equation:
\begin{subequations}
\begin{gather}
\dot{\delta}_i = \omega_i\\
M_i \dot{\omega}_i = p^\text{Bus}_i - D_i \Delta\omega_i - a^\text{F}_{i}(\omega_i) - \sum_{j=1}^n B_{ij}\sin{(\Delta\delta)}
\end{gather}
\end{subequations}
where $\dot{\delta}$ and $\dot{\omega}$ represent the time derivatives $d\delta/dt$ and $d\omega/dt$, respectively; $M$ denotes the inertia constant; $D=\frac{1}{R} + L$ is the combined frequency response coefficient from SGs and load, where $\frac{1}{R}$ and $L$ denote speed droop response coefficient and load damping coefficient, respectively; $\sum_{j=1}^n B_{ij}\sin(\Delta\delta)$ denotes the electrical power $p_{e, i}$ at each node $i$; the mechanical power $p_{m, i}$ is expressed as $p^\text{Gen}_{i} - \frac{\omega_i}{R_i}$; the bus power injection $p^\text{Bus}_i$ is defined as $p^\text{Gen}_{i} - p^\text{Load}_{i}$. Other constraints include limits on line capacity, actions, rate of change of frequency (RoCoF) $\omega_{\text{RoCoF}}$, nadir $\omega_{\text{Nadir}}$, and steady-state deviation (SSD) $\omega_{\text{SSD}}$:
\begin{subequations}
\begin{gather}
|p_{ij}| \le \overline{p}_{ij} ~~~ \underline{\bm a}^\text{F} \le \bm a^\text{F}(\bm \omega)\le \overline{\bm a}^\text{F}\\
\bm a^\text{F}(\bm{\omega}) ~ \text{is stabilizing}\\
|\omega_{\text{RoCoF}}| \leq \overline{\omega}_{\text{RoCoF}} ~~~ \underline{\omega} \leq |\omega_{\text{Nadir}}| \leq  \overline{\omega} ~~~ |\omega_{\text{SSD}}| \leq \overline{\omega}_{\text{SSD}}
\end{gather}
\end{subequations}
where $p_{ij}$ denotes the active power of branch $ij$; the requirement that $\bm{a}^\text{F}(\bm{\omega})$ must be stabilizing is defined using various methods, such as Lyapunov stability \cite{cui2022reinforcement}. The system frequency variations caused by sudden load increases or generator outages are shown in Fig. \ref{Fig_Frequency}.

\paragraph{Other Stability Control by Safe RL}\mbox{}
In addition to frequency stability control, there are many other types of stability control problems summarized in Table \ref{Table_StabilityControl}. Given the wide variety of stability issues covered, individual models are not provided here. However, detailed methodologies are available in the referenced papers in Table \ref{Table_StabilityControl}, which offer further insights into each type of stability control.

\subsection{Real-Time Operation}\label{sec4dispatch}

Real-time power system operation refers to the continuous monitoring, control, and optimization of the power grid to ensure secure and economical operation while adapting to rapidly changing conditions. It involves managing various constraints, ranging from simplified formulations to comprehensive security constraints, including economic dispatch, DC-OPF, AC-OPF, and SCOPF. The real-time operation of a power system must meet both security and economic requirements. Among these, AC-OPF is widely used for considering credible contingencies \cite{li2021online, kocuk2016strong}. However, most existing methods for solving OPF rely on analytical approaches, which pose significant computational challenges due to the large-scale nature of these problems. SCOPF extends the standard OPF by enforcing $N-1$ security constraints for contingency scenarios, which greatly increases problem size and solution times \cite{yan2022hybrid}. To address this, methods such as DC-PF approximations \cite{marano2012exploiting}, convex PF approximations \cite{yan2020convex}, and convex security constraint approximations \cite{su2023analytic} have been proposed. While these methods can speed up computation, their accuracy has been questioned, and they remain time-intensive for large-scale systems. To overcome these limitations, RL has been applied to improve both speed and solution quality, but conventional RL often struggles with safety constraints. As a result, safe RL has been increasingly adopted, offering a promising approach to managing both computational efficiency and adherence to security constraints in real-time power system operations.

\subsubsection{Economic Dispatch}\mbox{}\label{sec4economic}

\begin{figure}[!htb]
    \centering
    \includegraphics[width=8.5cm]{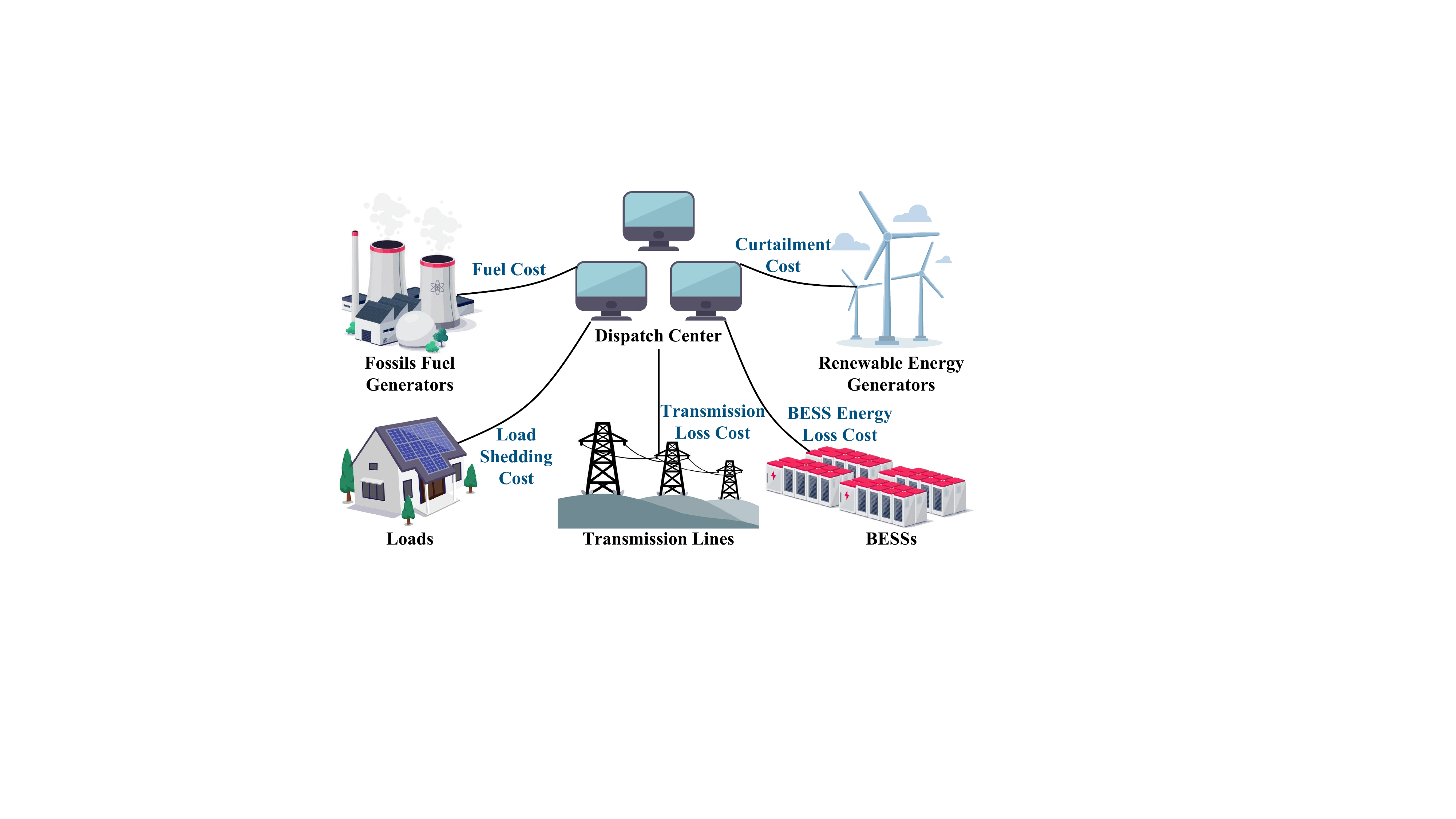}
    \caption{Schematic diagram of economic dispatch. The dispatch center coordinates power generation and consumption to minimize the total operational cost, which includes fuel costs from fossil fuel generators, curtailment costs from RESs, energy loss costs from BESSs, transmission loss costs, and load shedding costs.}
\label{Fig_Economic_Dispatch}
\end{figure}

Economic dispatch focuses on determining the optimal output of generating units to meet system demand at the lowest cost while satisfying operational constraints such as generator capacity limits and power balance. It plays a critical role in ensuring the economic efficiency of power system operation, particularly under varying load conditions and increasing integration of RESs. The schematic diagram of the power system economic dispatch is shown in Fig. \ref{Fig_Economic_Dispatch}.

A summary of existing papers applying safe RL to economic dispatch in power systems is shown in Table \ref{Table_Dispatch}.
\renewcommand{\arraystretch}{1.1}
\begin{table*}[htbp]
\caption{Safe RL Applications in Economic Dispatch}
\label{Table_Dispatch}
\centering
\begin{adjustbox}{center, max width=\textwidth}
\begin{tabular}{|>{\centering\arraybackslash}m{0.7cm} >{\centering\arraybackslash}m{6.0cm} >{\centering\arraybackslash}m{3.2cm} >
{\centering\arraybackslash}m{2.2cm} >{\centering\arraybackslash}m{4.0cm} |}
    \toprule %
    Ref. & Problem/Objective & Constraint & Constraint Type & Safety Techniques\\
    \midrule %
    \cite{li2022learning} & Total operating cost & System and devices & Cum/Soft & CPO \\ 
    \hline
    \cite{zhang2020multi} & Cost of microgrid & Global and local constraints & Cum/Soft & Lagrangian and projection \\ 
    \hline
    \cite{ye2023safe} & Costs of DGs production and RES curtailment & Distribution network & Cum/Soft & IPO \\ 
    \hline
    \cite{yi2023model} & Overall operation cost & Branch power flow security & Cum/Soft & Lagrangian relaxation and RRL \\ 
    \hline
    \cite{yan2022hybrid} & Total generation cost & Physical operation & Cum/Soft & Primal-dual method \\ 
    \hline
    \cite{wu2023constrained} & Fuel costs and power loss of BESSs & Physical constraints & Ins/Hard & Projection and primal-dual \\ 
    \hline
    \cite{sayed2023online} & Total operational cost & Gas and power systems & Cum/Soft & Lagrangian relaxation \\ 
    \hline
    \cite{yi2023real} & Operation cost & Operation & Ins/Hard & Safety layer, projection, and RRL \\ 
    \hline
    \cite{hong2023robust} & Total energy cost & Energy demand satisfaction & Cum/Soft & Lagrangian relaxation and RRL \\ 
    \hline
    \cite{liu2018distributed} & Total energy cost & Power system & Cum/Soft & Lagrange and logarithmic barrier \\ 
    \hline
    \cite{hao2024lyapunov} & Economics of microgrid & Physical constraints & Cum/Hard & Lyapunov method and RRL \\ %
    \hline
    \cite{wu2024real} & Real-time OPF & OPF & Cum/Soft & Lagrangian and action masking \\ 
    \hline
    \cite{sayed2022feasibility} & Generator fuel cost & Power system & Ins/Hard & Safety layer \\ 
    \hline
    \cite{chen2023improved} & Operating cost & Power system & Ins/Hard & Safety layer \\ 
    \hline
    \cite{zhang2024networked} & Operational cost & Operation and power & Cum/Hard & CPO and invalid action masking \\ 
    \hline
    \cite{shengren2023optimal} & Operating cost for the whole horizon & Operation & Ins/Hard & MILP formulation \\ 
    \hline
    \cite{yan2023real} & Total generation cost & Linguistic stipulation & Ins/Soft & Primal-dual and GPT \\ 
    \hline
    \cite{yan2020real} & Total operation cost & Operational constraints & Cum/Soft & Lagrangian relaxation \\ 
    \hline
    \cite{ceusters2023safe} & Multi-energy management & Thermal energy balance & Ins/Hard & Shielding method \\ 
    \hline
    \cite{wang2023secure} & Cost of electricity net, DG and gas & Power and gas networks & Ins/Hard & Safety layer \\ 
    \bottomrule %
\end{tabular}
\end{adjustbox}
\end{table*}
From Table \ref{Table_Dispatch}, it is evident that the primary objective functions across studies include minimizing operating costs, fuel costs, and RES curtailment costs. These objectives are pursued while ensuring reliable power supply and maintaining operational safety. While the GP method has not yet been applied in any domain and the Lyapunov method is not particularly suitable for economic dispatch, other safe RL techniques have been extensively utilized. Economic dispatch also stands out as the category with the largest number of publications. The studies also demonstrate integration with other NNs and optimization techniques, including edge-conditioned convolutional networks \cite{ye2023safe}, long short-term memory networks \cite{ye2023safe}, MILP formulations \cite{shengren2023optimal}, and GPT LLM \cite{yan2023real}, to address the complexities inherent in economic dispatch. Future research in economic dispatch should focus on addressing two critical challenges. First, the uncertainty from high RES penetration requires robust safe RL frameworks that can accommodate variability and unpredictability in generation and demand \cite{yi2023model}. Second, deploying safe RL in large‐scale power systems remains challenging because real‐world applications demand high computational efficiency and scalability.

In the following, we summarize the core framework for applying safe RL to economic dispatch using an example that includes SGs, RESs and BESSs while enforcing strict physics-based constraints such as AC/DC-PF. These equations can be easily extended to additional power system devices. The state, action, reward, and constraints are outlined as follows:

\paragraph{Safe RL for Economic Dispatch with AC-PF} AC-PF constraints describe the basic physics of power systems, which have been widely considered in OPF, voltage control, unit commitments, etc \cite{wu2023constrained}.

\textbf{State:} 
The states include active and reactive loads and voltage:
\begin{equation}
\label{Eq_AC_State}
\bm{s}^\text{AC}_t \triangleq \left(\bm{v}_t, \bm{p}^\text{Load}_t, \bm{q}^\text{Load}_t \right)
\end{equation}

\textbf{Action:}
The control actions encompass both active and reactive power generation of SGs, active power generation of RESs, alongside power charging or discharging of BESSs:
\begin{equation}
\label{Eq_AC_Action}
\bm{a}^\text{AC}_t \triangleq \left(\bm{p}^\text{SG}_t, \bm{q}^\text{SG}_t, \bm{p}^\text{RES}_t, \bm{p}^\text{BESS}_{\text{ch},t}, \bm{p}^\text{BESS}_{\text{dis}, t} \right)
\end{equation}

\textbf{Reward:} The reward includes SG generation cost, RES curtailment cost, and BESS operating cost:
\begin{subequations}
\label{Eq_AC_Reward}
\begin{alignat}{2}
\max_{\pi_\theta \in \Pi_{S}} & \mathbb{E}_{\tau \sim \pi} \left[ \sum_{t=0}^{\infty} \gamma^t R(\bm{s}_t, \bm{a}_t, \bm{s}_{t+1}) \right] \\
R^\text{AC}(\bm{s}, \bm{a}) &= -\left| \sum_{\forall i\in \mathcal{G}} \left(a^\text{SG}_i (p^\text{SG}_{i, t})^2 + b^\text{SG}_i p^\text{SG}_{i, t} + c^\text{SG}_i\right) \right| \notag \\
&\quad -\sum_{\forall i\in \mathcal{R}} c^\text{RES}_i \left|\hat{p}^\text{RES}_{i, t} - p^\text{RES}_{i, t}\right| \notag \\
&\quad -\sum_{\forall i\in \mathcal{B}} c^\text{BESS}_{\text{dis}, i}p^\text{BESS}_{\text{dis}, i, t} + \sum_{\forall i\in \mathcal{B}} c^\text{BESS}_{\text{ch}, i} p^\text{BESS}_{\text{ch}, i, t} \\
\bm{s}^\text{AC}_{t} &= f_t(\bm{s}^\text{AC}_{t-1}, \bm{a}^\text{AC}_{t-1} ) ~~~ \bm{a}^\text{AC}_t \sim \pi(\bm a^\text{AC}_t|\bm s^\text{AC}_{t-1})
\end{alignat}
\end{subequations}
where $a^\text{SG}$, $b^\text{SG}$, and $c^\text{SG}$ denote the quadratic, linear, and fixed fuel cost coefficients of SG, respectively; $c^\text{RES}$ and $c^\text{BESS}$ denote the cost coefficients of RES and BESS, respectively; $\hat{p}^\text{RES}$ denotes the predicted maximum RES output based on weather conditions.

\textbf{Constraint:}
The control actions derived from DRL must adhere to physics-hard constraints. AC-OPF constraints include bus active and reactive power balance constraints, SG active and reactive power generation constraints, RES active power generation constraints, voltage constraints, and branch apparent power constraints:
\begin{subequations}
\label{Eq_AC}
\begin{gather}
\mathbf{M}^\text{BESS}\bm{p}^\text{BESS}_{\text{dis}, t} - \mathbf{M}^\text{BESS}\bm{p}^\text{BESS}_{\text{ch},t} + \mathbf{M}^\text{SG} \bm{p}_t^\text{SG} + \notag \\
\mathbf{M}^\text{RES} \bm{p}_t^\text{RES} - \bm{p}^\text{Load}_t = \Re\{\mathbb{D}(\bm{v}_t \bm{v}_t^\mathcal{H} \mathbf{Y}^\mathcal{H})\}\\
\mathbf{M}^\text{SG} \bm{q}_t^\text{SG} - \bm{q}^\text{Load}_t = \Im\{\mathbb{D}(\bm{v}_t \bm{v}_t^\mathcal{H} \mathbf{Y}^\mathcal{H})\} \label{ACPF_cst2}\\
\underline{\bm{p}}^\text{SG} \le \bm{p}^\text{SG}_t \le \overline{\bm{p}}^\text{SG} ~~~ \underline{\bm{q}}^\text{SG} \le \bm{q}^\text{SG}_t \le \overline{\bm{q}}^\text{SG}\\
\underline{\bm{p}}^\text{RES} \le \bm{p}^\text{RES}_t \le \overline{\bm{p}}^\text{RES} ~~~ \underline{\bm{v}} \le |{\bm{v}}| \le \overline{\bm{v}} ~~~ |{s}_{ij}|\le \overline{s}_{ij}
\end{gather}
\end{subequations}
where $\Re$ and $\Im$ return a complex number’s real and imaginary parts, respectively; $\mathbb{D}$ returns a vector consisting of the diagonal elements of a matrix; $\mathcal{H}$ denotes Hermitian conjugate of a vector or matrix; $\mathbf{Y}$ is the admittance matrix; $G$ and $N$ denote cardinality of the set $\mathcal{G}$ and $\mathcal{N}$, respectively; $\mathbf{M}^\text{SG}$ denotes the matrix $\{0,1\}^{N\times G}$ that maps the SG generation vector $\bm{p}_t^\text{SG} \in \mathbb{R}^{G}$ to $\mathbb{R}^{N}$:
\begin{subequations}
\begin{gather}
[\mathbf{M}^\text{SG} \bm{p}_t^\text{SG}]_i = 0~~~ [\mathbf{M}^\text{SG} \bm{q}_t^\text{SG}]_i = 0, ~~\forall i \in \mathcal{N} \setminus \mathcal{G}\\
[\mathbf{M}^\text{SG} \bm{p}_t^\text{SG}]_i = p^\text{SG}_j~~~ [\mathbf{M}^\text{SG} \bm{q}_t^\text{SG}]_i = q^\text{SG}_j, ~~\forall i \in \mathcal{N}, \forall j \in \mathcal{G}
\end{gather}
\end{subequations}

\paragraph{Safe RL for Economic Dispatch with DC-PF} DC-PF constraints represent the linear relaxations of AC-PF, which are commonly included in DC-OPF and electricity market considerations \cite{wu2023pesgm}.

\textbf{State:}
The voltage and reactive power are overlooked in DC-PF.
\begin{equation}
\bm{s}^\text{DC}_t \triangleq \left(\bm{\vartheta}_t, \bm{p}^\text{Load}_t \right)
\end{equation}
where $\vartheta$ is the grid state in the DC-PF approximation.

\textbf{Action:}
The action involves only the generation or consumption of active power.
\begin{equation}
\bm{a}^\text{DC}_t \triangleq \left(\bm{p}^\text{SG}_t, \bm{p}^\text{RES}_t, \bm{p}^\text{BESS}_{\text{ch},t}, \bm{p}^\text{BESS}_{\text{dis}, t} \right) 
\end{equation}

\textbf{Reward:}
The reward is similar to the AC-PF \eqref{Eq_AC_Reward}.

\textbf{Constraint:}
The DC-OPF constraints are a simplification of the AC-OPF constraints, retaining only the active power components and disregarding voltage issues \cite{wu2023pesgm}.
\begin{subequations}
\label{Eq_DC}
\begin{gather}
\mathbf{M}^\text{BESS}\bm{p}^\text{BESS}_{\text{dis}, t} - \mathbf{M}^\text{BESS}\bm{p}^\text{BESS}_{\text{ch},t} + \mathbf{M}^\text{SG} \bm{p}_t^\text{SG} + \notag\\
\mathbf{M}^\text{RES} \bm{p}_t^\text{RES} - \bm{p}^\text{Load}_t = \mathbf{B} \bm{\vartheta}_t\\
\underline{\bm{p}}^\text{SG} \le \bm{p}^\text{SG}_t \le \overline{\bm{p}}^\text{SG} ~~ \underline{\bm{p}}^\text{RES} \le \bm{p}^\text{RES}_t \le \overline{\bm{p}}^\text{RES} ~~ 
|{p}_{ij}|\le \overline{p}_{ij}
\end{gather}
\end{subequations}
where $\mathbf{B}$ is the susceptance matrix. It is important to note that \eqref{Eq_AC} and \eqref{Eq_DC} are suitable for transmission networks and three-phase balanced distribution networks. However, for application in three-phase unbalanced distribution networks, they need to be extended to incorporate three-phase modeling.

\textbf{BESS Constraints}:
The BESS constraints include charging and discharging constraints, and SoC constraints:
\begin{subequations}
\label{Eq_BESS}
\begin{gather}
0 \le \bm{p}^\text{BESS}_{\text{ch}, t} \le \overline{\bm{p}}^\text{BESS}_\text{ch} ~~~ 0\le \bm{p}^\text{BESS}_{\text{dis}, t} \le \overline{\bm{p}}^\text{BESS}_\text{dis}\\
\underline{\bm{SoC}}^\text{BESS} \le \bm{SoC}^\text{BESS}_t \le \overline{\bm{SoC}}^\text{BESS}\\
\bm{SoC}^\text{BESS}_t = \bm{SoC}^\text{BESS}_{t-1} + \frac{\Delta t}{E_\text{cap}^\text{BESS}} \Big(\eta^\text{BESS}_\text{ch} \bm{p}^\text{BESS}_{\text{ch}, t} - \frac{\bm{p}^\text{BESS}_{\text{dis}, t}}{ \eta^\text{BESS}_\text{dis}} \Big)
\end{gather}
\end{subequations}
where $\eta^\text{BESS}_\text{ch}$ and $\eta^\text{BESS}_\text{dis}$ denote the efficiency of charging and discharging of BESS, respectively; $E_\text{cap}^\text{BESS}$ denotes the energy capacity of BESS.

\subsubsection{System Restoration}\mbox{}

\begin{figure}[!htb]
    \centering
    \includegraphics[width=8.5cm]{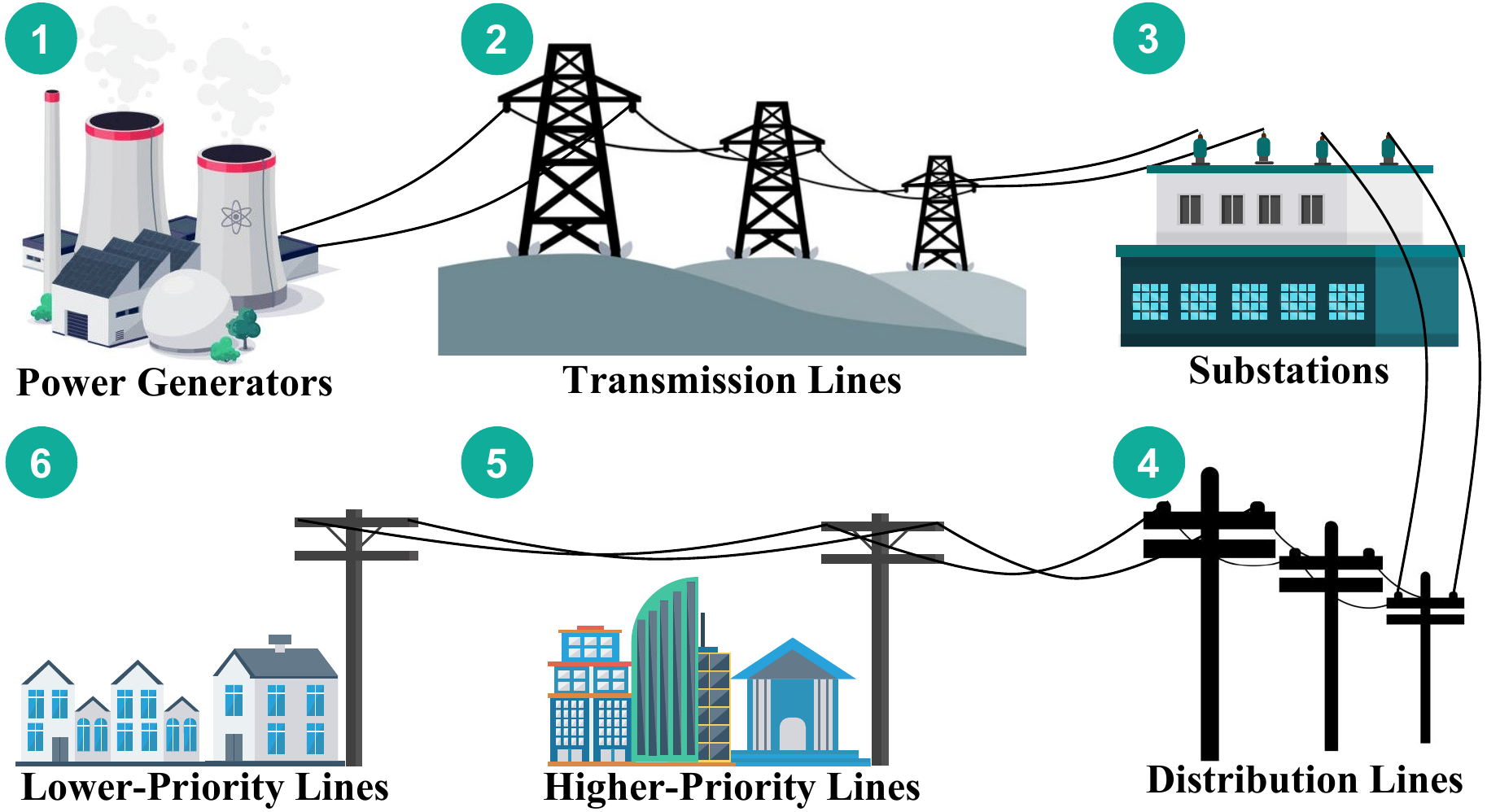}
    \caption{System restoration: Sequential recovery from generation facilities, transmission lines, substations, distribution lines, high-priority lines, to low-priority lines.}
\label{Fig_Restoration}
\end{figure}

System restoration is another critical aspect of power flow dispatch that involves swiftly recovering the power system from an impacted or blackout state to normal operation following extreme events, such as natural disasters or system-wide failures \cite{liang2024power}. Its primary objective is to re-energize affected areas quickly and safely, minimizing downtime and its economic and societal repercussions. This process typically involves complex decision-making to coordinate generation, transmission, and load recovery while maintaining system stability and adhering to operational constraints. The main process of system restoration is illustrated in Fig. \ref{Fig_Restoration}, which includes the restoration of power system equipment based on a prioritized sequence.

Studies such as \cite{du2022deep, zhang2023primal} have developed system restoration strategies using safe RL, either by controlling local DERs or by transferring load to safe areas, as shown in Table \ref{Table_OtherControl}. However, research on system restoration using safe RL remains limited, and further exploration is needed to address the unique challenges associated with these scenarios. In particular, extreme natural weather events, which are characterized by high impact but low probability, pose significant challenges to system restoration. These events often lead to severe disruptions, complex recovery conditions, and the need for robust, adaptive strategies to handle the heightened uncertainty and variability \cite{bie2017battling, xu2021resilience}. Future research should focus on developing safe RL frameworks specifically tailored for such extreme scenarios. This includes improving the generalization and robustness of safe RL algorithms to handle rare and unpredictable events. It also involves integrating real-time weather forecasting and system data to enhance situational awareness, and developing scalable solutions for large-scale systems with interconnected networks.

In the following, we provide an example of system restoration using safe RL. The state, action, reward, and constraints are shown as follows:

\textbf{State:}
The state includes the future RES output forecasting $\bm p_{t+1}^\text{RES}$, past restored loads $\bm p_{t-1}^\text{Load}$, current SoC of the BESSs $\bm {SoC}_{t}^\text{BESS}$, and remaining reserves of various types of generators $\overline{\bm p}_{t}^\text{Gen} - \bm p_{t}^\text{Gen}$.
\begin{equation}
\bm{s}^\text{Restoration}_t \triangleq \left(\bm p_{t+1}^\text{RES}, \bm p_{t-1}^\text{Load}, \bm {SoC}_{t}^\text{BESS}, \overline{\bm p}_{t}^\text{Gen} - \bm p_{t}^\text{Gen} \right) 
\end{equation}

\textbf{Action:}
The action includes the restored load $\bm p_{\text{restored}, t}^\text{Load}$, active power output of all kinds of generators $\bm p_{t}^\text{Gen}$ and BESSs $\bm p_{t}^\text{BESS}$. 
\begin{equation}
\bm{a}^\text{Restoration}_t \triangleq \left(\bm p_{\text{restored}, t}^\text{Load}, \bm p_{t}^\text{Gen}, \bm p_{t}^\text{BESS} \right) 
\end{equation}

\textbf{Reward:}
The reward is to maximize the sum of restored loads $\sum \bm p_{\text{restored}, t}^\text{Load}$.
\begin{equation}
R^\text{Restoration}(\bm{s}, \bm{a})= \sum \bm p_{\text{restored}, t}^\text{Load}
\end{equation}

\textbf{Constraint:}
System restoration requires adherence to fundamental power system operational constraints and equipment constraints, including AC-PF constraints \eqref{Eq_AC}, DC-PF constraints \eqref{Eq_DC}, BESS constraints \eqref{Eq_BESS}, etc., all of which have been detailed above. In addition, it is necessary to add constraints to ensure that the load is restored monotonically:
\begin{equation}
\label{Eq_load_restoration}
\bm p_{\text{restored}, t}^\text{Load} \le \bm p_{\text{restored}, t+1}^\text{Load}
\end{equation}

\renewcommand{\arraystretch}{1.1}
\begin{table*}[htbp]
\caption{Safe RL Applications in System Restoration, Unit Commitment, and Electricity Market}
\label{Table_OtherControl}
\centering
\begin{adjustbox}{center, max width=\textwidth}
\begin{tabular}{|>{\centering\arraybackslash}m{0.7cm} >{\centering\arraybackslash}m{4.0cm} >{\centering\arraybackslash}m{3.9cm} >
{\centering\arraybackslash}m{2.4cm} >{\centering\arraybackslash}m{4.7cm} |}
    \toprule %
    Ref. & Problem/Objective & Constraint & Constraint Type & Safety Techniques\\
    \midrule %
    \multicolumn{5}{|l|}{\textbf{System Restoration}} \\
    \hline
    \cite{du2022deep} & Service restoration & Power flow and voltage & Cum+Ins/Hard+Soft & Action clipping and penalty term \\ 
    \hline
    \cite{zhang2023primal} & Critical load restoration & Loads, DERs, ESSs & Cum/Soft & Primal-dual differentiable programming \\ 
    \hline
    \cite{vu2023safe} & Load restoration & Restoration & Ins/Hard & Invalid action masking \\
    \midrule
    \multicolumn{5}{|l|}{\textbf{Unit Commitment}} \\
    \hline
    \cite{ajagekar2022deep} & Unit commitment & Scheduling & Ins/Hard & Clipping \\ 
    \hline
    \cite{li2022risk} & Reserve scheduling & Voltage, RESs, tie line, and ESSs & Cum/Soft & Primal-dual method \\ 
    \midrule
    \multicolumn{5}{|l|}{\textbf{Electricity Market}} \\
    \hline
    \cite{shi2023augmented} & Scheduling of EV aggregators & EVs and driver's energy demand & Cum/Soft & Lagrangian relaxation \\ 
    \hline
    \cite{zhu2023budget} & V2G market & Maximum incentive & Cum/Soft & Primal-dual theories \\ 
    \hline
    \cite{yang2023dynamic} & Pricing strategy for congestion & Charging station, operator, grid & Cum/Soft & Adaptive constraint cost \\ 
    \hline
    \cite{lu2024sma} & Industrial parks energy trading & Market clearing mechanism & Cum/Soft & Lagrangian relaxation \\    
    \bottomrule %
\end{tabular}
\end{adjustbox}
\end{table*}

\subsection{Operational Planning}
Operational planning in power systems focuses on strategic, slow-timescale decision-making to ensure long-term system reliability, efficiency, and resilience. It encompasses tasks designed to anticipate and address future uncertainties, such as variations in demand, RES integration, and equipment availability, typically over day-ahead or even longer horizons. Unlike security control or real-time operation, which deal with immediate system stability and fast-paced adjustments, operational planning prioritizes systematic optimization and resource allocation over extended time frames. Operational planning involves tasks like unit commitment and electricity market.

\subsubsection{Unit Commitment}\mbox{}

Unit commitment schedules generating units to meet anticipated demand over a specified time horizon, typically day-ahead or longer \cite{zheng2014stochastic}. It determines each unit’s on/off status and output levels while minimizing operational costs, including fuel, start-up/shut-down and maintenance expenses \cite{abdi2021profit}. At the same time, unit commitment must satisfy constraints such as generator capacity limits, minimum up/down times, ramping limits, and reserve requirements. This problem is inherently complex due to its combinatorial nature, involving both discrete decisions (e.g., unit on/off states) and continuous variables (e.g., generation levels) \cite{yang2021comprehensive}. Traditionally, unit commitment has been addressed using mathematical optimization techniques such as MILP and dynamic programming \cite{putz2021comparison}. However, both methods face significant challenges when applied to large-scale power systems.

\begin{figure}[!htb]
    \centering
    \includegraphics[width=8.5cm]{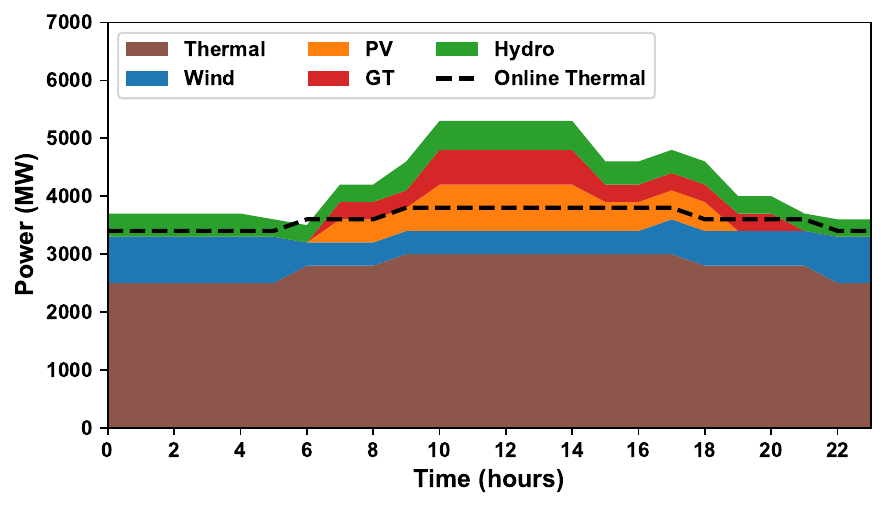}
    \caption{Example of unit commitment. The stacked areas represent actual dispatch levels of different units, while the dashed line indicates the total capacity of the committed thermal generators as determined in the unit commitment decision.}
\label{Fig_UC}
\end{figure}

Studies such as \cite{ajagekar2022deep, li2022risk} utilize safe RL to develop strategies for unit commitment and coordinated tie-line energy storage management, respectively, as shown in Table \ref{Table_OtherControl}. However, the current research on applying safe RL to unit commitment remains limited and requires further expansion. On one hand, there is a need to develop more advanced safe RL methods that effectively integrate domain knowledge and power system-specific techniques. On the other hand, addressing challenges associated with the integration of RESs is crucial, as their inherent uncertainty and variability can significantly impact the reliability of unit commitment decisions \cite{quan2015computational, quan2016integration}. Future efforts should focus on creating robust frameworks capable of managing these uncertainties while leveraging the strengths of safe RL to enhance the adaptability and efficiency of the power system \cite{wang2022distributionally}. An example of unit commitment is shown in Fig. \ref{Fig_UC}, which includes various types of generators and their power output distribution over a 24-hour period.

Subsequently, we illustrate how safe RL applies to unit commitment and reserve scheduling. The state, action, reward, and constraints are shown as follows:

\textbf{State:}
The state includes the historical and current net load forecasts $\bm p^\text{Load}_\text{his/pre}$, start-up, shut-down, and commitment decisions at the previous stage:
\begin{equation}
\bm{s}^\text{Reserve}_t \triangleq \left(\bm p^\text{Load}_\text{his}, \bm p^\text{Load}_\text{pre}, \bm u_{\text{start}, t-1}, \bm u_{\text{shut}, t-1}, \bm u_{\text{com}, t-1} \right) 
\end{equation}
where $\bm u_\text{start}$, $\bm u_\text{shut}$ and $\bm u_\text{com}$ denote the startup, shutdown and commitment status of generators, respectively.

\textbf{Action:} The action includes the current start-up, shut-down, and commitment decisions $\bm u_{\text{start/shut/com}, t}$, power output of generator $\bm p^\text{Gen}_t$:
\begin{equation}
\bm{a}^\text{Reserve}_t \triangleq \left(\bm u_{\text{start}, t}, \bm u_{\text{shut}, t}, \bm u_{\text{com}, t}, \bm p^\text{Gen}_t \right)
\end{equation}

\textbf{Reward:}
The reward is to minimize the overall costs, including the cost of power generation $R^\text{Gen}_\text{cost}$, commitment costs $R^\text{Com}_\text{cost}$, and start-up and shut-down costs $R^\text{Start/Shut}_\text{cost}$:
\begin{equation}
R^\text{Reserve}(\bm{s}, \bm{a})= -(R^\text{Gen}_\text{cost} + R^\text{Com}_\text{cost} + R^\text{Start}_\text{cost} + R^\text{Shut}_\text{cost})
\end{equation}

\textbf{Constraint:} The constraints include generator limits, minimum up-time and down-time constraints, logical relationship between the generator commitment decisions and start-up/shut-down decisions, power generation and reserve constraints, ramp-up and ramp-down limits of generators, and integrality requirement of commitment and start-up/shut-down decisions. For more details, refer to \cite{ajagekar2022deep}.

\subsubsection{Electricity Market}\mbox{}

\begin{figure}[!htb]
    \centering
    \includegraphics[width=8.5cm]{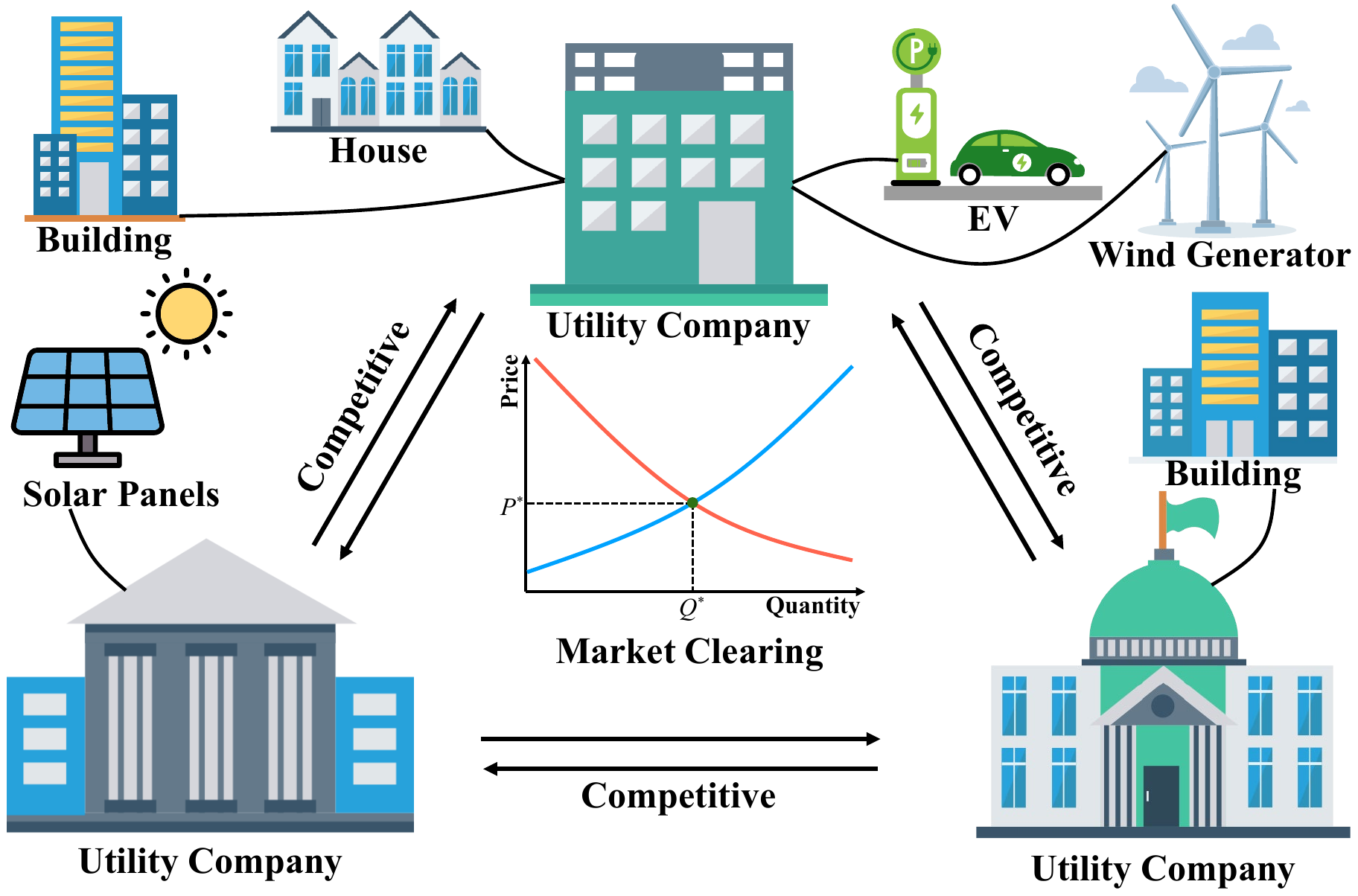}
    \caption{Competitive electricity market model. In a competitive electricity market framework, various entities interact through market mechanisms, acting as buyers or sellers of electricity by submitting bids or offers to a centralized utility or market operator. The market clearing process determines the equilibrium price and quantity based on the intersection of supply and demand curves.}
\label{Fig_Electricity_Market}
\end{figure}

The electricity market enables efficient resource allocation by facilitating electricity trading among generators, suppliers, and consumers \cite{karthikeyan2013review}. It operates on the principles of supply and demand, aiming to achieve economic efficiency while maintaining grid reliability. Electricity markets are typically structured into different timeframes, including day-ahead, intraday, and real-time markets, with each serving specific operational needs \cite{bjarghov2021developments}. These markets involve tasks such as determining electricity prices, scheduling generation, and ensuring sufficient reserves to meet demand fluctuations \cite{weron2014electricity}. Traditional methods for electricity market operations rely on optimization techniques such as MILP or MINLP \cite{alemany2018effects, canizes2013mixed}. These methods aim to minimize total operational costs while satisfying constraints such as power balance, generator limits, and transmission capacities. Additionally, game-theoretic approaches are employed to model the strategic interactions between market participants, providing insights into bidding strategies and market equilibrium \cite{hong2023bilevel}. With the increasing integration of RESs and the growing complexity of power systems, advanced techniques such as stochastic optimization and robust optimization have been introduced to account for uncertainties in generation and demand \cite{hakimi2021stochastic}. The competitive electricity market model is illustrated in Fig. \ref{Fig_Electricity_Market}.

In addition, safe RL has been applied in electricity markets. For example, \cite{yang2023dynamic} employs safe RL to formulate dynamic pricing strategies for controlling shiftable loads such as EVs, and HVAC systems. While some have used NNs to predict the optimal marginal prices of the OPF, such as in \cite{zhang2021predicting}, these approaches do not derive a stochastic policy. A summary of safe RL applications in electricity markets is presented in Table \ref{Table_OtherControl}. From Table \ref{Table_OtherControl}, it is evident that the current research on applying safe RL to electricity market operations remains limited, highlighting the need for further exploration. One of the key challenges lies in adapting safe RL frameworks to align with existing market rules and regulatory structures, which often vary significantly across regions and market types \cite{tsaousoglou2022market}. Safe RL must also address the complexity of making decisions in stochastic and dynamic environments, where uncertainties in demand, RES generation, and market conditions play a critical role \cite{hong2023bilevel}. Another significant challenge is the representation of certain constraints that are inherently data-driven and cannot be easily expressed in an analytical form, such as bidding behaviors in electricity markets \cite{tang2022multi}. These behaviors are influenced by the strategic interactions of market participants and can vary widely depending on historical data, participant strategies, and market conditions \cite{wu2022strategic}. Effectively integrating these data-driven constraints into safe RL frameworks requires innovative approaches, such as the incorporation of data-driven models, behavioral models, uncertainty quantification, or game-theoretic principles to capture the complexities of market dynamics \cite{ren2023reinforcement, qiu2022strategic, tang2022multi}.

The following example outlines how safe RL can optimize electricity pricing strategies in electricity markets. The state, action, reward, and constraints are as follows \cite{shi2023augmented, yang2023dynamic, zhu2023budget}:

\textbf{State:}
The state includes observed status information of the charging station (CS) and the distribution system operator (DSO), including the total cost of EV CSs $\bm s_\text{cost}^\text{CS}$ and the total cost of DSO $\bm s_\text{cost}^\text{DSO}$.
\begin{equation}
\bm{s}^\text{Market}_t \triangleq \left(\bm s_\text{cost}^\text{CS}, \bm s_\text{cost}^\text{DSO} \right) 
\end{equation}

\textbf{Action:}
The action denotes the incentive electricity price of different EV CSs $\bm\Lambda^\text{CS}$.
\begin{equation}
\bm{a}^\text{Market}_t \triangleq \left(\bm\Lambda^\text{CS} \right) 
\end{equation}

\textbf{Reward:}
The reward is to minimize the cost of EV users and maximize the profits of CSs and DSOs by setting different electricity prices.
\begin{equation}
R^\text{Market}(\bm{s}, \bm{a})= -R^\text{User} + R^\text{CS} + R^\text{DSO}
\end{equation}

\textbf{Constraint:}
In the electricity market, EVs are key participants, and their model is presented in Section \ref{sec4EV}.

\subsection{Emerging Areas}\label{sec4Emerging}
In recent years, some emerging areas have surfaced in power systems, where safe RL has been utilized to address challenges arising from the stochastic, dynamic, and complex nature of modern power systems. Among these diverse applications, two prominent areas stand out: EV charging and building energy management.

\subsubsection{EV Charging}\label{sec4EV}\mbox{}

\begin{figure}[!htb]
    \centering
    \includegraphics[width=8.5cm]{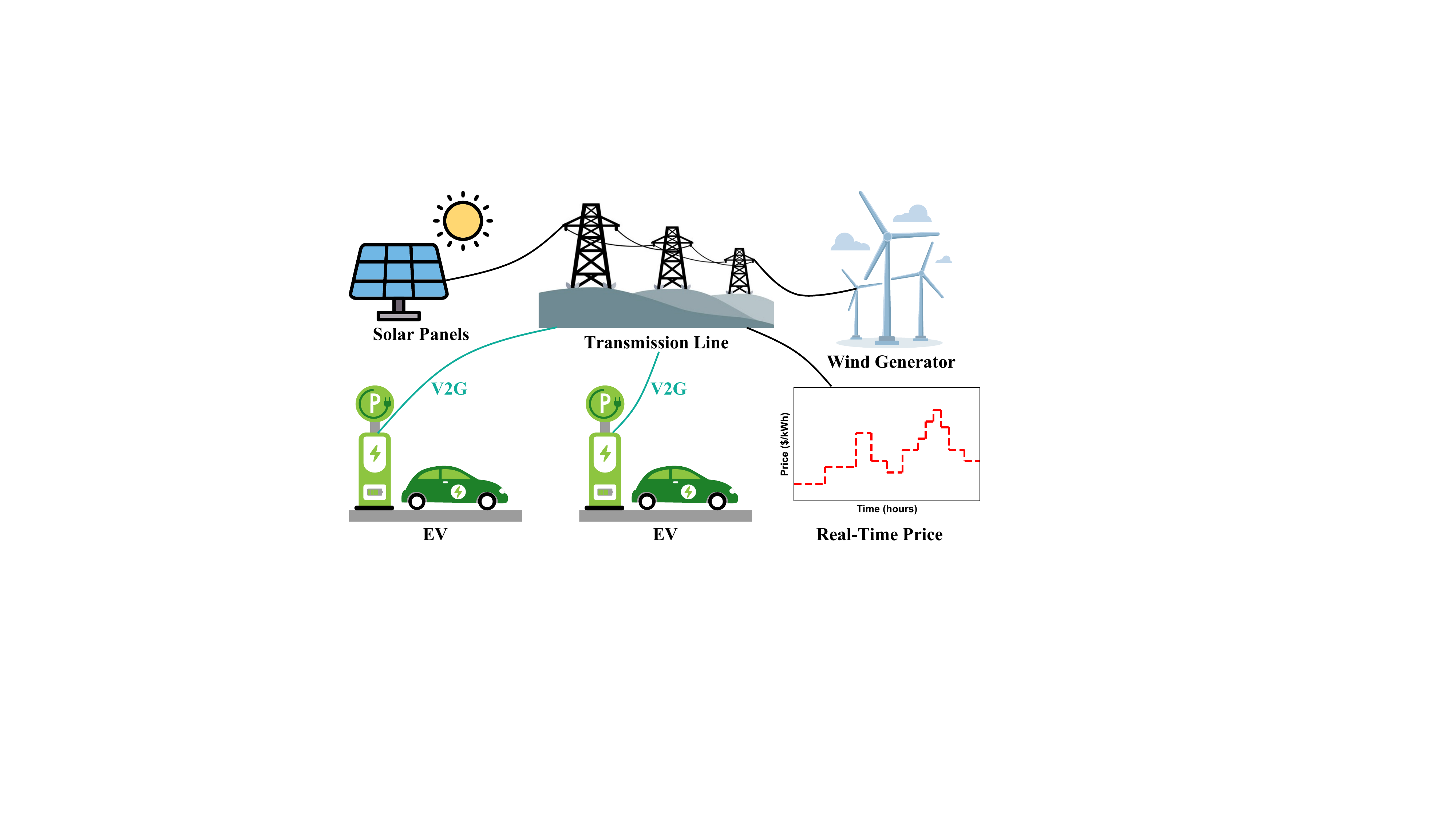}
    \caption{EV charging model based on real-time prices. In V2G mode, EVs can both draw energy from the grid and inject stored energy back into it, depending on real-time electricity prices.}
\label{Fig_EV}
\end{figure}

The Paris Agreement highlights EVs as a key means of reducing carbon emissions, spurring rapid global adoption. EVs' penetration reached almost 30 million in 2022 and is expected to grow to about 240 million by 2030 in the stated policies scenario, achieving an average annual growth rate of about 30\%. Based on this trend, EVs will account for over 10\% of the road vehicle fleet by 2030 \cite{iea2023global}. A real-time price-based EV charging model is illustrated in Fig. \ref{Fig_EV}.
However, the stochastic nature of EV charging can introduce unpredictable peak loads and voltage deviations in the power system. To address these issues, demand response for EVs has been proposed to mitigate grid peak loads and reduce charging costs. The complexity of optimizing EV charging lies in managing uncertainties related to charging demand, electricity prices, required charging energy, and V2G operations where EVs can sell electricity back to the grid.

Safe RL has shown effectiveness in tackling these challenges by training charging strategies that minimize costs, align SoC targets, and mitigate grid impacts \cite{chen2022deep, zhang2023safe, li2019constrained}. Additionally, safe RL methods have been used to design dynamic pricing strategies that balance supply and demand, reduce operational costs for distribution system operators, and incentivize consumer participation \cite{yang2023dynamic, shi2023augmented, zhu2023budget}. A summary of safe RL applications in EV charging is provided in Table \ref{Table_EVCharging}.
\renewcommand{\arraystretch}{1.1}
\begin{table*}[htbp]
\caption{Safe RL Applications in EV Charging}
\label{Table_EVCharging}
\centering
\begin{adjustbox}{center, max width=\textwidth}
\begin{tabular}{|>{\centering\arraybackslash}m{0.7cm} >{\centering\arraybackslash}m{6.0cm} >{\centering\arraybackslash}m{3.2cm} >
{\centering\arraybackslash}m{2.2cm} >{\centering\arraybackslash}m{4.0cm} |}
    \toprule %
    Ref. & Problem/Objective & Constraint & Constraint Type & Safety Techniques\\
    \midrule %
    \cite{wu2023network} & Optimal EV charging control & EV & Ins/Hard & Lagrangian and projection \\ 
    \hline
    \cite{jiang2021data} & Smooth out the load profile of a parking lot & EV charging and bound & Ins/Hard & Penalty function and projection \\ 
    \hline
    \cite{chen2022deep} & Minimize the EV charging cost & Entropy and SoC deviation & Cum/Soft & Lagrangian relaxation \\ 
    \hline
    \cite{zhang2023safe} & Maximize the total profit & Power and demands & Cum/Soft & Lagrangian relaxation \\ 
    \hline
    \cite{li2019constrained} & Maximize the revenue of electricity selling & EV charging & Cum/Soft & Lagrangian relaxation \\ 
    \hline
    \cite{zhang2020deep} & Energy management for plug-in hybrid EV & Physical components & Cum/Soft & Lagrangian relaxation \\ 
    \hline
    \cite{liessner2019safe} & Minimize the vehicle energy consumption & Battery power bound & Ins/Hard & Shielding method \\ 
    \hline
    \cite{guan2024rule} & Minimize the charging costs & Voltage and EV security & Ins/Hard & Shielding method \\ 
    \bottomrule %
\end{tabular}
\end{adjustbox}
\end{table*}
In Table \ref{Table_EVCharging}, the primary goals of these studies include minimizing charging costs, maximizing profits from electricity sales, smoothing load profiles \cite{jiang2021data}, and managing energy distribution with EVs \cite{zhang2020deep}. Some of these applications also incorporate advanced features such as V2G operations \cite{zhang2023safe}, non-linear charging behaviors, and stochastic factors like arrival and departure times  \cite{li2019constrained}, remaining energy, and real-time electricity prices \cite{chen2022deep}. In terms of specific safe RL technologies, most papers employ methods based on Lagrangian relaxation, projection methods, and shielding methods. Further exploration is needed to expand the application of other safe RL techniques in EV charging scenarios. Key future research areas include enhancing the scalability of safe RL for large-scale EV networks, incorporating real-time data (e.g., electricity prices, traffic, weather) for adaptability, and using multi-agent RL to coordinate distributed charging stations for efficient grid utilization. Addressing uncertainties in EV behavior, such as arrival and departure times, through probabilistic modeling is important. Using hybrid frameworks that combine traditional optimization with safe RL will improve both efficiency and interpretability \cite{abdullah2021reinforcement, nimalsiri2019survey}.

The following example outlines how safe RL can be configured with safety constraints for EV charging to minimize charging costs. The state, action, reward, and constraints are shown as follows:

\textbf{State:}
The state includes SoC $\bm{SoC}^{\text{EV}}_t$, remaining demand $\bm e^{\text{EV}}_t$, residual parking time $\bm t^{\text{EV}}_p$, charging price $\bm\Lambda^{\text{EV}}_{\text{ch}, t}$, V2G selling price $\bm\Lambda^{\text{EV}}_{\text{dis}, t}$, RES generation $\bm{p}^{\text{RESs}}_{t}$, and other load demand $\bm{p}^{\text{Load}}_{t}$ \cite{zhang2023safe, chen2022deep}:
\begin{equation}
\bm{s}^\text{EV}_t \triangleq \left(\bm{SoC}^{\text{EV}}_t, \bm e^{\text{EV}}_t, \bm t^{\text{EV}}_p, \bm\Lambda^{\text{EV}}_{\text{ch}, t}, \bm\Lambda^{\text{EV}}_{\text{dis}, t}, \bm{p}^{\text{RESs}}_{t}, \bm{p}^{\text{Load}}_{t} \right) 
\end{equation}

\textbf{Action:}
The action primarily includes the charging power $\bm p^{\text{EV}}_{\text{ch}, t}$ and discharging power $\bm p^{\text{EV}}_{\text{dis}, t}$ \cite{li2019constrained, zhang2023safe, chen2022deep}:
\begin{equation}
\bm{a}^\text{EV}_t \triangleq \left(\bm p^{\text{EV}}_{\text{ch}, t}, \bm p^{\text{EV}}_{\text{dis}, t} \right) 
\end{equation}

\textbf{Reward:}
The reward includes minimizing the charging cost associated with the time-varying electricity prices \eqref{Eq_EV_reward_cost}, maximizing the revenue in V2G mode \eqref{Eq_EV_reward_revenue}, and aligning the SoC closely with the target value \eqref{Eq_EV_reward_soc} \cite{chen2022deep, zhang2023safe}:
\begin{subequations}
\begin{gather}
R^\text{EV}(\bm{s}, \bm{a}) = -R^\text{EV}_\text{cost} + R^\text{EV}_\text{V2G} - R^\text{EV}_{\text{SoC}}\\
R^\text{EV}_\text{cost} = \bm\Lambda^{\text{EV}}_{\text{ch}, t} \bm p^\text{EV}_{\text{ch}, t} \label{Eq_EV_reward_cost}\\
R^\text{EV}_\text{V2G} = \bm\Lambda^{\text{EV}}_{\text{dis}, t} \bm p^\text{EV}_{\text{dis}, t} \label{Eq_EV_reward_revenue}\\
R^\text{EV}_{\text{SoC}} = |\bm{SoC}^{\text{EV}}_t - \bm{SoC}^\text{EV}_\text{target}|, \label{Eq_EV_reward_soc}
\end{gather}
\end{subequations}

\textbf{Constraint:}
Generally, EVs act as controllable loads within the electrical grid, with specific requirements for charging. When considering the V2G mode, the modeling of EVs is similar to that of BESS, as shown in \eqref{Eq_BESS} \cite{zhang2023safe}. Also, most EVs require a target SoC at a specified time $t$:
\begin{equation}
\label{Eq_EV_soc_target}
\bm{SoC}^{\text{EV}}_t \ge \bm{SoC}^\text{EV}_\text{target}
\end{equation}

\subsubsection{Building Energy Management}\mbox{}\label{sec4Building}

\begin{figure}[!htb]
    \centering
    \includegraphics[width=8.5cm]{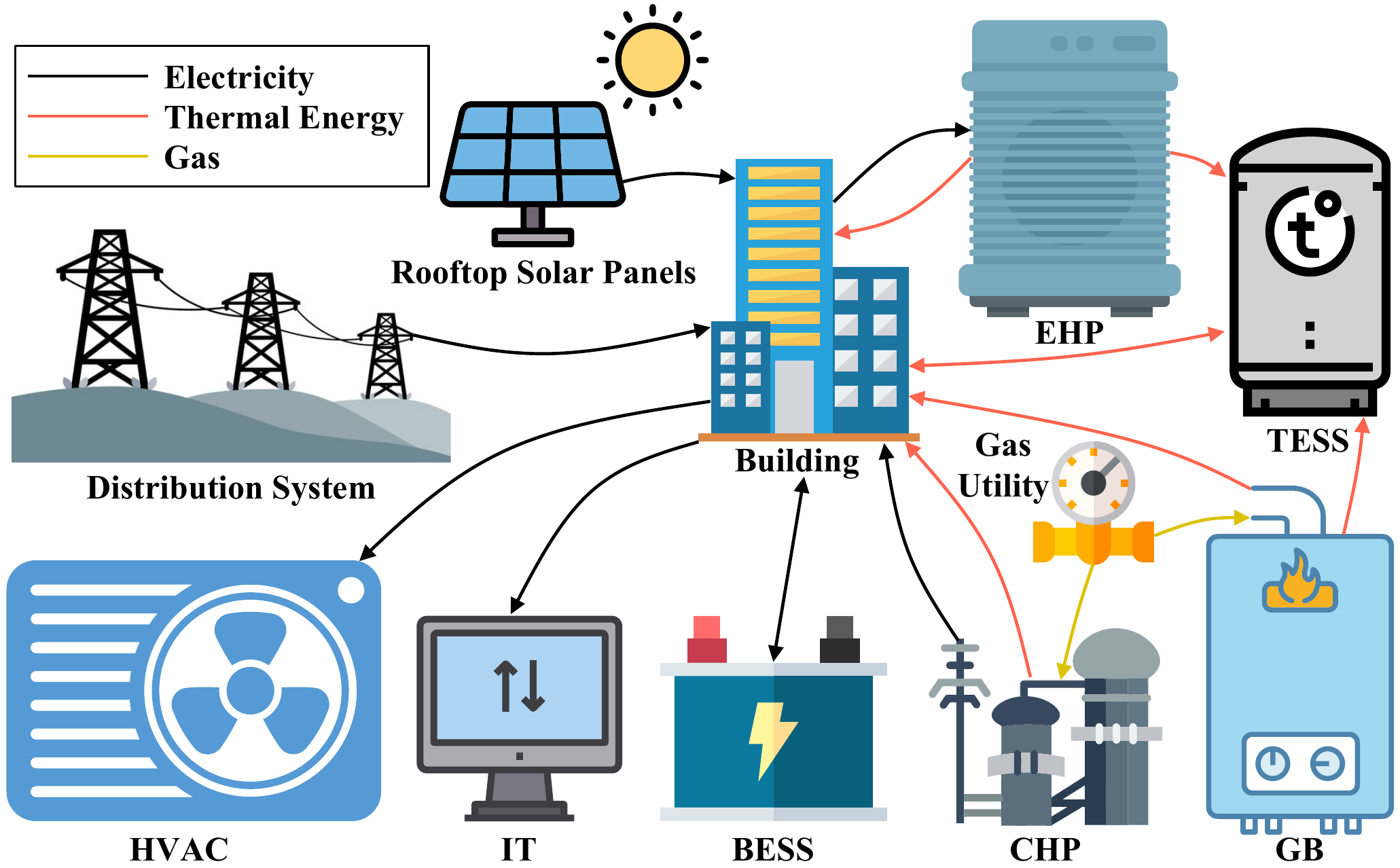}
    \caption{Building energy management structure. The building interacts with multiple forms of energy, such as electricity, thermal, and gas, through coordinated management strategies. It incorporates energy storage and enables energy conversion across different carriers, thereby improving energy efficiency, reducing operational costs, and enhancing flexibility in energy utilization.}
\label{Fig_Building}
\end{figure}

In 2022, the global buildings sector was a major energy consumer, accounting for 30\% of the final energy demand, primarily for operational needs like heating and cooling \cite{global2023un}. Energy hubs connect to both the electric grid and the natural gas network and meet electrical, heating and cooling demands by controlling RESs, ESSs, electric heat pumps (EHPs), gas boilers (GBs) and HVAC systems \cite{garmroodi2023optimal}. Therefore, effective control of cooling or HVAC systems for buildings and energy hubs is necessary. Traditional cooling control methods often rely on feedback control strategies, which, while effective in steady-state scenarios, lack the flexibility to adapt to dynamic and uncertain environments. In contrast, RL has emerged as a powerful tool for building energy management due to its ability to self-learn and adapt in complex and uncertain operational contexts. The primary objective of building energy management is to minimize energy consumption while ensuring that critical constraints are met. These constraints encompass thermal-related equipment, such as HVAC, EHP, and GB, along with electricity and heat demands, as well as environmental factors like temperature and humidity. Fig. \ref{Fig_Building} illustrates a typical system architecture, highlighting key electrical and thermal components.

A summary of safe RL applications in building energy management is provided in Table \ref{Table_BuildingControl}.
\renewcommand{\arraystretch}{1.1}
\begin{table*}[htbp]
\caption{Safe RL Applications in Building Energy Management}
\label{Table_BuildingControl}
\centering
\begin{adjustbox}{center, max width=\textwidth}
\begin{tabular}{|>{\centering\arraybackslash}m{0.7cm} >{\centering\arraybackslash}m{6.0cm} >{\centering\arraybackslash}m{3.2cm} >
{\centering\arraybackslash}m{2.2cm} >{\centering\arraybackslash}m{4.0cm} |}
    \toprule %
    Ref. & Problem/Objective & Constraint & Constraint Type & Safety Techniques\\
    \midrule %
    \cite{lin2023reinforcement} & Energy savings in building energy systems & Indoor temperature demand & Ins/Hard & Shielding \\ 
    \hline
    \cite{zhang2022residual} & Data center building cooling & Zone temperature & Ins/Hard & Shielding \\ 
    \hline
    \cite{garmroodi2023optimal} & Optimal dispatch of an energy hub & Energy and equipment & Cum/Soft & Primal-dual \\ 
    \hline
    \cite{ding2022safe} & Multi-energy management of smart home & Components in smart home & Cum/Soft & PDO \\ 
    \hline
    \cite{le2021deep} & Tropical air free-cooled data center control & Temperature and humidity & Cum/Soft & Lagrangian relaxation \\ 
    \hline
    \cite{yu2023district} & District cooling system control & Power requirement & Ins/Hard & Safety layer \\ 
    \hline
    \cite{zhang2022safe} & Safe building HVAC control & Building & Cum/Soft & Safety-aware objective \\ 
    \hline
    \cite{liang2021safe} & Resilient proactive scheduling of building & Components of building & Cum/Soft & Adaptive reward \\ 
    \hline
    \cite{qiu2022safe} & Real-time control in a smart energy-hub & Energy hub & Cum/Soft & Safety-guided function\\ 
    \hline
    \cite{sun2024energy} & Energy management for smart buildings & Voltage safety & Cum/Soft & Lagrangian relaxation \\ 
    \hline
    \cite{wang2025safe} & Building energy management & Operative temperature & Ins/Hard & MPC \\
    \hline
    \cite{wang2022green} & Data center cooling control & Thermal safety & Ins/Hard & Safety layer \\ 
    \hline
    \cite{wan2023safecool} & Cooling management in data centers  & Thermal safety & Cum/Soft & MPC \\
    \hline
    \cite{cao2023toward} & Data center cooling control & Rack cooling index & Ins/Hard & Lyapunov and projection \\ 
    \bottomrule %
\end{tabular}
\end{adjustbox}
\end{table*}
In Table \ref{Table_BuildingControl}, the studies cover diverse building types, including residential buildings, data centers, and energy hubs, and address challenges such as energy savings, thermal comfort, equipment safety, cooling control, and resilience to environmental uncertainties. Many approaches combine data-driven models with physical principles or empirical knowledge to improve decision-making under uncertainty. Examples include risk-based methods for handling extreme weather \cite{liang2021safe}, and MPC techniques to enhance safety and adaptability in dynamic environments \cite{wang2025safe, wan2023safecool}. Building energy management poses relatively lower systemic risks for individual buildings compared to other grid-scale applications, offering a safer environment for real-world deployment and experimentation. This makes buildings an ideal testbed for developing and refining new techniques \cite{lin2023reinforcement}. Given the smaller capacity of individual buildings, they are particularly sensitive to localized demand changes and stochastic behavior. Future work could focus on developing adaptive and robust safe RL algorithms capable of managing these uncertainties effectively \cite{golpira2019multi}.

Subsequently, an example is provided to demonstrate how safe RL is applied to building energy management. The state, action, reward, and constraints are outlined as follows:

\textbf{State:}
The state of the building, in relation to HVAC systems, includes indoor and outdoor temperature $T^{I/O}$, humidity $H$, actual airflow rate $\bm s^\text{air}$, and actual ventilation rate $\bm s^\text{ven}$ \cite{liu2022safe}. Additionally, it covers BESS SoC $\bm{SoC}^\text{BESS}$, TESS SoC $\bm{SoC}^\text{TESS}$, combined heat and power system (CHP) state $\bm s^\text{CHP}$, GB state $\bm s^\text{GB}$, EHP state $\bm s^\text{EHP}$, core operational equipment state such as information technology (IT) equipment temperature $T^\text{IT}$, human satisfaction indicators $\bm s^\text{Human}$, and exogenous state such as electricity prices $\bm\Lambda^{\text{Ele}}$, gas price $\bm\Lambda^{\text{Gas}}$ and carbon price $\bm\Lambda^{\text{Car}}$ \cite{lin2023reinforcement, liu2022safe, qiu2022safe}.
\begin{equation}
\begin{aligned}
\bm{s}^\text{Building}_t \triangleq (T^{I}, T^{O}, H, \bm s^\text{air}, \bm s^\text{ven}, \bm{SoC}^\text{BESS}, \bm{SoC}^\text{TESS}, &\\ 
\bm s^\text{CHP}, \bm s^\text{GB}, \bm s^\text{EHP}, T^\text{IT}, \bm s^\text{Human}, \bm\Lambda^{\text{Ele}}, \bm\Lambda^{\text{Gas}}, \bm\Lambda^{\text{Car}}&) 
\end{aligned}
\end{equation}

\textbf{Action:}
The action includes temperature setpoint $T_\text{set}$, humidity setpoint $H_\text{set}$, airflow rate $\bm a^\text{air}$, ventilation rate $\bm a^\text{ven}$, BESS charge or discharge power $\bm p^\text{BESS}_{\text{ch/dis}}$, TESS charge or discharge power $\bm h^\text{TESS}_{\text{ch/dis}}$, electricity generated by CHP $\bm p^\text{CHP}$, heat generated by CHP $\bm h^\text{CHP}$, GB $\bm h^\text{GB}$ and EHP $\bm h^\text{EHP}$, and RESs output $\bm p^\text{RES}$ \cite{zhang2022safe}.
\begin{equation}
\begin{aligned}
\bm{a}^\text{Building}_t \triangleq (T_\text{set}, H_\text{set}, \bm a^\text{air}, \bm a^\text{ven}, \bm p^\text{BESS}_{\text{ch}/\text{dis}}, &\\ 
\bm h^\text{TESS}_{\text{ch}/\text{dis}}, \bm p^\text{CHP}, \bm h^\text{CHP}, \bm h^\text{GB}, \bm h^\text{EHP}, \bm p^\text{RES}&) 
\end{aligned}
\end{equation}

\textbf{Reward:}
The reward is to minimize the total energy cost, including electricity, natural gas, heat, and long-term device degradation, especially for BESSs and TESSs. When specific room-temperature ranges must be maintained, temperature deviations are often included in the reward.
\begin{equation}
R^\text{Building}(\bm{s}, \bm{a})= -(R_\text{cost} + R_\text{degrade} + \Delta T)
\end{equation}
where $R_\text{cost}$, $R_\text{degrade}$ and $\Delta T$ represent the rewards for cost, device degradation, and temperature deviation, respectively.

\textbf{Constraint:}
The generation and consumption of electrical and thermal energy are equal, complying with the electrical and thermal balance equations \cite{qiu2022safe, ding2022safe}.
\begin{subequations}
\label{Eq_electrical_balance}
\begin{gather}
\bm p^{\text{Grid}}_t + \bm p^{\text{RESs}}_t + \bm p^{\text{BESS}}_{\text{dis}, t} + \bm p^{\text{CHP}}_{t} = \notag \\
\bm p^{\text{HVAC}}_t + \bm p^{\text{Load}}_t + \bm p^{\text{EV}}_t + \bm p^{\text{BESS}}_{\text{ch}, t} + \bm p^{\text{EHP}}_{t}\\
\bm h^{\text{CHP}}_t + \bm h^{\text{GB}}_t + \bm h^{\text{TESS}}_{\text{dis}, t} + \bm h^{\text{EHP}}_t = \bm h^{\text{TL}}_t + \bm h^{\text{TESS}}_{\text{ch}, t}
\end{gather}
\end{subequations}
where $\bm p$ and $\bm h$ denote the vectors of electrical and thermal energy generation or demand, respectively; $\text{TL}$ denotes thermal load. The constraints of BESS have already been shown in \eqref{Eq_BESS}. The constraints of TESS are formulated in a similar manner, following the structure of \eqref{Eq_BESS}.

CHP, utilizing gas for coupled heat and electricity generation, is a single-input-multi-output converter with high electrical and thermal energy efficiency, governed by the following constraints \cite{qiu2022safe}:
\begin{subequations}
\begin{gather}
\bm{p}^{\text{CHP}}_{t} = \eta^{\text{CHP}}_{p} \bm{g}^{\text{CHP}}_{t} ~~~ \bm{h}^{\text{CHP}}_{h} = \eta^{\text{CHP}}_{h} \bm{g}^{\text{CHP}}_{t} \label{Eq_CHP_PH}\\
\bm 0\le \bm{p}^{\text{CHP}}_{t} \le \overline{\bm{p}}^{\text{CHP}} ~~~ \bm 0\le \bm{h}^{\text{CHP}}_{h} \le \overline{\bm{h}}^{\text{CHP}} \label{Eq_CHP_max}
\end{gather}
\end{subequations}
where $\bm{g}^{\text{CHP}}_{t}$ denotes gas input of CHP; $\eta^{\text{CHP}}_{p}$ and $\eta^{\text{CHP}}_{h}$ denote the electrical and thermal energy efficiency of CHP, respectively; \eqref{Eq_CHP_PH} indicates the efficiency of converting natural gas into electric power $\bm{p}^{\text{CHP}}_{t}$ and heat power $\bm{h}^{\text{CHP}}_{h}$; \eqref{Eq_CHP_max} represents the range of $\bm{p}^{\text{CHP}}_{t}$ and $\bm{h}^{\text{CHP}}_{h}$.

GB and EHP respectively convert natural gas and electricity into heat to meet the heating demand, which can be represented as follows \cite{garmroodi2023optimal}:
\begin{subequations}
\begin{gather}
\bm{h}^{\text{GB}}_{h} = \eta^{\text{GB}} \bm{g}^{\text{GB}}_{t} ~~~ \bm{h}^{\text{EHP}}_{t} = \eta^{\text{EHP}} \bm{p}^{\text{EHP}}_{t} \label{Eq_GB_EHP_PH}\\
\bm 0\le \bm{h}^{\text{GB}}_{h} \le \overline{\bm{h}}^{\text{GB}} ~~~ \bm 0\le \bm{h}^{\text{EHP}}_{t} \le \overline{\bm{h}}^{\text{EHP}} \label{Eq_GB_EHP_max}
\end{gather}
\end{subequations}
where $\bm{g}^{\text{GB}}_{t}$ denotes gas input of GB; $\eta^{\text{GB}}$ and $\eta^{\text{EHP}}$ denote the efficiency of GB and EHP, respectively; \eqref{Eq_GB_EHP_PH} indicates the conversion of natural gas and electricity to heat with different efficiency; \eqref{Eq_GB_EHP_max} is the range of $\bm{h}^{\text{GB}}_{h}$ and $\bm{h}^{\text{EHP}}_{t}$.

HVAC systems play a crucial role in monitoring and regulating indoor temperature to maintain it within specified bounds \cite{ding2022safe, liang2021safe}.
\begin{subequations}
\begin{gather}
T^{I}_{t} = \epsilon T^{I}_{t-1} + (1 - \epsilon) \left( T^{O}_{t-1} - \frac{\eta^{\text{HVAC}} E^{\text{HVAC}}_{t-1}}{A} \right) \label{Eq_HVAC}\\
\underline{E}^{\text{HVAC}} \le E^{\text{HVAC}}_{t} \le \overline{E}^{\text{HVAC}} ~~~ \underline{T}^{I} \le T^{I}_{t} \le \overline{T}^{I} \label{Eq_HVAC_max}
\end{gather}
\end{subequations}
where $\epsilon$ and $A$ denotes the inertia parameter of temperature and thermal conductivity of HVAC, respectively; $\eta^{\text{HVAC}}$ denotes the efficiency of HVAC; \eqref{Eq_HVAC} indicates the temperature change of the room; \eqref{Eq_HVAC_max} represents the limits of HVAC energy consumption $E^{\text{HVAC}}_{t}$ and indoor temperature $T^{I}_{t}$.

\subsection{Discussion on Suitable Application Areas in Power Systems}\label{sec4ApplicationArea}

Safe RL builds on conventional RL by integrating constraint‐handling techniques to maximize reward while ensuring safety. However, its reliance on data‐driven exploration can limit applicability when safety requirements are strict, data are scarce or inaccurate, or real‐time performance is critical. The application areas of safe RL in power systems include:
\subsubsection{Traditional Generator and Load Control} Safe RL can simultaneously optimize the objective function and constraints in traditional OPF. Rather than solving nonlinear programs online, safe RL directly outputs control actions through the forward inference of NNs, reducing computational complexity and accelerating response time  \cite{zhang2020multi, yi2023model, yan2022hybrid}.
\subsubsection{RES Integration and Power Control} RESs like wind and solar exhibit high volatility and uncertainty. Safe RL can adapt to these changes and optimize output control strategies while ensuring that constraints on voltage, frequency, and power balance are met \cite{hao2024lyapunov, ye2023safe}.
\subsubsection{Topology Optimization and System Restoration} Grid operation requires dynamic responses to changes in network topology. Safe RL can learn optimal network reconfiguration strategies, preventing overloads and voltage violations during topology switching \cite{zheng2021vulnerability, du2022deep, hao2024safe}.
\subsubsection{ESS and EV System Control} ESS and EV charge and discharge management involve multiple timescales and strict operational limits. Safe RL can optimize scheduling while ensuring battery capacity and other constraints \cite{zhang2020deep, li2022risk}.
\subsubsection{Dynamic Voltage and Frequency Control} Voltage and frequency control in power grids requires dynamic adjustments under uncertain RES fluctuations and load disturbances. Safe RL can optimize control strategies in real-time, ensuring voltage and frequency constraints are met \cite{cui2023online, xia2022safe}.
\subsubsection{High-Penetration Inverter-Integrated System Control} Inverter‐based resources exhibit fast dynamics and nonlinear behavior that challenge traditional model‐based controllers. Added stochastic RES fluctuations and uncertain inverter parameters further increase complexity \cite{hatziargyriou2020definition}. Safe RL can learn coordinated control policies for multiple inverters, adapting to dynamic conditions while enforcing safety constraints such as voltage, frequency, and harmonic stability \cite{tarle2023safe}.

\setcounter{subsubsection}{0}
The inapplicable areas of safe RL include: 
\subsubsection{Relay Protection and Safety Control with High Real-Time Requirements} Relay protection demands millisecond‐level responses across all fault types and locations, which traditional logic‐based schemes and model‐driven controllers achieve via predefined rules and optimized algorithms to isolate faults and halt propagation \cite{anderson2022power}. Safe RL, however, cannot guarantee coverage of every fault scenario during training and may suffer from insufficient generalization or delayed responses, making it ill‐suited for ultra‐fast fault protection and emergency controls \cite{xu2021resilience}.
\subsubsection{Core Power Grid Dispatch with High Safety Requirements} Safe RL may occasionally breach constraints, offering no absolute safety guarantee. For critical dispatch tasks, model‐based approaches (e.g., OPF, MPC) provide firmer assurances of constraint satisfaction and system security \cite{ademola2020frequency}.
\subsubsection{Scenarios with Highly Deterministic Parameters and Accurate Modeling} In scenarios where system parameters are well-known and accurately modeled, model-based methods are typically more efficient and reliable, reducing the relative benefit of Safe RL \cite{yang2017linearized}.
\subsubsection{Scenarios with Low Data Availability or Reliability} Safe RL relies on large amounts of high-quality data to learn optimal control strategies. However, in certain extreme or rare events, such as power grid operation under extreme weather conditions, historical data is often insufficient or unrepresentative, making it difficult to cover all potential operating states and disturbance conditions \cite{panteli2016power}. Limited measurement accuracy and coverage can leave gaps in capturing system dynamics, undermining the reliability and generalization of Safe RL policies \cite{duchesne2020recent}.

Although some applications may not currently be suitable for safe RL, future advances in sensing, communication, algorithms, and computing power may improve its applicability.


\section{Real-World Deployment Cases and Roadmap}\label{sec5}
\begin{figure}[!htb]
    \centering
    \includegraphics[width=8.5cm]{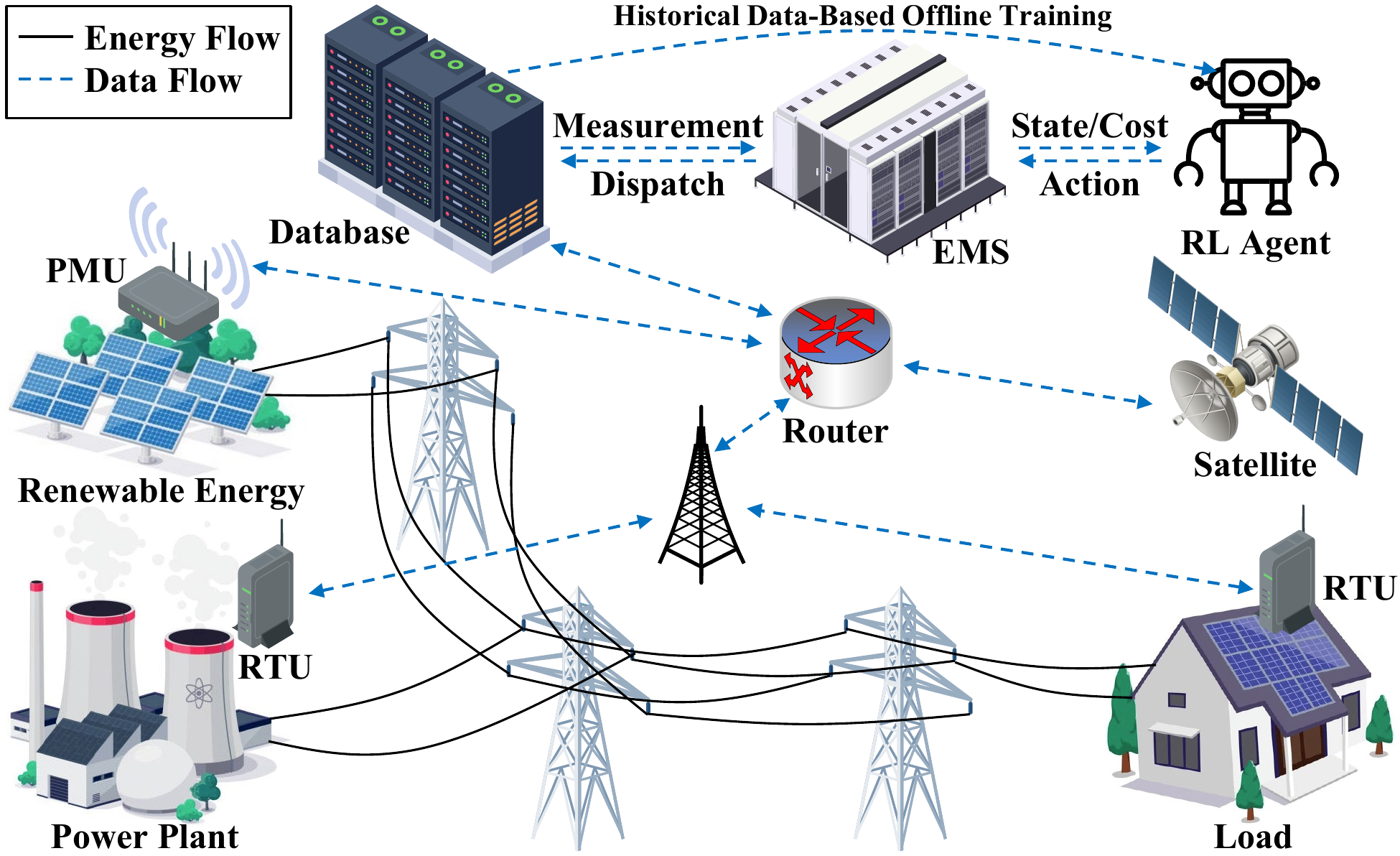}
    \caption{Integrated SCADA–EMS–RL for real-world power system control. RTUs and PMUs collect field measurements and transmit them to the central database and EMS. Both real-time and historical data are provided to the RL agent for training and decision-making. The RL agent then returns optimized control actions through the EMS and SCADA for field execution.}
\label{Fig_SCADA_EMS}
\end{figure}

The application of RL in power system optimization and control began a decade ago, while the use of safe RL in power systems started five years ago, with most research on safe RL emerging in the past three years. As a result, most existing studies remain at the theoretical exploration stage, and large-scale real-world deployment still requires further experimentation to gain practical experience and develop new algorithms leveraging emerging technologies. Fig. \ref{Fig_SCADA_EMS} illustrates the integration of safe RL with SCADA and EMS systems in a real-world deployment \cite{hu2017recent}.

In this section, we summarize the publicly available cases of RL deployment in real-world power systems and provide a detailed roadmap for its future development. Note that our examples do not distinguish between RL and safe RL because any RL method deployed in practice must inherently adhere to safe RL principles and avoid constraint violations.

\subsection{Real-World Deployment Cases}
\subsubsection{Gas Turbine Auto Tuner}
The Gas Turbine (GT) Auto Tuner, an AI-based solution for GTs, leverages a digital twin and RL to optimize turbine inlet temperature and reduce emissions. Co-developed by DEWA and Siemens Energy in 2019, it was the world’s first thermodynamic digital twin GT intelligent controller. Successfully deployed on four GTs, this innovative system has demonstrated its potential to lower NOx emissions and minimize the need for seasonal tuning by combining advanced AI techniques with enhanced turbine inlet temperature estimation. Upgrades for the GTs will enable interval extension between outages, providing increased operational flexibility, allowing for a higher number of starts and reduction of outages by approximately 25\% \cite{siemensGTautotuner}.

\subsubsection{Building Cooling Systems}
In 2014, U.S. data centers consumed approximately 70 billion kWh of electricity, accounting for about 1.8\% of the nation's total, highlighting the urgent energy challenge in the tech industry. To address this, DeepMind developed an RL algorithm for Google's data centers to optimize cooling efficiency. Every five minutes, the RL collects data from thousands of sensors, predicts the impact of various cooling actions using DNNs, and selects the optimal configurations to minimize energy consumption while adhering to safety constraints. These actions are then verified and implemented by the local control system \cite{googleAICooling, deepmindCoolingAI}. Building on this experience, DeepMind and Google applied RL to control commercial cooling systems while ensuring safety through a series of constraint-aware RL measures. Live experiments conducted at two real-world facilities demonstrated the effectiveness of this approach, achieving energy savings of approximately 9\% and 13\% at the respective sites \cite{luo2022controlling}. 
Additionally, TELUS and the Vector Institute have jointly launched the energy optimization system, a model-based RL solution designed to reduce operational costs and electricity use in commercial buildings, particularly data centers, across Canada. The system optimizes HVAC systems across TELUS network locations to enable energy-efficient temperature control. Approximately 40\% of the energy at these sites is used for cooling telecommunications equipment. The solution has shown promising results, as pilot tests at small data centers demonstrated nearly a 12\% reduction in annual electricity consumption and highlighted its significant potential to reduce environmental impact \cite{telusAIClimate}.
In addition, in 2020, SAB RL agents took control of a $15,000 m^2$ commercial building HVAC system located in Northern Europe. The building featured a modern building management system with advanced, state-of-the-art control sequences and already had a low energy consumption baseload of $32kWh/m^2\cdot\text{year}$, making further optimization particularly challenging. Within a few weeks, the SAB HVAC optimization workflow was introduced. This process included building characterization and energy data collection, along with interviews with facility managers to identify known pain points. A digital twin was then developed by creating a building model calibrated against actual energy data. Using this digital twin as a safe and accurate training environment, smart RL agents were trained. As a result, HVAC spending was reduced by 54\% without compromising thermal comfort or indoor air quality \cite{foobotAIHVAC}.

\subsubsection{DeepThermal}
DeepThermal is a model-based offline RL framework designed to optimize combustion control strategies for thermal power generating units. It utilizes historical operational data to address highly complex CMDP problems through purely offline training. Successfully deployed in four large coal-fired thermal power plants in China, DeepThermal has demonstrated significant improvements in combustion efficiency, showcasing its effectiveness in enhancing operational performance. Specifically, 1 to 2 years of historical operational data were used to train the models. The study considered more than 800 sensors and optimized approximately 100 control variables. A specially designed feature engineering process was applied to transform these sensor data into about 100 to 170 state variables and 30 to 50 action variables. The duration of the experiments ranged from 1 to 1.5 hours. The approach effectively improved combustion performance under all three load settings, with maximum increases in combustion efficiency of 0.52\%, 0.31\%, and 0.48\% within approximately 60 minutes compared to the initial values. This example explicitly employs safe RL techniques, specifically using the Lagrangian relaxation method to solve CMDP problems \cite{zhan2022deepthermal}.

\begin{table*}[htbp]
\caption{Analysis of the Real-World Applicability of Different Methods}
\label{Table_RealWorld_Applicability}
\centering
\begin{tabular}{|>{\centering\arraybackslash}m{2.5cm} >{\centering\arraybackslash}m{2.5cm} >{\raggedright\arraybackslash}m{11.8cm}|}
\toprule
Methods & Real-World Prospect & Key Features\\
\midrule
Lagrangian relaxation method & \ding{73}\ding{73}\ding{73}\ding{73}\ding{73} & Mature theory; easy implementation; high flexibility; risk of constraint violations; potential oscillations; challenging parameter/multiplier tuning; requires extensive practical experience \cite{ray2019benchmarking}. \\
\hline
Projection method & \ding{73}\ding{73}\ding{73}\ding{73} & Excellent safety guarantees; large projection overhead; necessity of well-defined projection operator; limited real-time performance in large-scale systems \cite{yang2020projection}. \\
\hline
Lyapunov method & \ding{73}\ding{73} & Strong theoretical stability guarantees; need for suitable Lyapunov function; challenging Lyapunov construction for large-scale systems; difficult practical application \cite{perkins2002lyapunov}. \\
\hline
GP method & \ding{73}\ding{73} & Enhances safety under uncertainty; extremely high computational complexity; poor scalability to large-scale power systems; requires dimensionality reduction or approximation for practicality \cite{sui2015safe}. \\
\hline
Shielding method & \ding{73}\ding{73}\ding{73} & Enforces hard constraints; limits exploration; complexity and rule count grow rapidly with scale; difficult to predefine shielding rules for diverse operating conditions \cite{alshiekh2018safe}. \\
\hline
Safety layer method & \ding{73}\ding{73}\ding{73}\ding{73} & Per-step constraint satisfaction; unsafe-to-safe action correction; computational efficiency sensitive to specific implementation; challenging policy adjustment for numerous real-world constraints \cite{dalal2018safe}. \\
\hline
Barrier function method & \ding{73}\ding{73}\ding{73}\ding{73} & Well-defined barrier function requirement; challenging barrier selection/tuning in high-dimensional real-world systems; potentially overly conservative \cite{wang2023safe}.\\
\hline
RRL & \ding{73}\ding{73}\ding{73}\ding{73} & Worst‐case hedge focus; uncertainty handling for practical power grid application; high training complexity; limited real‐world applicability \cite{pinto2017robust}. \\
\bottomrule
\end{tabular}
\end{table*}
\subsection{Real-World Deployment Challenges}
From the above real-world examples, it is evident that safe RL has been piloted and applied in low-risk, small-scale systems, but further efforts are needed to advance its development and deployment. Moreover, there remain significant challenges that need to be addressed when transitioning from simulation to real-world deployment:
\subsubsection{Model Uncertainty}
Practical power systems face diverse uncertainties, such as RES variability, load fluctuations, unexpected component failures, and parameter inaccuracies (especially in distribution networks), which make accurate environment modeling difficult. Methods like shielding or Lyapunov-based safe RL often rely heavily on accurate environment modeling or detailed system knowledge, so any mismatch can harm safety and performance. To address this gap, online calibration and uncertainty quantification techniques such as GP or RRL methods can be integrated into the learning loop to continuously update model parameters and estimate prediction confidence, thereby reducing the risk of policy mismatch in real-world deployment.
\subsubsection{Perception–Decision–Action Latency}
Real-world systems typically differ significantly from idealized simulation models. Factors such as sensor sampling latencies, communication jitter and computational lag in the perception–decision–action loop introduce nonnegligible latency that can cause an agent’s control commands to arrive too late to achieve their intended effect, particularly when preventing fast disturbances or transient faults. One way to mitigate these latencies is predictive buffering, where the agent forecasts future system states based on known latencies, precomputes control commands in advance and stores locally for immediate execution. Another approach is time‐compensated control, which proactively accounts for communication and computation latencies in the decision logic to offset their impact \cite{guo2023safe}.
\subsubsection{Execution Errors}
Execution errors stemming from actuator nonlinearities, tracking deviations, packet loss and hardware wear erode control fidelity. To guard against these failures, robust execution layers are needed to continuously monitor communication link integrity and actuator response accuracy. Real‐time detection algorithms compare expected outputs to actual measurements and can automatically trigger alarms or switch to redundant actuators when discrepancies arise, while command confirmation protocols ensure that every instruction is acknowledged and, if lost, retransmitted. Together, these mechanisms maintain safety margins and help sustain optimal performance even when individual components behave unpredictably.
\subsubsection{Computational Limitations}
Real-world deployment often requires executing near-instantaneous decision-making for complex, large-scale power systems. Safe RL often incurs extra computation, such as solving optimizations or safety projections at every step, which undermines real-time performance and scalability. Approaches relying on intensive calculations may struggle to scale, especially for emergency control requiring responses within 100–300 ms. Model simplification, approximation techniques, and parallelization can help, but ultra-high‐speed applications remain challenging.
\subsubsection{Scalability Issues}
Real-world power grids often involve extremely large-scale networks with numerous interconnected devices. As a result, methods that are computationally tractable in simplified or reduced-scale simulation models may face significant scalability and computational issues when applied to practical, full-scale scenarios.
\subsubsection{Continuous Adaptability}
Real power systems continuously evolve due to changes in network topology, component aging, regulatory requirements, and operational policies. Many safe RL methods, especially those with fixed or precomputed safety rules (e.g., shielding methods), may struggle to maintain effectiveness without frequent redesign or recalibration.
\subsubsection{Safety Assurance and Regulatory Acceptance}
Power systems require extremely high reliability, making regulatory approval and acceptance for deploying safe RL particularly challenging. Regulators typically demand strong theoretical guarantees, comprehensive validation, and high interpretability, which current safe RL methods struggle to satisfy and therefore require further research and experimental testing.

In the face of these challenges, different safe RL techniques exhibit varying degrees of suitability for real-world deployment. Table \ref{Table_RealWorld_Applicability} presents an analysis of each method’s applicability in real-world deployments. Similarly, this is only a general analysis, and specific evaluation and selection should be conducted based on the actual conditions.

\begin{table*}[htbp]
\caption{Performance Indicators for Real-World Deployment}
\label{Table_RealWorld_Indicators}
\centering
\begin{tabular}{|>{\centering\arraybackslash}m{2.5cm} >{\raggedright\arraybackslash}m{3.7cm} >{\raggedright\arraybackslash}m{10.6cm}|}
\toprule
Indicator Category & Definition & Example Indicators\\
\midrule
Learning & Overall policy effectiveness & Cumulative reward; mean episodic return \\
\hline
Safety & Degree of constraint satisfaction & Constraint violation rate; max/mean overshoot; worst‐case violation magnitude \\
\hline
Efficiency & Data/sample usage & Convergence speed; episodes to convergence; sample efficiency \\
\hline
Real-time capability & Decision execution efficiency & Action generation time; communication/inference latency; control loop period \\
\hline
Robustness & Resilience to disturbances & Reward degradation under disturbances; violation rate change \\
\hline
Economic benefit & Cost savings or profit gains	& Operational cost savings; market profit; social welfare \\
\hline
Domain-specific	& Application-specific indicators & Voltage/frequency deviation/violation rate; RoCof; frequency nadir; maximum rotor angle difference; energy/comfort metrics; load restored time; reserve shortfall duration; bid acceptance rate; average wait time per EV; peak load reduction; HVAC cycling count \\
\bottomrule
\end{tabular}
\end{table*}

\subsection{Real-World Deployment Roadmap}
Based on the above analysis, the roadmap for deploying safe RL in real-world applications can be outlined as follows:
\subsubsection{Algorithm Innovation}
\paragraph{Hybrid Approaches} Combine data-driven safe RL methods with physics-based models to enhance interpretability and enforce strict operational constraints \cite{wu2024real}.
\paragraph{Robust Models} Enhance robustness by utilizing robust and adversarial training techniques to ensure reliable operation under varying levels of RES penetration and extreme weather conditions \cite{hao2024lyapunov, hong2023robust}.
\paragraph{Constraint-Aware Learning} Develop algorithms capable of simultaneously handling multiple, non-linear, and interdependent constraints such as voltage, frequency, and thermal limits, ensuring real-time adaptability \cite{shengren2023optimal}.
\paragraph{Scalable Techniques} Design scalable safe RL frameworks suitable for large-scale power systems, incorporating decentralized and distributed learning approaches \cite{ma2024efficient}.

\subsubsection{Benchmarking and Testing Infrastructure}
\paragraph{Domain-Specific Environments} Develop standardized power system benchmark environments, similar to those described in Section \ref{sec3benchmark}, but designed to be more versatile and compatible with a broader range of scenarios and models. These environments should integrate realistic dynamic models and robust safety requirements to enable fair algorithm testing and comparison \cite{fan2022powergym, wolgast2024learning}.
\paragraph{Standardized Metrics} Develop universally accepted evaluation metrics for safe RL algorithms based on the aforementioned benchmark environments, including safety compliance rates, computational efficiency, and scalability \cite{garcia2015comprehensive}.
\paragraph{Domain-Specific Algorithms} Develop tailored safe RL algorithms for specific domains, such as Lyapunov-based methods for grid stability, robust optimization for renewable integration, and decentralized learning for demand response, ensuring each approach aligns with the unique characteristics and constraints of its application \cite{shengren2023optimal, nguyen2022three}.

\subsubsection{Real-World Deployment}
\paragraph{Pilot Projects in Low-Risk Applications} Initiate deployment with small-scale pilot projects and low-risk scenarios, such as microgrids, building energy management, or demand response programs, to validate algorithm performance and ensure safety under real-world conditions. Based on the summary of existing real-world deployments, this aligns with the current developmental stage of safe RL \cite{siemensGTautotuner, googleAICooling, deepmindCoolingAI, zhan2022deepthermal}.
\paragraph{Scaling to Regional Grids} Expand deployment to larger regional grids for tasks such as generator dispatch, voltage regulation, RES integration, and load balancing, using insights from earlier phases to enhance scalability and robustness \cite{li2022learning, liu2020two}.
\paragraph{Progressive Integration into Critical Applications} Collaborate with system operators to incorporate safe RL into critical applications such as frequency stability, and system restoration. Leverage advanced monitoring tools, digital twins, and SCADA systems to enable dynamic, real-time decision-making and ensure operational reliability at scale \cite{cui2022reinforcement, vu2021barrier}.
\paragraph{Coordinated Deployment Across Applications} Enable seamless integration of safe RL across compatible power system operations, such as planning, dispatch, and control, ensuring effective coordination between these domains to optimize overall performance.

\subsubsection{Policy and Regulatory Alignment}
\paragraph{Incentivizing Adoption}
Work with governments and industry stakeholders to create incentives, such as subsidies or regulatory benefits, that encourage the adoption of safe RL technologies in power systems.
\paragraph{Standard Development and Guidelines} Collaborate with policymakers and standardization bodies to establish detailed guidelines and practical standards for the development and deployment of safe RL, covering safety protocols, operational best practices, and validation methods.
\paragraph{Ethical and Social Considerations} Proactively tackle ethical concerns such as data privacy, accountability, and transparency, ensuring that safe RL applications align with societal expectations and promote trust in automated power system operations \cite{vishwanath2024reinforcement}.
\paragraph{Continuous Monitoring and Compliance}
Establish mechanisms for continuous monitoring and evaluation of deployed RL systems to ensure ongoing compliance with safety and operational standards, adapting to evolving grid conditions and technological advancements \cite{fulton2018safe}.


\subsection{Real-World Deployment Performance Indicators}
To assess and compare algorithm performance in real-world deployments, we have summarized key empirical indicators in Table \ref{Table_RealWorld_Indicators} that cover learning performance, safety, efficiency, real-time capability, robustness, economic benefit, and domain-specific indicators.

\section{Challenges and Outlook}\label{sec6}
The application of safe RL in power systems is still in its infancy, facing a variety of challenges, including scalability, distributed settings, uncertainty, rapid changes in power networks, multi-constraint handling, reward design, user-centric design, early and emergency response, and real-world deployment, etc. In addition, we further discuss the potential future research directions.
\subsection{Challenges in Safe RL}
Although the general challenges of RL have already been reviewed in \cite{chen2022reinforcement, li2023deep}, this subsection will explore the unique challenges faced by existing safe RL methods \cite{dulac2019challenges}.

\subsubsection{Scalability to Large-Scale Power Systems}
Real-world power systems encompass a vast number of buses and power lines. For instance, the Eastern Interconnection, a major North American power grid system, has been modeled with over 60,000 buses in certain simulations \cite{ingleson2005tracking}. Consequently, large-scale multi-agent systems face scalability issues in such environments for four primary reasons \cite{nguyen2020deep}. First, the state and action spaces expand dramatically with an increasing number of agents, a phenomenon known as the ``curse of dimensionality" \cite{lu2024overcoming}. This expansion results in an exponentially increasing search space for optimal actions. Secondly, as the number of buses grows, the number of power flow constraints and other physics-based constraints also escalates. Moreover, some research accounts for security constraints due to demand uncertainty in power systems, which further complicates the constraints in the safe RL training process \cite{hao2024safe}. Lastly, the high dimensionality and non-convexity of the power system optimization landscape make it challenging for safe RL to converge to feasible results using stochastic gradient descent.

To address these issues, methods such as reduced-order polytopal constraints and low-order elliptical constraints have been employed to approximate complex constraints in large-scale systems \cite{hreinsson2021new}. These techniques provide a practical way to integrate extensive constraints into safe RL frameworks by simplifying their representation while preserving the fidelity of the system's critical dynamics. Furthermore, model-based RL offers an efficient approach to mitigate scalability challenges by incorporating system dynamics models. By simulating the environment offline, model-based RL minimizes the need for extensive real-world exploration, thereby accelerating policy learning while maintaining safety \cite{cui2023online, vu2021barrier, feng2023bridging}. Meta-learning and transfer learning further enhance scalability. Meta-learning enables safe RL agents to generalize knowledge from small-scale systems to larger ones \cite{beck2023survey}. Transfer learning adapts pre-trained policies from simpler or related tasks to more complex environments, significantly reducing training time \cite{da2019survey}.

In addition to these methods, there are three approaches for partitioning large-scale problems. One promising technique is the use of factored action spaces, which decompose the action space into smaller, more manageable components \cite{tang2022leveraging}. This method has proven effective in other complex environments, such as StarCraft and Dota 2, showcasing its versatility in handling combinatorial and continuous control problems. Another effective strategy is HRL, which splits the decision-making process into multiple layers. High-level policies manage strategic decisions, while low-level policies handle tactical controls. This hierarchical decomposition reduces the effective action space at each layer, improving scalability and convergence in large systems \cite{xiong2022hisarl, xia2024hierarchical}. Additionally, parallel computing and distributed safe RL methods are increasingly being adopted. These techniques distribute the learning task across multiple agents or processing units, enabling simultaneous exploration and faster convergence \cite{zhu2024parallel}. In power systems, distributed RL can allocate localized control tasks (e.g., voltage regulation or frequency control) to individual agents, which are coordinated by a central policy to ensure overall system stability \cite{yang2023safety}.

\subsubsection{Distributed Safe RL}
In the previous challenge, distributed safe RL was discussed as a solution for scalability issues in large-scale power systems. However, the deployment of distributed safe RL also faces significant challenges that warrant careful consideration \cite{yang2023safety, verbraeken2020survey}. For instance, in a multi-agent setting, the actions of one agent can alter the environment experienced by others, making it difficult for agents to converge to stable policies \cite{lu2021decentralized}. Additionally, coordination among distributed agents often requires communication, which can introduce latency, scalability issues, and data privacy concerns in large-scale systems. Designing rewards that effectively balance local optimization objectives with global system stability is another challenge, particularly when agents have limited visibility of the overall system state \cite{zhang2021distributional}. Most importantly, for safe RL, ensuring safety and adherence to physical constraints in a distributed RL framework is highly complex, especially when agents operate with incomplete information. Since agents have access only to local information, deriving globally stable actions can be difficult or even unattainable \cite{chung2020distributed}.

Several solutions have been proposed to address these challenges. For example, during training, a centralized critic or coordinator can provide agents with global information, while execution remains decentralized to ensure safety and scalability \cite{lyu2021contrasting}. Additionally, agents can communicate only when significant changes in their local states occur, thereby reducing unnecessary communication \cite{chen2021communication}. Federated Learning offers another approach, enabling agents to learn collaboratively without sharing raw data, thus preserving privacy and reducing communication bandwidth requirements \cite{koursioumpas2024safe}. Combining distributed RL with HRL and physics-informed learning can further enhance performance by reducing the complexity of solving large-scale systems through HRL and ensuring safety by embedding power system knowledge into the RL framework \cite{chen2023distributed}.

\subsubsection{Uncertainty, Distribution Shift, and Data Sufficiency}\label{sec6A3} 
Modern power systems face increased uncertainty due to fluctuations in RESs and loads, leading to the issue of distribution shift during training and deployment. This means that the same action may result in different state outcomes under ambient uncertainties \cite{fujimoto2024assessing}. At the same time, the early-stage data insufficiency poses a critical challenge for safe RL methods, as the algorithm may struggle to identify effective or safe control strategies without sufficient historical or operational data. These combined factors necessitate solutions that simultaneously address uncertainty and data limitations to ensure robust and reliable safe RL applications.

Safe RL methods, such as Lagrangian, projection, safety layer, shielding, and Lyapunov methods, if not integrated with probabilistic approaches, may converge near constraint boundaries in pursuit of maximizing rewards. This behavior increases the risk of control failures when faced with uncertainties, as the absence of probabilistic considerations limits the ability of these methods to account for variability in system dynamics and external conditions. Early-stage data insufficiency further exacerbates these risks.

Several techniques can mitigate this issue. For instance, GP-based safe RL can incorporate uncertainty into its framework by setting different confidence intervals based on RES uncertainty. RRL can also generate control policies by considering the worst-case uncertainty scenarios. Beyond these probabilistic methods, meta-learning \cite{beck2023survey} and transfer learning \cite{da2019survey} improve the adaptability and generalization of safe RL by enabling rapid adaptation to new tasks and leveraging knowledge from related tasks, respectively. Additionally, online learning techniques can be applied to adopt conservative strategies in the early deployment stages while continuously learning and updating from real-time operational data \cite{nguyen2022three}. Data augmentation and stochastic perturbation can be employed during training to increase distribution diversity \cite{laskin2020reinforcement}. Physics-based models can also be integrated to improve interpretability and enhance policy validation, reducing the potential risks of policy failures \cite{chen2022physics}.

\subsubsection{Rapid Changes in Power Networks}
Rapid changes in power networks, such as topology modifications due to equipment outages, fault isolations, or grid reconfigurations, pose significant challenges for safe RL. These changes alter the system's state dynamics and constraints, requiring careful consideration of their reusability and universality of learned policies. In many cases, it may be necessary to rederive and redefine the system dynamics and constraints to adapt to the new operating conditions. Safe RL must address these issues by incorporating mechanisms for rapid adaptation, real-time decision-making, and robust handling of uncertainties. Possible solutions include: integrating model-based RL to simulate topology changes and recalibrate policies efficiently \cite{deng2024safety}; employing online learning to continuously adapt policies with real-time feedback \cite{nguyen2022three}; leveraging GNNs to dynamically update representations of the grid structure \cite{wu2023constrained}; using meta-learning to enable fast adaptation to new topologies with minimal retraining \cite{beck2023survey}; combining safe RL with optimization and expert systems to provide feasible and safe initial solutions \cite{du2024real}; utilizing data augmentation and scenario-based training to prepare agents for diverse topological conditions \cite{laskin2020reinforcement}. In addition, some of the methods for addressing uncertainties discussed in Challenge \ref{sec6A3} can also be applied to manage rapid changes in power networks.

\subsubsection{Multi-Constraint Handling}
Existing safe RL methods are already capable of handling multi-constraint problems. For instance, the Lagrangian relaxation method addresses multiple constraints by introducing multiple Lagrange multipliers \cite{huang2022constrained}. The projection method projects actions onto the combined constraint set to ensure compliance with all constraints \cite{kim2024trust}. The Lyapunov method constructs either a separate Lyapunov function for each stability-related constraint or a joint Lyapunov function for multiple constraints, providing rigorous mathematical stability proofs. GP method independently models each constraint to form a probabilistic representation of the joint safe region. The shielding method performs independent safety checks for each constraint and selects an alternative action that satisfies all constraints if a violation occurs. Safety layer method models multiple constraints jointly as a single optimization problem. The barrier function method constructs a barrier function for each constraint and combines them through weighted summation. RRL integrates multiple constraints into the objective function, optimizing for worst-case rewards while ensuring all constraints are met \cite{zhou2024multi}.

However, these methods share common challenges in multi-constraint scenarios, such as difficulties in updating multiple Lagrange multipliers and weights; the presence of multiple high-dimensional and conflicting constraints \cite{yao2024gradient}; the time-consuming nature of computing feasible alternative actions; overly conservative policies; and the infeasibility of real-time application in high-dimensional settings. To address these challenges, potential solutions include leveraging hierarchical optimization to prioritize critical constraints \cite{roza2023safe}, employing dimensionality reduction techniques to simplify high-dimensional constraints, integrating adaptive weight adjustment methods for conflicting constraints, developing efficient approximation algorithms for alternative action computation, and using parallel computing or model simplifications to enhance real-time applicability \cite{vu2021safe}.

\subsubsection{Reward Design}
Reward design is a critical component of safe RL, as it directly influences the learning process and the resulting policy's safety and efficiency. In safety-critical applications of power systems, designing rewards that effectively balance performance objectives with safety requirements presents unique challenges. For example, challenges include designing appropriate weights and priorities for multi-objective rewards; aligning individual rewards in multi-agent settings with both local and global objectives; resolving conflicts between safety and reward; balancing short-term rewards with long-term goals; dealing with sparse rewards resulting from rare constraint violations or critical failures. Moreover, non-stationarity caused by changes in system topology, load profiles, and RES generation can make the same reward signals correspond to different outcomes over time \cite{calvo2023state, cai2023safe}.

Potential solutions include designing adaptive weights to balance multiple objectives as well as objectives and constraints \cite{yang2023dynamic, liang2021safe}; introducing intermediate rewards to provide continuous feedback during learning; pre-training on simulation models that include more incident scenarios \cite{li2022learning, yi2023model}; dynamically adapting reward signals based on the current system state or operational conditions; integrating physical system models into the reward design to enhance interpretability and safety; and incorporating risk-aware rewards into the reward function to penalize high-risk actions \cite{nagabandi2018learning}.

\subsubsection{User-Centric Design}
The practical application of safe RL systems in power systems hinges on their usability and interpretability, particularly for system operators. However, despite technical advancements, research often overlooks user-centric design elements such as operator trust, seamless integration, and actionable insights. The inherent lack of interpretability in RL models, frequently seen as ``black boxes", undermines trust and complicates adoption. Moreover, the complexity of outputs can misalign with operators’ expertise, increasing the risk of implementation errors. Integration with existing frameworks also presents challenges, often requiring extensive modifications or additional training. In addition, safe RL systems often lack user-friendly interfaces, real-time feedback, and actionable explanations, making them inaccessible to non-technical users and limiting effective intervention. These gaps, combined with insufficient mechanisms for validation and accountability, erode trust in safety-critical scenarios, further impeding their effectiveness in real-world environments \cite{glanois2024survey}.

Potential solutions include developing explainable RL algorithms that provide interpretable decision-making processes \cite{wells2021explainable}. Combining RL with rule-based systems ensures outputs align with operators' existing knowledge, while human-in-the-loop systems enable operators to provide feedback and approve recommended actions during training and deployment \cite{retzlaff2024human, yan2023real}. Simplified interfaces and dashboards can translate RL decisions into actionable insights using visual tools such as heatmaps or risk indicators. Hybrid models that integrate physics-based approaches enhance interpretability by embedding system dynamics into RL policies, ensuring adherence to operational rules \cite{wu2024real}. Real-time monitoring and feedback mechanisms can explain the rationale behind decisions, allowing operators to explore alternative scenarios and outcomes \cite{zolfagharian2024smarla}. Incremental deployment, starting with low-risk tasks, builds operator trust and familiarity, complemented by tailored training programs that demonstrate RL behavior in simulated environments \cite{siemensGTautotuner, googleAICooling, deepmindCoolingAI, zhan2022deepthermal}. Additionally, regulatory and accountability mechanisms, such as logging decision-making processes for auditability, ensure compliance and foster trust in safety-critical applications.

\subsubsection{Early and Emergency Response to Potential Dangers}
When deploying safe RL in power systems, potential risks may arise, requiring early responses or emergency actions to ensure the system's safe operation. Addressing this challenge involves three key approaches. First, strict safety-guarantee safe RL algorithms, such as projection, Lyapunov, shielding, safety layer, and barrier function methods, can be employed to tightly constrain the action space and ensure safety constraints are upheld throughout policy learning and execution \cite{sayed2022feasibility, cui2022decentralized, chen2022physics, wan2023adapsafe}. Second, a real-time risk assessment module can be introduced to evaluate the potential impact of an action on system safety before execution. Predictive models, such as MPC, can provide short-term rapid predictions to determine whether an action might cause harm \cite{wang2025safe, wan2023safecool}. Finally, traditional evaluation methods or physical knowledge can be incorporated to validate each action. If an action fails to meet safety standards, an emergency stop can be triggered, and the system can switch to a conservative strategy. This redundancy can be achieved by integrating a traditional controller as a safety backup. When potential risks are detected, the system transitions from safe RL to the traditional controller to ensure safety \cite{modares2023safe}.

\subsubsection{Real-World Deployment}
In Section \ref{sec5}, we summarized the existing real-world cases of RL deployment in power systems, some of which explicitly indicated the use of safe RL. From these real-world examples, it is evident that safe RL has been tested and deployed in some low-risk, localized systems \cite{lin2023reinforcement, siemensGTautotuner, googleAICooling, deepmindCoolingAI, zhan2022deepthermal}. However, it is still in its early stages and faces numerous challenges before achieving large-scale deployment. These challenges include learning from limited samples during early deployment in live systems, dealing with communication or controller delays, managing uncertainty and randomness introduced by RESs, addressing rare event scenarios, handling multi-constraint systems, ensuring scalability for large-scale power systems, operating under partially observable conditions, meeting real-time computation requirements, enabling offline learning, improving interpretability, satisfying some hard constraints, and adapting to topology changes \cite{dulac2019challenges}. Most of these challenges are discussed in greater detail in this section, along with potential solutions. However, there is still a long way to go before achieving large-scale application.

\subsection{Future Directions in Safe RL}
Based on the challenges discussed above, we outline several potential future research directions for applying safe RL in power systems.

\subsubsection{Exploring Offline Safe RL}
DRL algorithms are based on an online learning paradigm, which presents a significant hurdle to their widespread adoption in power systems. In general, such online interaction is not practical, due to the expense (e.g., in robotics, educational agents, or healthcare) and risk (e.g., in autonomous driving, power systems, or healthcare) associated with exploring control actions in a safety-critical system \cite{levine2020offline}. Even in domains where online interaction is viable, leveraging previously collected data is often preferable, especially in complex domains that require extensive datasets for effective generalization.

Safe RL endeavors to achieve a policy that maximizes rewards within defined constraints, demonstrating advantages in meeting safety requirements for real-world applications. Nonetheless, many deep safe RL approaches primarily address safety post-training, neglecting the costs associated with constraint violations during the training phase. The necessity of collecting online interaction samples poses challenges in ensuring training safety, as preventing the agent from executing unsafe behaviors during learning is non-trivial \cite{liu2023constrained}. Although carefully designed correction systems or human interventions can serve as safety mechanisms to filter unsafe actions during training, their application may prove costly due to the low sample efficiency of many RL approaches. 

It is important to add that it is reasonable to use a simulation environment as a digital twin to train. In fact, even if discrepancies between simulations and real-world conditions are unavoidable, high-fidelity simulations and model-based numerical optimization remain the core components of energy management systems and are the foundation for control actions currently used to manage the grid. If these models are accurate enough for decision systems used today to optimally select control actions, then it is reasonable to assume that are sufficiently accurate to train optimum policies. This is an important question to address in research since at the moment there is no comprehensive characterization of how the discrepancies between simulated and real environments affect performance and safety \cite{prudencio2023survey}.

\subsubsection{Emphasizing Privacy in the Learning Process}
As safe RL algorithms grow in popularity, so too do concerns about their privacy implications \cite{xue2024privacy}. The value or policy functions released are trained using reward signals and other inputs that often depend on sensitive data. In the domain of power systems, some rewards could inadvertently expose critical measurement data, such as voltage phasors and power demands, which in turn could lead to issues like false data injection. This historical data can potentially be deduced by recursively querying the released functions. One potential research direction is the development of differentially private algorithms for safe RL, which safeguard reward information from being compromised by techniques such as inverse RL \cite{wang2019privacy}. The issue of privacy becomes even more critical in the offline RL setting, which is arguably more relevant for applications handling sensitive data. For example, in the EV charging domain, online RL necessitates the continual execution of new exploratory policies for each arriving EV, involving sensitive data like arrival and departure times. In contrast, offline RL relies on historical data of EV charging behavior, which can be particularly sensitive \cite{qiao2024offline}. However, these differentially private mechanisms could introduce uncertainty into safety constraints. Concurrently, differentially private AC-PF constrained OPF has been explored, with studies formulating it as robust optimization to ensure the feasibility of these safety constraints \cite{dvorkin2020differentially}. One potential approach is to develop robust formulation training for safe DRL.

\subsubsection{Integrating Federated Learning Mechanism}
To simultaneously address privacy and scalability issues, integrating federated learning into safe DRL could be a viable solution. In practical scenarios, RL faces challenges such as poor agent performance in large action and state spaces due to limited sample exploration and low sample efficiency impacting learning speed. Information exchange between agents can significantly boost learning rates. While distributed and parallel RL algorithms address these issues by centralizing data, parameters, or gradients for model training, this centralization can compromise privacy, leading to agent mistrust and data interception risks \cite{qi2021federated}.

Federated learning, however, enables information exchange without compromising privacy, helping agents adapt to diverse environments. It also addresses the simulation-reality gap often present in RL; while many RL algorithms depend on pre-training in simulation environments that do not perfectly mirror the real-world, federated learning can amalgamate insights from both to more accurately bridge this gap \cite{fan2021fault}. Additionally, federated learning is beneficial when agents only observe partial features, enabling effective aggregation of this limited information.
These considerations give rise to the idea of federated safe RL, which merges federated learning and safe RL within a privacy-preserving framework, adapting safe RL strategies for sequential decision-making tasks.

\subsubsection{Advancing Convex Insights}
Convex optimization is extensively explored for its ability to provide analytical convergence and optimality guarantees, which in turn yield more stable policies. In the context of safe DRL with convex or non-convex constraints, integrating convex insights can enhance these convergence guarantees. Advancing these insights into safe DRL, consider exploring the application of ICNNs. Rather than training a conventional policy that inputs data and outputs control actions, which must adhere to stringent physical constraints, ICNNs offer a promising alternative due to their superior generalization capabilities. This approach bridges the gap between model accuracy and control tractability by constructing networks that are convex relative to their inputs, as detailed by \cite{amos2017input} and further applied by \cite{chen2018optimal} to model complex physical systems accurately. Consequently, training an ICNN-based policy can more easily incorporate convex constraints to ensure feasible and safe optimal control actions with performance guarantees.

Additionally, using convex functions to approximate the policy function represents another viable strategy. Here, policy optimization can be formulated as a constrained optimization problem, where both the objective and constraints are initially nonconvex. By creating a series of surrogate convex-constrained optimization problems that locally substitute nonconvex functions with convex quadratic functions derived from policy gradient estimators as described by \cite{yu2019convergent}, this method allows for the practical application of theoretical insights to operational policies.
These strategies underscore the potential of convex optimization techniques in enhancing the robustness and effectiveness of safe DRL algorithms, particularly in applications that demand adherence to strict safety and physical constraints.

\subsubsection{Hybrid/Fused Methods}
In the application of safe RL in power systems, hybrid/fused methods combine multiple approaches to address challenges related to uncertainty, safety, and complexity. By integrating safe RL, optimization techniques, physical models, and other data-driven methods, these approaches enhance the efficiency, safety, and reliability of policies. Compared to conventional RL, safe RL places greater emphasis on hybrid/fused methods. For example, in \cite{sayed2023optimal}, a dynamic layer is embedded between the SAC-generated policy and the power system environment. This layer generates fully operable control actions by solving embedded power flow equations and ensuring that the control solutions satisfy various constraints, such as power flow and voltage limits. Similarly, in \cite{wu2024real}, the Jacobian matrix, which represents the sensitivity relationship between power injection and system voltage amplitude/phase, is utilized to mask action directions irrelevant to constraints, thereby reducing exploration risks. Moreover, in \cite{wu2023constrained}, a complex-valued spatio-temporal GCN is employed for the actor to capture the spatiotemporal correlations of the environment state in a modified TD3 framework using primal-dual methods to solve the stochastic dynamic OPF problem. In \cite{shengren2023optimal}, the action-value function, approximated through a DNN, is formulated as a MILP problem, enabling the incorporation of constraints directly into the action space. In addition, methods such as the Lyapunov method, barrier function method, and RRL draw inspiration from traditional optimization techniques for handling constraints and ensuring stability \cite{jin2020stability, gu2022recurrent, wan2023adapsafe}. These examples demonstrate the significant progress made in hybrid/fused methods-based safe RL. However, there remains a need to develop new algorithms, integrate additional traditional techniques, and incorporate emerging technologies to further advance this field.

\subsubsection{Developing LLM-in-the-loop RL}
Numerous practical objectives and constraints of power systems, such as those outlined in the security guideline and operation manual, are based on linguistic stipulations and are difficult to model. In actual power system operations, when these constraints are violated, system operators typically need to take corrective actions \cite{yan2023real}. Therefore, a human-in-the-loop approach has been proposed, where humans are integrated into the RL iteration process. This involvement allows humans to actively participate in constraint management, thereby enhancing the reliability of RL \cite{sun2023optimal, yang2020optimal}. Nonetheless, human-in-the-loop is limited by the availability and time constraints of human experts, making it unfeasible for tasks that require extensive amounts of training data or continuous adaptation.

With the advent of LLMs, the possibility of transitioning from human-in-the-loop to LLM-in-the-loop systems emerges as a viable alternative to address the aforementioned challenges \cite{cao2024survey}. LLMs, with their powerful learning capabilities and vast knowledge based on power system data and linguistic stipulations, can provide consistent, real-time, and potentially unbiased feedback compared to human experts \cite{Majumder2024}. For example, \cite{yan2023real} integrates the GPT LLM into the OPF framework with linguistic rules. This model quantifies natural language stipulations as objectives and constraints within the power system optimization problem for the first time. In the future, leveraging specialized knowledge in the power system domain to train dedicated LLMs will be crucial for extending their application across a broader spectrum of the power system industry. However, challenges remain in how LLMs can efficiently learn from power system knowledge bases, integrate with existing software tools, quantify uncertainties, and ensure the safety of constraints \cite{Majumder2024}.

\section{Conclusion}\label{sec7}
This paper represents the first comprehensive review of the application of safe RL in modern power systems. It begins by introducing the foundational concepts of safe RL. Next, it defines safe RL in the context of power system optimization and control, reviewing constraints, environments, and safety, while exploring motivations from a comparative perspective. It then summarizes existing safe RL algorithms, contrasts their applicability across different domains, and introduces current benchmark environments, algorithms, and software. Following this, the paper provides an extensive overview of almost all existing studies on safe RL applications in power systems, summarizing the key elements of state, action, reward, and constraint settings across various applications, analyzing suitable and unsuitable deployment areas, and outlining real-world deployment cases alongside a future roadmap. Finally, it discusses the challenges and outlook for safe RL development. As the application of safe RL in power systems is a relatively recent development, emerging mainly in the past three years, this paper provides a comprehensive summary and discussion to inspire future researchers and encourage practical deployment in suitable areas, integrating with traditional methods to serve modern power systems.

\bibliographystyle{IEEEtran}
\bibliography{ref.bib}

\begin{thebibliography}{100}
\providecommand{\url}[1]{#1}
\csname url@samestyle\endcsname
\providecommand{\newblock}{\relax}
\providecommand{\bibinfo}[2]{#2}
\providecommand{\BIBentrySTDinterwordspacing}{\spaceskip=0pt\relax}
\providecommand{\BIBentryALTinterwordstretchfactor}{4}
\providecommand{\BIBentryALTinterwordspacing}{\spaceskip=\fontdimen2\font plus
\BIBentryALTinterwordstretchfactor\fontdimen3\font minus \fontdimen4\font\relax}
\providecommand{\BIBforeignlanguage}[2]{{%
\expandafter\ifx\csname l@#1\endcsname\relax
\typeout{** WARNING: IEEEtran.bst: No hyphenation pattern has been}%
\typeout{** loaded for the language `#1'. Using the pattern for}%
\typeout{** the default language instead.}%
\else
\language=\csname l@#1\endcsname
\fi
#2}}
\providecommand{\BIBdecl}{\relax}
\BIBdecl

\bibitem{aien2016comprehensive}
M.~Aien, A.~Hajebrahimi, and M.~Fotuhi-Firuzabad, ``A comprehensive review on uncertainty modeling techniques in power system studies,'' \emph{Renew. Sustain. Energy Rev.}, vol.~57, pp. 1077--1089, May 2016.

\bibitem{roald2023power}
L.~A. Roald, D.~Pozo, A.~Papavasiliou, D.~K. Molzahn, J.~Kazempour, and A.~Conejo, ``Power systems optimization under uncertainty: A review of methods and applications,'' \emph{Electric Power Syst. Res.}, vol. 214, Jan. 2023, {Art.} no. 108725.

\bibitem{cheng2023survey}
G.~Cheng, Y.~Lin, A.~Abur, A.~G{\'o}mez-Exp{\'o}sito, and W.~Wu, ``A survey of power system state estimation using multiple data sources: {PMUs, SCADA, AMI}, and beyond,'' \emph{IEEE Trans. Smart Grid}, vol.~15, no.~1, pp. 1129--1151, Jan. 2024.

\bibitem{li2024artificial}
Y.~Li, Y.~Ding, S.~He, F.~Hu, J.~Duan, G.~Wen, H.~Geng, Z.~Wu, H.~B. Gooi, Y.~Zhao \emph{et~al.}, ``Artificial intelligence-based methods for renewable power system operation,'' \emph{Nature Rev. Electr. Eng.}, vol.~1, no.~3, pp. 163--179, Feb. 2024.

\bibitem{chen2022reinforcement}
X.~Chen, G.~Qu, Y.~Tang, S.~Low, and N.~Li, ``Reinforcement learning for selective key applications in power systems: Recent advances and future challenges,'' \emph{IEEE Trans. Smart Grid}, vol.~13, no.~4, pp. 2935--2958, Jul. 2022.

\bibitem{li2023deep}
Y.~Li, C.~Yu, M.~Shahidehpour, T.~Yang, Z.~Zeng, and T.~Chai, ``Deep reinforcement learning for smart grid operations: Algorithms, applications, and prospects,'' \emph{Proc. IEEE}, vol. 111, no.~9, pp. 1055--1096, Sep. 2023.

\bibitem{nguyen2020deep}
T.~T. Nguyen, N.~D. Nguyen, and S.~Nahavandi, ``Deep reinforcement learning for multiagent systems: A review of challenges, solutions, and applications,'' \emph{IEEE Trans. Cybern.}, vol.~50, no.~9, pp. 3826--3839, Sep. 2020.

\bibitem{li2017deep}
Y.~Li, ``Deep reinforcement learning: An overview,'' \emph{arXiv preprint arXiv:1701.07274}, 2017.

\bibitem{gu2024review}
S.~Gu, L.~Yang, Y.~Du, G.~Chen, F.~Walter, J.~Wang, and A.~Knoll, ``A review of safe reinforcement learning: Methods, theories and applications,'' \emph{IEEE Trans. Pattern Anal. Mach. Intell.}, vol.~46, no.~12, pp. 11\,216--11\,235, Dec. 2024.

\bibitem{garcia2015comprehensive}
J.~Garc{\i}a and F.~Fern{\'a}ndez, ``A comprehensive survey on safe reinforcement learning,'' \emph{J. Mach. Learn. Res.}, vol.~16, no.~1, pp. 1437--1480, 2015.

\bibitem{zhao2022deep}
J.~Zhao, F.~Li, S.~Mukherjee, and C.~Sticht, ``Deep reinforcement learning-based model-free on-line dynamic multi-microgrid formation to enhance resilience,'' \emph{IEEE Trans. Smart Grid}, vol.~13, no.~4, pp. 2557--2567, Jul. 2022.

\bibitem{zhang2023data}
M.~Zhang, G.~Guo, S.~Magn{\'u}sson, R.~C. Pilawa-Podgurski, and Q.~Xu, ``Data driven decentralized control of inverter based renewable energy sources using safe guaranteed multi-agent deep reinforcement learning,'' \emph{IEEE Trans. Sustain. Energy}, vol.~15, no.~2, pp. 1288--1299, Apr. 2024.

\bibitem{yan2023multi}
R.~Yan, Q.~Xing, and Y.~Xu, ``Multi agent safe graph reinforcement learning for {PV} inverter s based real-time decentralized {Volt/Var} control in zoned distribution networks,'' \emph{IEEE Trans. Smart Grid}, vol.~15, no.~1, pp. 299--311, Jan. 2024.

\bibitem{arulkumaran2017deep}
K.~Arulkumaran, M.~P. Deisenroth, M.~Brundage, and A.~A. Bharath, ``Deep reinforcement learning: A brief survey,'' \emph{IEEE Signal Process. Mag.}, vol.~34, no.~6, pp. 26--38, Nov. 2017.

\bibitem{wells2021explainable}
L.~Wells and T.~Bednarz, ``Explainable {AI} and reinforcement learning—a systematic review of current approaches and trends,'' \emph{Front. Artif. Intell.}, vol.~4, May 2021, {Art.} no. 550030.

\bibitem{prudencio2023survey}
R.~F. Prudencio, M.~R. Maximo, and E.~L. Colombini, ``A survey on offline reinforcement learning: Taxonomy, review, and open problems,'' \emph{IEEE Trans. Neural Netw. Learn. Syst.}, vol.~35, no.~8, pp. 10\,237--10\,257, Aug. 2024.

\bibitem{zhang2019deep}
Z.~Zhang, D.~Zhang, and R.~C. Qiu, ``Deep reinforcement learning for power system applications: An overview,'' \emph{CSEE J. Power Energy Syst.}, vol.~6, no.~1, pp. 213--225, Mar. 2020.

\bibitem{glavic2019deep}
M.~Glavic, ``(deep) reinforcement learning for electric power system control and related problems: A short review and perspectives,'' \emph{Annu. Rev. Control}, vol.~48, pp. 22--35, 2019.

\bibitem{cao2020reinforcement}
D.~Cao, W.~Hu, J.~Zhao, G.~Zhang, B.~Zhang, Z.~Liu, Z.~Chen, and F.~Blaabjerg, ``Reinforcement learning and its applications in modern power and energy systems: A review,'' \emph{J. Modern Power Syst. Clean Energy}, vol.~8, no.~6, pp. 1029--1042, Nov. 2020.

\bibitem{zhao2023state}
W.~Zhao, T.~He, R.~Chen, T.~Wei, and C.~Liu, ``State-wise safe reinforcement learning: A survey,'' in \emph{Proc. Int. Joint Conf. Artif. Intell.}, Aug. 2023, pp. 6814--6822.

\bibitem{wang2023safe}
\BIBentryALTinterwordspacing
X.~Wang, R.~Wang, and Y.~Cheng, ``Safe reinforcement learning: A survey,'' \emph{Acta Automatica Sinica}, vol.~49, no.~9, pp. 1813--1835, Sep. 2023. [Online]. Available: \url{http://www.aas.net.cn/article/doi/10.16383/j.aas.c220631}
\BIBentrySTDinterwordspacing

\bibitem{li2023research}
J.~Li, X.~Wang, S.~Chen, and D.~Yan, ``Research and application of safe reinforcement learning in power system,'' in \emph{Proc. Asia Conf. Power Electr. Eng.}, 2023, pp. 1977--1982.

\bibitem{su2024review}
T.~Su, T.~Wu, J.~Zhao, A.~Scaglione, and L.~Xie, ``A review of safe reinforcement learning methods for modern power systems,'' \emph{arXiv preprint arXiv:2407.00304}, 2024.

\bibitem{yu2024safe}
P.~Yu, Z.~Wang, H.~Zhang, and Y.~Song, ``Safe reinforcement learning for power system control: A review,'' \emph{arXiv preprint arXiv:2407.00681}, 2024.

\bibitem{bui2025critical}
V.-H. Bui, S.~Mohammadi, S.~Das, A.~Hussain, G.~V. Hollweg, and W.~Su, ``A critical review of safe reinforcement learning strategies in power and energy systems,'' \emph{Eng. Appl. Artif. Intell.}, vol. 143, Mar. 2025, {Art.} no. 110091.

\bibitem{bui2024critical}
V.-H. Bui, S.~Das, A.~Hussain, G.~V. Hollweg, and W.~Su, ``A critical review of safe reinforcement learning techniques in smart grid applications,'' \emph{arXiv preprint arXiv:2409.16256}, 2024.

\bibitem{su2025github}
\BIBentryALTinterwordspacing
T.~Su, T.~Wu, J.~Zhao, A.~Scaglione, and L.~Xie, ``{SafeRL-Power-System},'' Accessed: Mar. 24, 2025. [Online]. Available: \url{https://github.com/eetongsu/SafeRL-Power-System}
\BIBentrySTDinterwordspacing

\bibitem{sutton2018reinforcement}
R.~S. Sutton and A.~G. Barto, \emph{Reinforcement learning: An introduction}.\hskip 1em plus 0.5em minus 0.4em\relax Cambridge, MA, USA: MIT Press, 2018.

\bibitem{krasowski2023provably}
H.~Krasowski, J.~Thumm, M.~M{\"u}ller, L.~Sch{\"a}fer, X.~Wang, and M.~Althoff, ``Provably safe reinforcement learning: Conceptual analysis, survey, and benchmarking,'' \emph{Trans. Mach. Learn. Res.}, Sep. 2024.

\bibitem{liu2021policy}
Y.~Liu, A.~Halev, and X.~Liu, ``Policy learning with constraints in model-free reinforcement learning: A survey,'' in \emph{Proc. Int. Joint Conf. Artif. Intell.}, Aug. 2021, pp. 1--8.

\bibitem{wachi2024survey}
A.~Wachi, X.~Shen, and Y.~Sui, ``A survey of constraint formulations in safe reinforcement learning,'' \emph{arXiv preprint arXiv:2402.02025}, 2024.

\bibitem{achiam2017constrained}
J.~Achiam, D.~Held, A.~Tamar, and P.~Abbeel, ``Constrained policy optimization,'' in \emph{Proc. Int. Conf. Mach. Learn.}, vol.~70, no.~10, Aug. 2017, pp. 22--31.

\bibitem{li2022learning}
H.~Li and H.~He, ``Learning to operate distribution networks with safe deep reinforcement learning,'' \emph{IEEE Trans. Smart Grid}, vol.~13, no.~3, pp. 1860--1872, May 2022.

\bibitem{zhang2020multi}
Q.~Zhang, K.~Dehghanpour, Z.~Wang, F.~Qiu, and D.~Zhao, ``Multi-agent safe policy learning for power management of networked microgrids,'' \emph{IEEE Trans. Smart Grid}, vol.~12, no.~2, pp. 1048--1062, Mar. 2021.

\bibitem{ye2023safe}
Y.~Ye, H.~Wang, P.~Chen, Y.~Tang, and G.~Strbac, ``Safe deep reinforcement learning for microgrid energy management in distribution networks with leveraged spatial-temporal perception,'' \emph{IEEE Trans. Smart Grid}, vol.~14, no.~5, pp. 3759--3775, Sep. 2023.

\bibitem{yi2023model}
Z.~Yi, Y.~Xu, and C.~Wu, ``Model-free economic dispatch for virtual power plants: An adversarial safe reinforcement learning approach,'' \emph{IEEE Trans. Power Syst.}, vol.~39, no.~2, pp. 3153--3168, Mar. 2024.

\bibitem{liu2020two}
H.~Liu and W.~Wu, ``Two-stage deep reinforcement learning for inverter-based {Volt-VAR} control in active distribution networks,'' \emph{IEEE Trans. Smart Grid}, vol.~12, no.~3, pp. 2037--2047, May 2021.

\bibitem{lin2023reinforcement}
X.~Lin, D.~Yuan, and X.~Li, ``Reinforcement learning with dual safety policies for energy savings in building energy systems,'' \emph{Buildings}, vol.~13, no.~3, p. 580, 2023.

\bibitem{vu2023multi}
L.~Vu, T.~Vu, T.~L. Vu, and A.~Srivastava, ``Multi-agent deep reinforcement learning for distributed load restoration,'' \emph{IEEE Trans. Smart Grid}, vol.~15, no.~2, pp. 1749--1760, Mar. 2024.

\bibitem{yan2022hybrid}
Z.~Yan and Y.~Xu, ``A hybrid data-driven method for fast solution of security-constrained optimal power flow,'' \emph{IEEE Trans. Power Syst.}, vol.~37, no.~6, pp. 4365--4374, Nov. 2022.

\bibitem{cao2022model}
D.~Cao, J.~Zhao, W.~Hu, F.~Ding, N.~Yu, Q.~Huang, and Z.~Chen, ``Model-free voltage control of active distribution system with {PVs} using surrogate model-based deep reinforcement learning,'' \emph{Appl. Energy}, vol. 306, Jan. 2022, {Art.} no. 117982.

\bibitem{su2025safe}
T.~Su, J.~Zhao, Y.~Yao, A.~Selim, and F.~Ding, ``Safe reinforcement learning-based transient stability control for islanded microgrids with topology reconfiguration,'' \emph{IEEE Trans. Smart Grid}, 2025.

\bibitem{kundur1994power}
P.~Kundur, \emph{Power System Stability and Control}.\hskip 1em plus 0.5em minus 0.4em\relax New York, NY, USA: McGraw-Hill, 1994.

\bibitem{kundur2004definition}
P.~Kundur \emph{et~al.}, ``Definition and classification of power system stability {IEEE/CIGRE} joint task force on stability terms and definitions,'' \emph{IEEE Trans. Power Syst.}, vol.~19, no.~3, pp. 1387--1401, Aug. 2004.

\bibitem{prostejovsky2016distribution}
A.~M. Prostejovsky, O.~Gehrke, A.~M. Kosek, T.~Strasser, and H.~W. Bindner, ``Distribution line parameter estimation under consideration of measurement tolerances,'' \emph{IEEE Trans. Ind. Inform.}, vol.~12, no.~2, pp. 726--735, Apr. 2016.

\bibitem{tang2022multi}
Q.~Tang, H.~Guo, and Q.~Chen, ``Multi-market bidding behavior analysis of energy storage system based on inverse reinforcement learning,'' \emph{IEEE Trans. Power Syst.}, vol.~37, no.~6, pp. 4819--4831, Nov. 2022.

\bibitem{lockwood2022review}
O.~Lockwood and M.~Si, ``A review of uncertainty for deep reinforcement learning,'' in \emph{Proc. AAAI Conf. Artif. Intell. Interactive Digit. Entertainment}, vol.~18, no.~1, 2022, pp. 155--162.

\bibitem{glanois2024survey}
C.~Glanois, P.~Weng, M.~Zimmer, D.~Li, T.~Yang, J.~Hao, and W.~Liu, ``A survey on interpretable reinforcement learning,'' \emph{Mach. Learn.}, pp. 1--44, Apr. 2024.

\bibitem{zanon2020safe}
M.~Zanon and S.~Gros, ``Safe reinforcement learning using robust {MPC},'' \emph{IEEE Trans. Autom. Control}, vol.~66, no.~8, pp. 3638--3652, Aug. 2021.

\bibitem{altman1999constrained}
E.~Altman, \emph{Constrained {Markov} decision processes}.\hskip 1em plus 0.5em minus 0.4em\relax London, U.K.: Chapman and Hall, Mar. 1999.

\bibitem{chow2018risk}
Y.~Chow, M.~Ghavamzadeh, L.~Janson, and M.~Pavone, ``Risk-constrained reinforcement learning with percentile risk criteria,'' \emph{J. Mach. Learn. Res.}, vol.~18, no. 167, pp. 1--51, 2018.

\bibitem{bertsekas2015convex}
D.~Bertsekas, \emph{Convex optimization algorithms}.\hskip 1em plus 0.5em minus 0.4em\relax Athena Scientific, 2015.

\bibitem{ray2019benchmarking}
A.~Ray, J.~Achiam, and D.~Amodei, ``Benchmarking safe exploration in deep reinforcement learning,'' \emph{arXiv preprint arXiv:1910.01708}, 2019.

\bibitem{gu2021multi}
S.~Gu, J.~G. Kuba, M.~Wen, R.~Chen, Z.~Wang, Z.~Tian, J.~Wang, A.~Knoll, and Y.~Yang, ``Multi-agent constrained policy optimisation,'' \emph{arXiv preprint arXiv:2110.02793}, 2021.

\bibitem{stooke2020responsive}
A.~Stooke, J.~Achiam, and P.~Abbeel, ``Responsive safety in reinforcement learning by {PID} {Lagrangian} methods,'' in \emph{Proc. Int. Conf. Mach. Learn.}, 2020, pp. 9133--9143.

\bibitem{wu2023constrained}
T.~Wu, A.~Scaglione, and D.~Arnold, ``Constrained reinforcement learning for predictive control in real-time stochastic dynamic optimal power flow,'' \emph{IEEE Trans. Power Syst.}, vol.~39, no.~3, pp. 5077--5090, May 2024.

\bibitem{wang2019safe}
W.~Wang, N.~Yu, Y.~Gao, and J.~Shi, ``Safe off-policy deep reinforcement learning algorithm for {Volt-VAR} control in power distribution systems,'' \emph{IEEE Trans. Smart Grid}, vol.~11, no.~4, pp. 3008--3018, Jul. 2020.

\bibitem{wu2023network}
T.~Wu, A.~Scaglione, A.~P. Surani, D.~Arnold, and S.~Peisert, ``Network-constrained reinforcement learning for optimal {EV} charging control,'' in \emph{Proc. IEEE Int. Conf. Smart Grid Commun.}, 2023, pp. 1--6.

\bibitem{sayed2023online}
A.~R. Sayed, X.~Zhang, Y.~Wang, G.~Wang, J.~Qiu, and C.~Wang, ``Online operational decision-making for integrated electric-gas systems with safe reinforcement learning,'' \emph{IEEE Trans. Power Syst.}, vol.~39, no.~2, pp. 2893--2906, Mar. 2024.

\bibitem{ding2020natural}
D.~Ding, K.~Zhang, T.~Basar, and M.~Jovanovic, ``Natural policy gradient primal-dual method for constrained {Markov} decision processes,'' in \emph{Proc. Adv. Neural Inf. Process. Syst.}, vol.~33, 2020, pp. 8378--8390.

\bibitem{yang2020projection}
T.-Y. Yang, J.~Rosca, K.~Narasimhan, and P.~J. Ramadge, ``Projection-based constrained policy optimization,'' in \emph{Proc. Int. Conf. Learn. Representations}, 2019, pp. 1--24.

\bibitem{zhang2020first}
Y.~Zhang, Q.~Vuong, and K.~Ross, ``First order constrained optimization in policy space,'' in \emph{Proc. Adv. Neural Inf. Process. Syst.}, vol.~33, 2020, pp. 15\,338--15\,349.

\bibitem{yang2022constrained}
L.~Yang, J.~Ji, J.~Dai, L.~Zhang, B.~Zhou, P.~Li, Y.~Yang, and G.~Pan, ``Constrained update projection approach to safe policy optimization,'' in \emph{Proc. Adv. Neural Inf. Process. Syst.}, vol.~35, 2022, pp. 9111--9124.

\bibitem{jiang2021data}
Y.~Jiang, Q.~Ye, B.~Sun, Y.~Wu, and D.~H. Tsang, ``Data-driven coordinated charging for electric vehicles with continuous charging rates: A deep policy gradient approach,'' \emph{IEEE Internet Things J.}, vol.~9, no.~14, pp. 12\,395--12\,412, Jul. 2021.

\bibitem{wang2019volt}
W.~Wang, N.~Yu, J.~Shi, and Y.~Gao, ``{Volt-VAR} control in power distribution systems with deep reinforcement learning,'' in \emph{Proc. IEEE Int. Conf. Commun. Control Comput. Technol. Smart Grids}, Oct. 2019, pp. 1--7.

\bibitem{sepulchre2012constructive}
R.~Sepulchre, M.~Jankovic, and P.~V. Kokotovic, \emph{Constructive nonlinear control}.\hskip 1em plus 0.5em minus 0.4em\relax Springer Science \& Business Media, 2012.

\bibitem{perkins2002lyapunov}
T.~J. Perkins and A.~G. Barto, ``Lyapunov design for safe reinforcement learning,'' \emph{J. Mach. Learn. Res.}, vol.~3, pp. 803--832, Dec 2002.

\bibitem{cui2022reinforcement}
W.~Cui, Y.~Jiang, and B.~Zhang, ``Reinforcement learning for optimal primary frequency control: A {Lyapunov} approach,'' \emph{IEEE Trans. Power Syst.}, vol.~38, no.~2, pp. 1676--1688, Mar. 2023.

\bibitem{cui2022decentralized}
W.~Cui, J.~Li, and B.~Zhang, ``Decentralized safe reinforcement learning for inverter-based voltage control,'' \emph{Electric Power Syst. Res.}, vol. 211, Oct. 2022, {Art.} no. 108609.

\bibitem{shi2022stability}
Y.~Shi, G.~Qu, S.~Low, A.~Anandkumar, and A.~Wierman, ``Stability constrained reinforcement learning for real-time voltage control,'' in \emph{Proc. Amer. Control Conf.}, 2022, pp. 2715--2721.

\bibitem{williams2006gaussian}
C.~K. Williams and C.~E. Rasmussen, \emph{{Gaussian} processes for machine learning}.\hskip 1em plus 0.5em minus 0.4em\relax Cambridge, MA, USA: MIT Press, 2006, vol.~2, no.~3.

\bibitem{akametalu2014reachability}
A.~K. Akametalu, J.~F. Fisac, J.~H. Gillula, S.~Kaynama, M.~N. Zeilinger, and C.~J. Tomlin, ``Reachability-based safe learning with {Gaussian} processes,'' in \emph{Proc. IEEE Conf. Decis. Control}, Dec. 2014, pp. 1424--1431.

\bibitem{sui2015safe}
Y.~Sui, A.~Gotovos, J.~Burdick, and A.~Krause, ``Safe exploration for optimization with {Gaussian} processes,'' in \emph{Proc. Int. Conf. Mach. Learn.}, 2015, pp. 997--1005.

\bibitem{turchetta2016safe}
M.~Turchetta, F.~Berkenkamp, and A.~Krause, ``Safe exploration in finite {Markov} decision processes with {Gaussian} processes,'' in \emph{Proc. Adv. Neural Inf. Process. Syst.}, vol.~29, 2016, pp. 4312-- 4320.

\bibitem{cowen2022samba}
A.~I. Cowen-Rivers, D.~Palenicek, V.~Moens, M.~A. Abdullah, A.~Sootla, J.~Wang, and H.~Bou-Ammar, ``{SAMBA}: Safe model-based \& active reinforcement learning,'' \emph{Mach. Learn.}, vol. 111, no.~1, pp. 173--203, 2022.

\bibitem{deisenroth2011pilco}
M.~Deisenroth and C.~E. Rasmussen, ``{PILCO}: A model-based and data-efficient approach to policy search,'' in \emph{Proc. Int. Conf. Mach. Learn.}, 2011, pp. 465--472.

\bibitem{alshiekh2018safe}
M.~Alshiekh, R.~Bloem, R.~Ehlers, B.~K{\"o}nighofer, S.~Niekum, and U.~Topcu, ``Safe reinforcement learning via shielding,'' in \emph{Proc. AAAI Conf. Artif. Intell.}, vol.~32, no.~1, Apr. 2018, p. 2661–2669.

\bibitem{chen2022physics}
P.~Chen, S.~Liu, X.~Wang, and I.~Kamwa, ``Physics-shielded multi-agent deep reinforcement learning for safe active voltage control with photovoltaic/battery energy storage systems,'' \emph{IEEE Trans. Smart Grid}, vol.~14, no.~4, pp. 2656--2667, Jul. 2023.

\bibitem{zhang2022residual}
Q.~Zhang, M.~H.~B. Mahbod, C.-B. Chng, P.-S. Lee, and C.-K. Chui, ``Residual physics and post-posed shielding for safe deep reinforcement learning method,'' \emph{IEEE Trans. Cybern.}, vol.~54, no.~2, pp. 865--876, Feb. 2024.

\bibitem{ajagekar2022deep}
A.~Ajagekar and F.~You, ``Deep reinforcement learning based unit commitment scheduling under load and wind power uncertainty,'' \emph{IEEE Trans. Sustain. Energy}, vol.~14, no.~2, pp. 803--812, Apr. 2023.

\bibitem{dalal2018safe}
G.~Dalal, K.~Dvijotham, M.~Vecerik, T.~Hester, C.~Paduraru, and Y.~Tassa, ``Safe exploration in continuous action spaces,'' \emph{arXiv preprint arXiv:1801.08757}, 2018.

\bibitem{yi2023real}
Z.~Yi, X.~Wang, C.~Yang, C.~Yang, M.~Niu, and W.~Yin, ``Real-time sequential security-constrained optimal power flow: A hybrid knowledge-data-driven reinforcement learning approach,'' \emph{IEEE Trans. Power Syst.}, vol.~39, no.~1, pp. 1664--1680, Jan. 2024.

\bibitem{gao2022model}
Y.~Gao and N.~Yu, ``Model-augmented safe reinforcement learning for {Volt-VAR} control in power distribution networks,'' \emph{Appl. Energy}, vol. 313, May 2022, {Art.} no. 118762.

\bibitem{du2022deep}
Y.~Du and D.~Wu, ``Deep reinforcement learning from demonstrations to assist service restoration in islanded microgrids,'' \emph{IEEE Trans. Sustain. Energy}, vol.~13, no.~2, pp. 1062--1072, Apr. 2022.

\bibitem{wang2023enforcing}
Y.~Wang, S.~S. Zhan, R.~Jiao, Z.~Wang, W.~Jin, Z.~Yang, Z.~Wang, C.~Huang, and Q.~Zhu, ``Enforcing hard constraints with soft barriers: Safe reinforcement learning in unknown stochastic environments,'' in \emph{Proc. Int. Conf. Mach. Learn.}, 2023, pp. 36\,593--36\,604.

\bibitem{liu2020ipo}
Y.~Liu, J.~Ding, and X.~Liu, ``{IPO}: Interior-point policy optimization under constraints,'' in \emph{Proc. AAAI Conf. Artif. Intell.}, vol.~34, no.~04, 2020, pp. 4940--4947.

\bibitem{cui2023online}
H.~Cui, Y.~Ye, J.~Hu, Y.~Tang, Z.~Lin, and G.~Strbac, ``Online preventive control for transmission overload relief using safe reinforcement learning with enhanced spatial-temporal awareness,'' \emph{IEEE Trans. Power Syst.}, vol.~39, no.~1, pp. 517--532, Jan. 2024.

\bibitem{vu2021barrier}
T.~L. Vu, S.~Mukherjee, R.~Huang, and Q.~Huang, ``Barrier function-based safe reinforcement learning for emergency control of power systems,'' in \emph{Proc. IEEE Conf. Decis. Control}, 2021, pp. 3652--3657.

\bibitem{li2021safe}
Y.~Li, N.~Li, H.~E. Tseng, A.~Girard, D.~Filev, and I.~Kolmanovsky, ``Safe reinforcement learning using robust action governor,'' in \emph{Proc. Learn. Dyn. Control}, 2021, pp. 1093--1104.

\bibitem{kordabad2022safe}
A.~B. Kordabad, R.~Wisniewski, and S.~Gros, ``Safe reinforcement learning using {Wasserstein} distributionally robust {MPC} and chance constraint,'' \emph{IEEE Access}, vol.~10, pp. 130\,058--130\,067, 2022.

\bibitem{pfrommer2022safe}
S.~Pfrommer, T.~Gautam, A.~Zhou, and S.~Sojoudi, ``Safe reinforcement learning with chance-constrained model predictive control,'' in \emph{Proc. Learn. Dyn. Control}, 2022, pp. 291--303.

\bibitem{coulson2021distributionally}
J.~Coulson, J.~Lygeros, and F.~D{\"o}rfler, ``Distributionally robust chance constrained data-enabled predictive control,'' \emph{IEEE Trans. Autom. Control}, vol.~67, no.~7, pp. 3289--3304, Jul. 2022.

\bibitem{yu2021robust}
J.~Yu, C.~Gehring, F.~Sch{\"a}fer, and A.~Anandkumar, ``Robust reinforcement learning: A constrained game-theoretic approach,'' in \emph{Proc. Learn. Dyn. Control}, 2021, pp. 1242--1254.

\bibitem{rajeswaran2020game}
A.~Rajeswaran, I.~Mordatch, and V.~Kumar, ``A game theoretic framework for model based reinforcement learning,'' in \emph{Proc. Int. Conf. Mach. Learn.}, 2020, pp. 7953--7963.

\bibitem{asheralieva2019hierarchical}
A.~Asheralieva and D.~Niyato, ``Hierarchical game-theoretic and reinforcement learning framework for computational offloading in {UAV}-enabled mobile edge computing networks with multiple service providers,'' \emph{IEEE Internet Things J.}, vol.~6, no.~5, pp. 8753--8769, Oct. 2019.

\bibitem{tessler2019action}
C.~Tessler, Y.~Efroni, and S.~Mannor, ``Action robust reinforcement learning and applications in continuous control,'' in \emph{Proc. Int. Conf. Mach. Learn.}, 2019, pp. 6215--6224.

\bibitem{ni2019multistage}
Z.~Ni and S.~Paul, ``A multistage game in smart grid security: A reinforcement learning solution,'' \emph{IEEE Trans. Neural Netw. Learn. Syst.}, vol.~30, no.~9, pp. 2684--2695, Sep. 2019.

\bibitem{guo2021reinforcement}
Y.~Guo, L.~Wang, Z.~Liu, and Y.~Shen, ``Reinforcement-learning-based dynamic defense strategy of multistage game against dynamic load altering attack,'' \emph{Int. J. Electr. Power Energy Syst.}, vol. 131, Oct. 2021, {Art.} no. 107113.

\bibitem{bui2022dynamic}
V.-H. Bui, A.~Hussain, and W.~Su, ``A dynamic internal trading price strategy for networked microgrids: A deep reinforcement learning-based game-theoretic approach,'' \emph{IEEE Trans. Smart Grid}, vol.~13, no.~5, pp. 3408--3421, Sep. 2022.

\bibitem{surani2024competitive}
A.-P. Surani, T.~Wu, and A.~Scaglione, ``Competitive reinforcement learning for real-time pricing and scheduling control in coupled {EV} charging stations and power networks,'' in \emph{Proc. Int. Conf. Syst. Sci.}, 2024.

\bibitem{peng2022model}
B.~Peng, J.~Duan, J.~Chen, S.~E. Li, G.~Xie, C.~Zhang, Y.~Guan, Y.~Mu, and E.~Sun, ``Model-based chance-constrained reinforcement learning via separated proportional-integral {Lagrangian},'' \emph{IEEE Trans. Neural Netw. Learn. Syst.}, vol.~35, no.~1, pp. 466--478, Jan. 2024.

\bibitem{hassan2018optimal}
A.~Hassan, R.~Mieth, M.~Chertkov, D.~Deka, and Y.~Dvorkin, ``Optimal load ensemble control in chance-constrained optimal power flow,'' \emph{IEEE Trans. Smart Grid}, vol.~10, no.~5, pp. 5186--5195, Sep. 2019.

\bibitem{ciftci2019data}
O.~Ciftci, M.~Mehrtash, and A.~Kargarian, ``Data-driven nonparametric chance-constrained optimization for microgrid energy management,'' \emph{IEEE Trans. Ind. Inform.}, vol.~16, no.~4, pp. 2447--2457, Apr. 2020.

\bibitem{liang2024joint}
J.~Liang, W.~Jiang, C.~Lu, and C.~Wu, ``Joint chance-constrained unit commitment: Statistically feasible robust optimization with learning-to-optimize acceleration,'' \emph{IEEE Trans. Power Syst.}, vol.~39, no.~5, pp. 6508--6521, Sep. 2024.

\bibitem{yu2019convergent}
M.~Yu, Z.~Yang, M.~Kolar, and Z.~Wang, ``Convergent policy optimization for safe reinforcement learning,'' in \emph{Proc. Adv. Neural Inf. Process. Syst.}, vol.~32, Sep. 2019, pp. 3127--3139.

\bibitem{chow2018lyapunov}
Y.~Chow, O.~Nachum, E.~Duenez-Guzman, and M.~Ghavamzadeh, ``A {Lyapunov}-based approach to safe reinforcement learning,'' in \emph{Proc. Adv. Neural Inf. Process. Syst.}, vol.~31, 2018, pp. 8103--8112.

\bibitem{berkenkamp2017safe}
F.~Berkenkamp, M.~Turchetta, A.~Schoellig, and A.~Krause, ``Safe model-based reinforcement learning with stability guarantees,'' in \emph{Proc. Adv. Neural Inform. Process. Syst.}, vol.~30, 2017, pp. 908--919.

\bibitem{politowicz2024safety}
A.~Politowicz, S.~Mazumder, and B.~Liu, ``Safety through permissibility: Shield construction for fast and safe reinforcement learning,'' \emph{arXiv preprint arXiv:2405.19414}, 2024.

\bibitem{moos2022robust}
J.~Moos, K.~Hansel, H.~Abdulsamad, S.~Stark, D.~Clever, and J.~Peters, ``Robust reinforcement learning: A review of foundations and recent advances,'' \emph{Mach. Learn. Knowl. Extraction}, vol.~4, no.~1, pp. 276--315, Mar. 2022.

\bibitem{tessler2018reward}
C.~Tessler, D.~J. Mankowitz, and S.~Mannor, ``Reward constrained policy optimization,'' in \emph{Proc. Int. Conf. Learn. Representations}, New Orleans, LA, USA, May 2019, pp. 1--15.

\bibitem{paternain2019constrained}
S.~Paternain, L.~Chamon, M.~Calvo-Fullana, and A.~Ribeiro, ``Constrained reinforcement learning has zero duality gap,'' in \emph{Proc. Adv. Neural Inf. Process. Syst.}, vol.~32, 2019, p. 7553–7563.

\bibitem{gros2020safe}
S.~Gros, M.~Zanon, and A.~Bemporad, ``Safe reinforcement learning via projection on a safe set: How to achieve optimality?'' \emph{IFAC-PapersOnLine}, vol.~53, no.~2, pp. 8076--8081, 2020.

\bibitem{pinto2017robust}
L.~Pinto, J.~Davidson, R.~Sukthankar, and A.~Gupta, ``Robust adversarial reinforcement learning,'' in \emph{Proc. Int. Conf. Mach. Learn.}, 2017, pp. 2817--2826.

\bibitem{gu2024saferlbaselines}
\BIBentryALTinterwordspacing
S.~Gu, L.~Yang, Y.~Du, G.~Chen, F.~Walter, J.~Wang, and A.~Knoll, ``{Safe-Reinforcement-Learning-Baselines},'' Accessed: Mar. 24, 2025. [Online]. Available: \url{https://github.com/chauncygu/Safe-Reinforcement-Learning-Baselines}
\BIBentrySTDinterwordspacing

\bibitem{ray2019gym}
\BIBentryALTinterwordspacing
A.~Ray, J.~Achiam, and D.~Amodei, ``{Safety-Gym}: Tools for accelerating safe exploration research,'' Accessed: Mar. 24, 2025. [Online]. Available: \url{https://github.com/openai/safety-gym}
\BIBentrySTDinterwordspacing

\bibitem{ray2019algorithm}
\BIBentryALTinterwordspacing
------, ``{Safety Starter Agents: Basic constrained RL agents},'' Accessed: Mar. 24, 2025. [Online]. Available: \url{https://github.com/openai/safety-starter-agents}
\BIBentrySTDinterwordspacing

\bibitem{ji2023paper}
J.~Ji, B.~Zhang, J.~Zhou, X.~Pan, W.~Huang, R.~Sun, Y.~Geng, Y.~Zhong, J.~Dai, and Y.~Yang, ``{Safety-Gymnasium}: A unified safe reinforcement learning benchmark,'' in \emph{Proc. Adv. Neural Inform. Process. Syst.}, vol.~36, 2023.

\bibitem{ji2023gymnasium}
\BIBentryALTinterwordspacing
------, ``{Safety-Gymnasium: A Unified Safe Reinforcement Learning Benchmark},'' Accessed: Mar. 24, 2025. [Online]. Available: \url{https://github.com/PKU-Alignment/safety-gymnasium}
\BIBentrySTDinterwordspacing

\bibitem{ji2023algorithm}
\BIBentryALTinterwordspacing
------, ``{Safe Policy Optimization: A benchmark repository for safe reinforcement learning algorithms},'' Accessed: Mar. 24, 2025. [Online]. Available: \url{https://github.com/PKU-Alignment/Safe-Policy-Optimization}
\BIBentrySTDinterwordspacing

\bibitem{ji2023omnisafe}
J.~Ji, J.~Zhou, B.~Zhang, J.~Dai, X.~Pan, R.~Sun, W.~Huang, Y.~Geng, M.~Liu, and Y.~Yang, ``{OmniSafe}: An infrastructure for accelerating safe reinforcement learning research,'' \emph{J. Mach. Learn. Res.}, vol.~25, no. 285, pp. 1--6, 2024.

\bibitem{ji2023omnisafealgorithm}
\BIBentryALTinterwordspacing
------, ``{OmniSafe}: An infrastructural framework for accelerating safe {RL} research,'' Accessed: Mar. 24, 2025. [Online]. Available: \url{https://github.com/PKU-Alignment/omnisafe}
\BIBentrySTDinterwordspacing

\bibitem{zhang2022penalized}
L.~Zhang, L.~Shen, L.~Yang, S.~Chen, B.~Yuan, X.~Wang, and D.~Tao, ``Penalized proximal policy optimization for safe reinforcement learning,'' \emph{arXiv preprint arXiv:2205.11814}, 2022.

\bibitem{xu2021crpo}
T.~Xu, Y.~Liang, and G.~Lan, ``{CRPO}: A new approach for safe reinforcement learning with convergence guarantee,'' in \emph{Proc. Int. Conf. Mach. Learn.}, 2021, pp. 11\,480--11\,491.

\bibitem{sikchi2022learning}
H.~Sikchi, W.~Zhou, and D.~Held, ``Learning off-policy with online planning,'' in \emph{Proc. Conf. Robot Learn.}, 2022, pp. 1622--1633.

\bibitem{wen2018constrained}
M.~Wen and U.~Topcu, ``Constrained cross-entropy method for safe reinforcement learning,'' \emph{IEEE Trans. Autom. Control}, vol.~66, no.~7, pp. 3123--3137, Jul. 2021.

\bibitem{liu2020constrained}
Z.~Liu, H.~Zhou, B.~Chen, S.~Zhong, M.~Hebert, and D.~Zhao, ``Constrained model-based reinforcement learning with robust cross-entropy method,'' \emph{arXiv preprint arXiv:2010.07968}, 2020.

\bibitem{ma2022conservative}
Y.~J. Ma, A.~Shen, O.~Bastani, and J.~Dinesh, ``Conservative and adaptive penalty for model-based safe reinforcement learning,'' in \emph{Proc. AAAI Conf. Artif. Intell.}, vol.~36, no.~5, 2022, pp. 5404--5412.

\bibitem{fujimoto2019off}
S.~Fujimoto, D.~Meger, and D.~Precup, ``Off-policy deep reinforcement learning without exploration,'' in \emph{Proc. Int. Conf. Mach. Learn.}, 2019, pp. 2052--2062.

\bibitem{wang2020critic}
Z.~Wang, A.~Novikov, K.~Zolna, J.~S. Merel, J.~T. Springenberg, S.~E. Reed, B.~Shahriari, N.~Siegel, C.~Gulcehre, N.~Heess \emph{et~al.}, ``Critic regularized regression,'' in \emph{Proc. Adv. Neural Inf. Process. Syst.}, vol.~33, 2020, pp. 7768--7778.

\bibitem{lee2022coptidice}
J.~Lee, C.~Paduraru, D.~J. Mankowitz, N.~Heess, D.~Precup, K.-E. Kim, and A.~Guez, ``{COptiDICE}: Offline constrained reinforcement learning via stationary distribution correction estimation,'' in \emph{Proc. Int. Conf. Learn. Representations}, Jan. 2022.

\bibitem{sun2021safe}
H.~Sun, Z.~Xu, M.~Fang, Z.~Peng, J.~Guo, B.~Dai, and B.~Zhou, ``Safe exploration by solving early terminated {MDP},'' \emph{arXiv preprint arXiv:2107.04200}, 2021.

\bibitem{sootla2022saute}
A.~Sootla, A.~I. Cowen-Rivers, T.~Jafferjee, Z.~Wang, D.~H. Mguni, J.~Wang, and H.~Ammar, ``Saut{\'e} {RL}: Almost surely safe reinforcement learning using state augmentation,'' in \emph{Proc. Int. Conf. Mach. Learn.}, 2022, pp. 20\,423--20\,443.

\bibitem{sootla2022effects}
A.~Sootla, A.~I. Cowen-Rivers, J.~Wang, and H.~B. Ammar, ``Effects of safety state augmentation on safe exploration,'' in \emph{Proc. Int. Conf. Neural Inform. Process. Syst.}, Nov. 2022, pp. 34\,464--34\,477.

\bibitem{bam2005power}
L.~Bam and W.~Jewell, ``Review: Power system analysis software tools,'' in \emph{Proc. IEEE Power Energy Soc. General Meeting}, 2005, pp. 139--144.

\bibitem{vogt2018survey}
M.~Vogt, F.~Marten, and M.~Braun, ``A survey and statistical analysis of smart grid co-simulations,'' \emph{Appl. Energy}, vol. 222, pp. 67--78, Jul. 2018.

\bibitem{dugan2011open}
R.~C. Dugan and T.~E. McDermott, ``An open source platform for collaborating on smart grid research,'' in \emph{Proc. IEEE Power Energy Soc. General Meeting}, 2011, pp. 1--7.

\bibitem{chassin2014gridlab}
D.~P. Chassin, J.~C. Fuller, and N.~Djilali, ``{GridLAB-D}: An agent-based simulation framework for smart grids,'' \emph{J. Appl. Math.}, vol. 2014, no.~1, Jun. 2014, {Art.} no. 492320.

\bibitem{zimmerman2010matpower}
R.~D. Zimmerman, C.~E. Murillo-S{\'a}nchez, and R.~J. Thomas, ``{MATPOWER}: Steady-state operations, planning, and analysis tools for power systems research and education,'' \emph{IEEE Trans. Power Syst.}, vol.~26, no.~1, pp. 12--19, Feb. 2011.

\bibitem{thurner2018pandapower}
L.~Thurner, A.~Scheidler, F.~Sch{\"a}fer, J.-H. Menke, J.~Dollichon, F.~Meier, S.~Meinecke, and M.~Braun, ``Pandapower—an open-source {Python} tool for convenient modeling, analysis, and optimization of electric power systems,'' \emph{IEEE Trans. Power Syst.}, vol.~33, no.~6, pp. 6510--6521, Nov. 2018.

\bibitem{PyPSA}
T.~Brown, J.~H\"orsch, and D.~Schlachtberger, ``{PyPSA}: {Python} for power system analysis,'' \emph{J. Open Res. Softw.}, vol.~6, no.~4, Jan. 2018.

\bibitem{coffrin2018powermodels}
C.~Coffrin, R.~Bent, K.~Sundar, Y.~Ng, and M.~Lubin, ``{PowerModels.jl}: An open-source framework for exploring power flow formulations,'' in \emph{Proc. Power Syst. Comput. Conf.}, Jun. 2018, pp. 1--8.

\bibitem{chow1992toolbox}
J.~H. Chow and K.~W. Cheung, ``A toolbox for power system dynamics and control engineering education and research,'' \emph{IEEE Trans. Power Syst.}, vol.~7, no.~4, pp. 1559--1564, Nov. 1992.

\bibitem{milano2005open}
F.~Milano, ``An open source power system analysis toolbox,'' \emph{IEEE Trans. Power Syst.}, vol.~20, no.~3, pp. 1199--1206, Aug. 2005.

\bibitem{lara2021powersystems}
J.~D. Lara, C.~Barrows, D.~Thom, D.~Krishnamurthy, and D.~Callaway, ``{PowerSystems.jl}—a power system data management package for large scale modeling,'' \emph{SoftwareX}, vol.~15, Jul. 2021, {Art.} no. 100747.

\bibitem{fobes2020powermodelsdistribution}
D.~M. Fobes, S.~Claeys, F.~Geth, and C.~Coffrin, ``{PowerModelsDistribution.jl}: An open-source framework for exploring distribution power flow formulations,'' \emph{Electric Power Syst. Res.}, vol. 189, Dec. 2020, {Art.} no. 106664.

\bibitem{cui2020hybrid}
H.~Cui, F.~Li, and K.~Tomsovic, ``Hybrid symbolic-numeric framework for power system modeling and analysis,'' \emph{IEEE Trans. Power Syst.}, vol.~36, no.~2, pp. 1373--1384, Mar. 2021.

\bibitem{lara2023powersimulationsdynamics}
J.~D. Lara, R.~Henriquez-Auba, M.~Bossart, D.~S. Callaway, and C.~Barrows, ``{PowerSimulationsDynamics.jl}--an open source modeling package for modern power systems with inverter-based resources,'' \emph{arXiv preprint arXiv:2308.02921}, 2023.

\bibitem{guironnet2018towards}
A.~Guironnet, M.~Saugier, S.~Petitrenaud, F.~Xavier, and P.~Panciatici, ``Towards an open-source solution using modelica for time-domain simulation of power systems,'' in \emph{Proc. IEEE PES Innovative Smart Grid Technol. Conf. Europe}, 2018, pp. 1--6.

\bibitem{su2024survey}
T.~Su, J.~Peng, A.~Selim, J.~Zhao, and J.~Tan, ``A survey of open-source power system dynamic simulators with grid-forming inverter for machine learning applications,'' \emph{arXiv preprint arXiv:2412.08065}, 2024.

\bibitem{heid2020omg}
\BIBentryALTinterwordspacing
S.~Heid, D.~Weber, H.~Bode, E.~Hüllermeier, and O.~Wallscheid, ``{OMG}: A scalable and flexible simulation and testing environment toolbox for intelligent microgrid control,'' \emph{J. Open Source Softw.}, vol.~5, no.~54, p. 2435, 2020. [Online]. Available: \url{https://doi.org/10.21105/joss.02435}
\BIBentrySTDinterwordspacing

\bibitem{huang2019adaptive}
Q.~Huang, R.~Huang, W.~Hao, J.~Tan, R.~Fan, and Z.~Huang, ``Adaptive power system emergency control using deep reinforcement learning,'' \emph{IEEE Trans. Smart Grid}, vol.~11, no.~2, pp. 1171--1182, Mar. 2020.

\bibitem{huang2019RLGC}
\BIBentryALTinterwordspacing
------, ``{RLGC},'' Accessed: Mar. 24, 2025. [Online]. Available: \url{https://github.com/RLGC-Project/RLGC}
\BIBentrySTDinterwordspacing

\bibitem{fan2022powergym}
T.-H. Fan, X.~Y. Lee, and Y.~Wang, ``{PowerGym}: A reinforcement learning environment for {Volt-Var} control in power distribution systems,'' in \emph{Proc. Annual Learn. Dyn. Control Conf.}, 2022, pp. 21--33.

\bibitem{wolgast2024learning}
T.~Wolgast and A.~Nie{\ss}e, ``Learning the optimal power flow: Environment design matters,'' \emph{Energy AI}, vol.~17, Sep. 2024, {Art.} no. 100410.

\bibitem{wolgast2024opfgym}
\BIBentryALTinterwordspacing
------, ``{OPF-Gym},'' Accessed: Mar. 24, 2025. [Online]. Available: \url{https://github.com/Digitalized-Energy-Systems/opfgym}
\BIBentrySTDinterwordspacing

\bibitem{eichelbeck2024commonpower}
M.~Eichelbeck, H.~Markgraf, and M.~Althoff, ``{CommonPower}: Supercharging machine learning for smart grids,'' \emph{arXiv preprint arXiv:2406.03231}, 2024.

\bibitem{eichelbeck2024github}
\BIBentryALTinterwordspacing
------, ``{CommonPower}: A framework for safe data-driven smart grid control,'' Accessed: Mar. 24, 2025. [Online]. Available: \url{https://github.com/TUMcps/commonpower}
\BIBentrySTDinterwordspacing

\bibitem{balu1992line}
N.~Balu, T.~Bertram, A.~Bose, V.~Brandwajn, G.~Cauley, D.~Curtice, A.~Fouad, L.~Fink, M.~G. Lauby, B.~F. Wollenberg \emph{et~al.}, ``On-line power system security analysis,'' \emph{Proc. IEEE}, vol.~80, no.~2, pp. 262--282, Feb. 1992.

\bibitem{petinrin2016impact}
J.~Petinrin and M.~Shaabanb, ``Impact of renewable generation on voltage control in distribution systems,'' \emph{Renew. Sustain. Energy Rev.}, vol.~65, pp. 770--783, Nov. 2016.

\bibitem{bastos2020machine}
A.~F. Bastos, S.~Santoso, V.~Krishnan, and Y.~Zhang, ``Machine learning-based prediction of distribution network voltage and sensor allocation,'' in \emph{Proc. IEEE Power Energy Soc. Gen. Meeting}, 2020, pp. 1--5.

\bibitem{sun2019review}
H.~Sun, Q.~Guo, J.~Qi, V.~Ajjarapu, R.~Bravo, J.~Chow, Z.~Li, R.~Moghe, E.~Nasr-Azadani, U.~Tamrakar \emph{et~al.}, ``Review of challenges and research opportunities for voltage control in smart grids,'' \emph{IEEE Trans. Power Syst.}, vol.~34, no.~4, pp. 2790--2801, Jul. 2019.

\bibitem{murray2021voltage}
W.~Murray, M.~Adonis, and A.~Raji, ``Voltage control in future electrical distribution networks,'' \emph{Renew. Sustain. Energy Rev.}, vol. 146, Aug. 2021, {Art.} no. 111100.

\bibitem{ruan2020distributed}
H.~Ruan, H.~Gao, Y.~Liu, L.~Wang, and J.~Liu, ``Distributed voltage control in active distribution network considering renewable energy: A novel network partitioning method,'' \emph{IEEE Trans. Power Syst.}, vol.~35, no.~6, pp. 4220--4231, Nov. 2020.

\bibitem{wu2022reinforcement}
T.~Wu, A.~Scaglione, and D.~Arnold, ``Reinforcement learning using physics inspired graph convolutional neural networks,'' in \emph{Proc. Annu. Allerton Conf. Commun., Control, Comput.}, Sep. 2022, pp. 1--8.

\bibitem{roberts2021deep}
C.~Roberts, S.-T. Ngo, A.~Milesi, A.~Scaglione, S.~Peisert, and D.~Arnold, ``Deep reinforcement learning for mitigating cyber-physical {DER} voltage unbalance attacks,'' in \emph{Proc. Amer. Control Conf.}, 2021, pp. 2861--2867.

\bibitem{carreno2024voltage}
I.~L. Carre{\~n}o, A.~Scaglione, D.~Arnold, and T.~Wu, ``Voltage security region of a three-phase unbalanced distribution power system with dynamics,'' \emph{IEEE Trans. Power Syst.}, vol.~39, no.~5, pp. 6441--6455, Sep. 2024.

\bibitem{photovoltaics2018ieee}
\emph{IEEE Standard for Interconnection and Interoperability of Distributed Energy Resources With Associated Electric Power Systems Interfaces}, IEEE Std. 1547-2018, Apr. 2018.

\bibitem{liu2021online}
H.~Liu and W.~Wu, ``Online multi-agent reinforcement learning for decentralized inverter-based {Volt-VAR} control,'' \emph{IEEE Trans. Smart Grid}, vol.~12, no.~4, pp. 2980--2990, Jul. 2021.

\bibitem{chen2022saver}
Y.~Chen, Y.~Shi, D.~Arnold, and S.~Peisert, ``{SAVER}: Safe learning-based controller for real-time voltage regulation,'' in \emph{Proc. IEEE Power Energy Soc. Gen. Meeting}, 2022, pp. 1--5.

\bibitem{zhang2023dnn}
M.~Zhang, G.~Guo, T.~Zhao, and Q.~Xu, ``{DNN} assisted projection based deep reinforcement learning for safe control of distribution grids,'' \emph{IEEE Trans. Power Syst.}, vol.~39, no.~4, pp. 5687--5698, Jul. 2024.

\bibitem{kou2020safe}
P.~Kou, D.~Liang, C.~Wang, Z.~Wu, and L.~Gao, ``Safe deep reinforcement learning-based constrained optimal control scheme for active distribution networks,'' \emph{Appl. Energy}, vol. 264, Apr. 2020, {Art.} no. 114772.

\bibitem{guo2023safe}
G.~Guo, M.~Zhang, Y.~Gong, and Q.~Xu, ``Safe multi-agent deep reinforcement learning for real-time decentralized control of inverter based renewable energy resources considering communication delay,'' \emph{Appl. Energy}, vol. 349, Nov. 2023, {Art.} no. 121648.

\bibitem{nguyen2022three}
H.~T. Nguyen and D.-H. Choi, ``Three-stage inverter-based peak shaving and {Volt-VAR} control in active distribution networks using online safe deep reinforcement learning,'' \emph{IEEE Trans. Smart Grid}, vol.~13, no.~4, pp. 3266--3277, Jul. 2022.

\bibitem{deng2024safety}
Y.~Deng, M.~Zhou, M.~Chen, and Z.~Yang, ``Safety deep reinforcement learning approach to voltage control in flexible network topologies,'' in \emph{Proc. Conf. Fully Actuat. Syst. Theory Appl.}, 2024, pp. 395--400.

\bibitem{zhao2023explicit}
X.~Zhao and Q.~Xu, ``Explicit reinforcement learning safety layer for computationally efficient inverter-based voltage regulation,'' in \emph{Proc. Amer. Control Conf.}, 2023, pp. 4501--4506.

\bibitem{feng2023bridging}
J.~Feng, W.~Cui, J.~Cort{\'e}s, and Y.~Shi, ``Bridging transient and steady-state performance in voltage control: A reinforcement learning approach with safe gradient flow,'' \emph{IEEE Control Syst. Lett.}, vol.~7, pp. 2845--2850, 2023.

\bibitem{machowski2020power}
J.~Machowski, Z.~Lubosny, J.~W. Bialek, and J.~R. Bumby, \emph{Power system dynamics: Stability and control}.\hskip 1em plus 0.5em minus 0.4em\relax Hoboken, NJ, USA: John Wiley Sons, 2020.

\bibitem{wehenkel2004preventive}
L.~Wehenkel and M.~Pavella, ``Preventive vs. emergency control of power systems,'' in \emph{Proc. IEEE PES Power Syst. Conf. Expo.}, 2004, pp. 1665--1670.

\bibitem{hatziargyriou2020definition}
N.~Hatziargyriou, J.~Milanovic, C.~Rahmann, V.~Ajjarapu, C.~Canizares, I.~Erlich, D.~Hill, I.~Hiskens, I.~Kamwa, B.~Pal \emph{et~al.}, ``Definition and classification of power system stability--revisited \& extended,'' \emph{IEEE Trans. Power Syst.}, vol.~36, no.~4, pp. 3271--3281, Jul. 2021.

\bibitem{zhang2023deep}
H.~Zhang, X.~Sun, M.~H. Lee, and J.~Moon, ``Deep reinforcement learning based active network management and emergency load-shedding control for power systems,'' \emph{IEEE Trans. Smart Grid}, vol.~15, no.~2, pp. 1423--1437, Mar. 2024.

\bibitem{xia2022safe}
Y.~Xia, Y.~Xu, Y.~Wang, S.~Mondal, S.~Dasgupta, A.~K. Gupta, and G.~M. Gupta, ``A safe policy learning-based method for decentralized and economic frequency control in isolated networked-microgrid systems,'' \emph{IEEE Trans. Sustain. Energy}, vol.~13, no.~4, pp. 1982--1993, Oct. 2022.

\bibitem{wan2023adapsafe}
X.~Wan, M.~Sun, B.~Chen, Z.~Chu, and F.~Teng, ``{AdapSafe}: Adaptive and safe-certified deep reinforcement learning-based frequency control for carbon-neutral power systems,'' in \emph{Proc. AAAI Conf. Artif. Intell.}, 2023.

\bibitem{tabas2022computationally}
D.~Tabas and B.~Zhang, ``Computationally efficient safe reinforcement learning for power systems,'' in \emph{Proc. Amer. Control Conf.}, 2022, pp. 3303--3310.

\bibitem{zhou2022coordinated}
Y.~Zhou, L.~Zhou, D.~Shi, and X.~Zhao, ``Coordinated frequency control through safe reinforcement learning,'' in \emph{Proc. IEEE Power Energy Soc. Gen. Meeting}, 2022, pp. 1--5.

\bibitem{gupta2021coordinated}
P.~Gupta, A.~Pal, and V.~Vittal, ``Coordinated wide-area damping control using deep neural networks and reinforcement learning,'' \emph{IEEE Trans. Power Syst.}, vol.~37, no.~1, pp. 365--376, Jan. 2022.

\bibitem{kwon2023risk}
K.-b. Kwon, S.~Mukherjee, T.~L. Vu, and H.~Zhu, ``Risk-constrained reinforcement learning for inverter-dominated power system controls,'' \emph{IEEE Control Syst. Lett.}, vol.~7, pp. 3854--3859, 2023.

\bibitem{tarle2023safe}
M.~Tarle, M.~Larsson, G.~Ingestr{\"o}m, L.~Nordstr{\"o}m, and M.~Bj{\"o}rkman, ``Safe reinforcement learning for mitigation of model errors in {FACTS} setpoint control,'' in \emph{Proc. Int. Conf. Smart Energy Syst. Technol.}, 2023, pp. 1--6.

\bibitem{jin2020stability}
M.~Jin and J.~Lavaei, ``Stability-certified reinforcement learning: A control-theoretic perspective,'' \emph{IEEE Access}, vol.~8, pp. 229\,086--229\,100, 2020.

\bibitem{gu2022recurrent}
F.~Gu, H.~Yin, L.~El~Ghaoui, M.~Arcak, P.~Seiler, and M.~Jin, ``Recurrent neural network controllers synthesis with stability guarantees for partially observed systems,'' in \emph{Proc. AAAI Conf. Artif. Intell.}, vol.~36, no.~5, 2022, pp. 5385--5394.

\bibitem{zhao2023barrier}
T.~Zhao, J.~Wang, and M.~Yue, ``A barrier-certificated reinforcement learning approach for enhancing power system transient stability,'' \emph{IEEE Trans. Power Syst.}, vol.~38, no.~6, pp. 5356--5366, Nov. 2023.

\bibitem{obaid2019frequency}
Z.~A. Obaid, L.~M. Cipcigan, L.~Abrahim, and M.~T. Muhssin, ``Frequency control of future power systems: Reviewing and evaluating challenges and new control methods,'' \emph{J. Modern Power Syst. Clean Energy}, vol.~7, no.~1, pp. 9--25, Jan. 2019.

\bibitem{bevrani2021power}
H.~Bevrani, H.~Golp{\^\i}ra, A.~R. Messina, N.~Hatziargyriou, F.~Milano, and T.~Ise, ``Power system frequency control: An updated review of current solutions and new challenges,'' \emph{Electr. Power Syst. Res.}, vol. 194, May 2021, {Art.} no. 107114.

\bibitem{li2021online}
H.~Li, Z.~Wang, L.~Li, and H.~He, ``Online microgrid energy management based on safe deep reinforcement learning,'' in \emph{Proc. IEEE Symp. Ser. Comput. Intell.}, 2021, pp. 1--8.

\bibitem{kocuk2016strong}
B.~Kocuk, S.~S. Dey, and X.~A. Sun, ``Strong {SOCP} relaxations for the optimal power flow problem,'' \emph{Oper. Res.}, vol.~64, no.~6, pp. 1177--1196, May 2016.

\bibitem{marano2012exploiting}
A.~Marano-Marcolini, F.~Capitanescu, J.~L. Martinez-Ramos, and L.~Wehenkel, ``Exploiting the use of {DC SCOPF} approximation to improve iterative {AC SCOPF} algorithms,'' \emph{IEEE Trans. Power Syst.}, vol.~27, no.~3, pp. 1459--1466, Aug. 2012.

\bibitem{yan2020convex}
M.~Yan, M.~Shahidehpour, A.~Paaso, L.~Zhang, A.~Alabdulwahab, and A.~Abusorrah, ``A convex three-stage {SCOPF} approach to power system flexibility with unified power flow controllers,'' \emph{IEEE Trans. Power Syst.}, vol.~36, no.~3, pp. 1947--1960, May 2021.

\bibitem{su2023analytic}
T.~Su, J.~Zhao, X.~Chen, and X.~Liu, ``Analytic input convex neural networks-based model predictive control for power system transient stability enhancement,'' in \emph{Proc. IEEE Power Energy Soc. Gen. Meeting}, 2023, pp. 1--5.

\bibitem{hong2023robust}
S.-H. Hong and H.-S. Lee, ``Robust energy management system with safe reinforcement learning using short-horizon forecasts,'' \emph{IEEE Trans. Smart Grid}, vol.~14, no.~3, pp. 2485--2488, May 2023.

\bibitem{liu2018distributed}
W.~Liu, P.~Zhuang, H.~Liang, J.~Peng, and Z.~Huang, ``Distributed economic dispatch in microgrids based on cooperative reinforcement learning,'' \emph{IEEE Trans. Neural Netw. Learn. Syst.}, vol.~29, no.~6, pp. 2192--2203, Jun. 2018.

\bibitem{hao2024lyapunov}
G.~Hao, Y.~Li, Y.~Li, L.~Jiang, and Z.~Zeng, ``Lyapunov-based safe reinforcement learning for microgrid energy management,'' \emph{IEEE Trans. Neural Netw. Learn. Syst.}, vol.~36, no.~6, pp. 9985--9999, Jun. 2025.

\bibitem{wu2024real}
P.~Wu, C.~Chen, D.~Lai, J.~Zhong, and Z.~Bie, ``Real-time optimal power flow method via safe deep reinforcement learning based on primal-dual and prior knowledge guidance,'' \emph{IEEE Trans. Power Syst.}, vol.~40, no.~1, pp. 597--611, Jan. 2025.

\bibitem{sayed2022feasibility}
A.~R. Sayed, C.~Wang, H.~Anis, and T.~Bi, ``Feasibility constrained online calculation for real-time optimal power flow: A convex constrained deep reinforcement learning approach,'' \emph{IEEE Trans. Power Syst.}, vol.~38, no.~6, pp. 5215--5227, Nov. 2023.

\bibitem{chen2023improved}
Y.~Chen, Q.~Du, H.~Liu, L.~Cheng, and M.~S. Younis, ``Improved proximal policy optimization algorithm for sequential security-constrained optimal power flow based on expert knowledge and safety layer,'' \emph{J. Modern Power Syst. Clean Energy}, vol.~12, no.~3, pp. 742--753, May 2024.

\bibitem{zhang2024networked}
J.~Zhang, L.~Sang, Y.~Xu, and H.~Sun, ``Networked multiagent-based safe reinforcement learning for low-carbon demand management in distribution networks,'' \emph{IEEE Trans. Sustain. Energy}, vol.~15, no.~3, pp. 1528--1545, Jul. 2024.

\bibitem{shengren2023optimal}
H.~Shengren, P.~P. Vergara, E.~M.~S. Duque, and P.~Palensky, ``Optimal energy system scheduling using a constraint-aware reinforcement learning algorithm,'' \emph{Int. J. Electr. Power Energy Syst.}, vol. 152, Oct. 2023, {Art.} no. 109230.

\bibitem{yan2023real}
Z.~Yan and Y.~Xu, ``Real-time optimal power flow with linguistic stipulations: Integrating {GPT}-agent and deep reinforcement learning,'' \emph{IEEE Trans. Power Syst.}, vol.~39, no.~2, pp. 4747--4750, Mar. 2024.

\bibitem{yan2020real}
------, ``Real-time optimal power flow: A {Lagrangian} based deep reinforcement learning approach,'' \emph{IEEE Trans. Power Syst.}, vol.~35, no.~4, pp. 3270--3273, Jul. 2020.

\bibitem{ceusters2023safe}
G.~Ceusters, L.~R. Camargo, R.~Franke, A.~Now{\'e}, and M.~Messagie, ``Safe reinforcement learning for multi-energy management systems with known constraint functions,'' \emph{Energy AI}, vol.~12, Apr. 2023, {Art.} no. 100227.

\bibitem{wang2023secure}
Y.~Wang, D.~Qiu, M.~Sun, G.~Strbac, and Z.~Gao, ``Secure energy management of multi-energy microgrid: A physical-informed safe reinforcement learning approach,'' \emph{Appl. Energy}, vol. 335, Apr. 2023, {Art.} no. 120759.

\bibitem{wu2023pesgm}
T.~Wu, A.~Scaglione, and D.~Arnold, ``Constrained reinforcement learning for stochastic dynamic optimal power flow control,'' in \emph{Proc. IEEE Power Energy Soc. Gen. Meeting}, 2023, pp. 1--5.

\bibitem{liang2024power}
K.~Liang, H.~Wang, D.~Pozo, and V.~Terzija, ``Power system restoration with large renewable penetration: State-of-the-art and future trends,'' \emph{Int. J. Electr. Power Energy Syst.}, vol. 155, Jan. 2024, {Art.} no. 109494.

\bibitem{zhang2023primal}
X.~Zhang, B.~Knueven, A.~Zamzam, M.~Reynolds, and W.~Jones, ``Primal-dual differentiable programming for distribution system critical load restoration,'' in \emph{Proc. IEEE Power Energy Soc. Gen. Meeting}, 2023, pp. 1--5.

\bibitem{bie2017battling}
Z.~Bie, Y.~Lin, G.~Li, and F.~Li, ``Battling the extreme: A study on the power system resilience,'' \emph{Proc. IEEE}, vol. 105, no.~7, pp. 1253--1266, Jul. 2017.

\bibitem{xu2021resilience}
L.~Xu, Q.~Guo, Y.~Sheng, S.~Muyeen, and H.~Sun, ``On the resilience of modern power systems: A comprehensive review from the cyber-physical perspective,'' \emph{Renew. Sustain. Energy Rev.}, vol. 152, Dec. 2021, {Art.} no. 111642.

\bibitem{vu2023safe}
L.~Vu, T.~Vu, T.-L. Vu, and A.~Srivastava, ``Safe exploration reinforcement learning for load restoration using invalid action masking,'' in \emph{Proc. IEEE Power Energy Soc. General Meeting}, 2023, pp. 1--5.

\bibitem{li2022risk}
X.~Li, X.~Han, and M.~Yang, ``Risk-based reserve scheduling for active distribution networks based on an improved proximal policy optimization algorithm,'' \emph{IEEE Access}, vol.~11, pp. 15\,211--15\,228, 2022.

\bibitem{shi2023augmented}
X.~Shi, Y.~Xu, G.~Chen, and Y.~Guo, ``An augmented {Lagrangian}-based safe reinforcement learning algorithm for carbon-oriented optimal scheduling of {EV} aggregators,'' \emph{IEEE Trans. Smart Grid}, vol.~15, no.~1, pp. 795--809, Jan. 2024.

\bibitem{zhu2023budget}
T.~Zhu, X.~Zhang, J.~Duan, Z.~Zhou, and X.~Chen, ``A budget-aware incentive mechanism for vehicle-to-grid via reinforcement learning,'' in \emph{Proc. IEEE Int. Symp. Qual. Service}, 2023, pp. 1--10.

\bibitem{yang2023dynamic}
H.~Yang, Y.~Xu, and Q.~Guo, ``Dynamic incentive pricing on charging stations for real-time congestion management in distribution network: An adaptive model-based safe deep reinforcement learning method,'' \emph{IEEE Trans. Sustain. Energy}, vol.~15, no.~2, pp. 1100--1113, Apr. 2024.

\bibitem{lu2024sma}
R.~Lu, N.~Wu, T.~Yang, Y.~Chen, M.~Sun, D.~Wang, and X.~Peng, ``{SMA-PDPPO}: Safe multiagent primal-dual deep reinforcement learning for industrial parks energy trading,'' \emph{IEEE Trans. Ind. Inform.}, vol.~21, no.~3, pp. 2640--2649, Mar. 2025.

\bibitem{zheng2014stochastic}
Q.~P. Zheng, J.~Wang, and A.~L. Liu, ``Stochastic optimization for unit commitment—a review,'' \emph{IEEE Trans. Power Syst.}, vol.~30, no.~4, pp. 1913--1924, Jul. 2015.

\bibitem{abdi2021profit}
H.~Abdi, ``Profit-based unit commitment problem: A review of models, methods, challenges, and future directions,'' \emph{Renew. Sustain. Energy Rev.}, vol. 138, Mar. 2021, {Art.} no. 110504.

\bibitem{yang2021comprehensive}
N.~Yang, Z.~Dong, L.~Wu, L.~Zhang, X.~Shen, D.~Chen, B.~Zhu, and Y.~Liu, ``A comprehensive review of security-constrained unit commitment,'' \emph{J. Modern Power Syst. Clean Energy}, vol.~10, no.~3, pp. 562--576, May 2022.

\bibitem{putz2021comparison}
D.~Putz, D.~Schwabeneder, H.~Auer, and B.~Fina, ``A comparison between mixed-integer linear programming and dynamic programming with state prediction as novelty for solving unit commitment,'' \emph{Int. J. Electr. Power Energy Syst.}, vol. 125, Feb. 2021, {Art.} no. 106426.

\bibitem{quan2015computational}
H.~Quan, D.~Srinivasan, A.~M. Khambadkone, and A.~Khosravi, ``A computational framework for uncertainty integration in stochastic unit commitment with intermittent renewable energy sources,'' \emph{Appl. Energy}, vol. 152, pp. 71--82, Aug. 2015.

\bibitem{quan2016integration}
H.~Quan, D.~Srinivasan, and A.~Khosravi, ``Integration of renewable generation uncertainties into stochastic unit commitment considering reserve and risk: A comparative study,'' \emph{Energy}, vol. 103, pp. 735--745, May 2016.

\bibitem{wang2022distributionally}
S.~Wang, C.~Zhao, L.~Fan, and R.~Bo, ``Distributionally robust unit commitment with flexible generation resources considering renewable energy uncertainty,'' \emph{IEEE Trans. Power Syst.}, vol.~37, no.~6, pp. 4179--4190, Nov. 2022.

\bibitem{karthikeyan2013review}
S.~P. Karthikeyan, I.~J. Raglend, and D.~P. Kothari, ``A review on market power in deregulated electricity market,'' \emph{Int. J. Electr. Power Energy Syst.}, vol.~48, pp. 139--147, Jun. 2013.

\bibitem{bjarghov2021developments}
S.~Bjarghov, M.~L{\"o}schenbrand, A.~I. Saif, R.~A. Pedrero, C.~Pfeiffer, S.~K. Khadem, M.~Rabelhofer, F.~Revheim, and H.~Farahmand, ``Developments and challenges in local electricity markets: A comprehensive review,'' \emph{IEEE Access}, vol.~9, pp. 58\,910--58\,943, 2021.

\bibitem{weron2014electricity}
R.~Weron, ``Electricity price forecasting: A review of the state-of-the-art with a look into the future,'' \emph{Int. J. Forecasting}, vol.~30, no.~4, pp. 1030--1081, Oct. 2014.

\bibitem{alemany2018effects}
J.~Alemany, L.~Kasprzyk, and F.~Magnago, ``Effects of binary variables in mixed integer linear programming based unit commitment in large-scale electricity markets,'' \emph{Electr. Power Syst. Res.}, vol. 160, pp. 429--438, Apr. 2018.

\bibitem{canizes2013mixed}
B.~Canizes, J.~Soares, P.~Faria, and Z.~Vale, ``Mixed integer non-linear programming and artificial neural network based approach to ancillary services dispatch in competitive electricity markets,'' \emph{Appl. Energy}, vol. 108, pp. 261--270, Aug. 2013.

\bibitem{hong2023bilevel}
Q.~Hong, F.~Meng, J.~Liu, and R.~Bo, ``A bilevel game-theoretic decision-making framework for strategic retailers in both local and wholesale electricity markets,'' \emph{Appl. Energy}, vol. 330, Jan. 2023, {Art.} no. 120311.

\bibitem{hakimi2021stochastic}
S.~M. Hakimi, A.~Hasankhani, M.~Shafie-khah, and J.~P. Catal{\~a}o, ``Stochastic planning of a multi-microgrid considering integration of renewable energy resources and real-time electricity market,'' \emph{Appl. Energy}, vol. 298, Sep. 2021, {Art.} no. 117215.

\bibitem{zhang2021predicting}
Z.~Zhang and M.~Wu, ``Predicting real-time locational marginal prices: A {GAN}-based approach,'' \emph{IEEE Trans. Power Syst.}, vol.~37, no.~2, pp. 1286--1296, Mar. 2022.

\bibitem{tsaousoglou2022market}
G.~Tsaousoglou, J.~S. Giraldo, and N.~G. Paterakis, ``Market mechanisms for local electricity markets: A review of models, solution concepts and algorithmic techniques,'' \emph{Renew. Sustain. Energy Rev.}, vol. 156, Mar. 2022, {Art.} no. 111890.

\bibitem{wu2022strategic}
J.~Wu, J.~Wang, and X.~Kong, ``Strategic bidding in a competitive electricity market: An intelligent method using multi-agent transfer learning based on reinforcement learning,'' \emph{Energy}, vol. 256, Oct. 2022, {Art.} no. 124657.

\bibitem{ren2023reinforcement}
K.~Ren, J.~Liu, X.~Liu, and Y.~Nie, ``Reinforcement learning-based bi-level strategic bidding model of gas-fired unit in integrated electricity and natural gas markets preventing market manipulation,'' \emph{Appl. Energy}, vol. 336, Apr. 2023, {Art.} no. 120813.

\bibitem{qiu2022strategic}
D.~Qiu, Z.~Dong, G.~Ruan, H.~Zhong, G.~Strbac, and C.~Kang, ``Strategic retail pricing and demand bidding of retailers in electricity market: A data-driven chance-constrained programming,'' \emph{Adv. Appl. Energy}, vol.~7, Sep. 2022, {Art.} no. 100100.

\bibitem{iea2023global}
\BIBentryALTinterwordspacing
{International Energy Agency}, ``{Global EV Outlook 2023},'' Accessed: Mar. 24, 2025. [Online]. Available: \url{https://www.iea.org/reports/global-ev-outlook-2023}
\BIBentrySTDinterwordspacing

\bibitem{chen2022deep}
G.~Chen, L.~Yang, and X.~Cao, ``A deep reinforcement learning-based charging scheduling approach with augmented {Lagrangian} for electric vehicle,'' \emph{Appl. Energy}, vol. 378, Jan. 2025, {Art.} no. 124706.

\bibitem{zhang2023safe}
S.~Zhang, R.~Jia, H.~Pan, and Y.~Cao, ``A safe reinforcement learning-based charging strategy for electric vehicles in residential microgrid,'' \emph{Appl. Energy}, vol. 348, Oct. 2023, {Art.} no. 121490.

\bibitem{li2019constrained}
H.~Li, Z.~Wan, and H.~He, ``Constrained {EV} charging scheduling based on safe deep reinforcement learning,'' \emph{IEEE Trans. Smart Grid}, vol.~11, no.~3, pp. 2427--2439, May 2020.

\bibitem{zhang2020deep}
H.~Zhang, J.~Peng, H.~Tan, H.~Dong, and F.~Ding, ``A deep reinforcement learning-based energy management framework with {Lagrangian} relaxation for plug-in hybrid electric vehicle,'' \emph{IEEE Trans. Transport. Electrific.}, vol.~7, no.~3, pp. 1146--1160, Sep. 2020.

\bibitem{liessner2019safe}
R.~Liessner, A.~M. Dietermann, and B.~B{\"a}ker, ``Safe deep reinforcement learning hybrid electric vehicle energy management,'' in \emph{Proc. Int. Conf. Agents Artif. Intell.}, 2019, pp. 161--181.

\bibitem{guan2024rule}
Y.~Guan, J.~Zhang, W.~Ma, and L.~Che, ``Rule-based shields embedded safe reinforcement learning approach for electric vehicle charging control,'' \emph{Int. J. Electr. Power Energy Syst.}, vol. 157, Jun. 2024, {Art.} no. 109863.

\bibitem{abdullah2021reinforcement}
H.~M. Abdullah, A.~Gastli, and L.~Ben-Brahim, ``Reinforcement learning based {EV} charging management systems--a review,'' \emph{IEEE Access}, vol.~9, pp. 41\,506--41\,531, 2021.

\bibitem{nimalsiri2019survey}
N.~I. Nimalsiri, C.~P. Mediwaththe, E.~L. Ratnam, M.~Shaw, D.~B. Smith, and S.~K. Halgamuge, ``A survey of algorithms for distributed charging control of electric vehicles in smart grid,'' \emph{IEEE Trans. Intell. Transp. Syst.}, vol.~21, no.~11, pp. 4497--4515, Nov. 2020.

\bibitem{global2023un}
I.~Hamilton, H.~Kennard, J.~Amorocho, S.~Steuwer, J.~Kockat, Z.~Toth, C.~Delmastro, R.~M. Gordon, and K.~Petrichenko, ``Global status report for buildings and construction,'' UN Environment Programme, Tech. Rep., 2024.

\bibitem{garmroodi2023optimal}
A.~D. Garmroodi, F.~Nasiri, and F.~Haghighat, ``Optimal dispatch of an energy hub with compressed air energy storage: A safe reinforcement learning approach,'' \emph{J. Energy Storage}, vol.~57, Jan. 2023, {Art.} no. 106147.

\bibitem{ding2022safe}
H.~Ding, Y.~Xu, B.~C.~S. Hao, Q.~Li, and A.~Lentzakis, ``A safe reinforcement learning approach for multi-energy management of smart home,'' \emph{Electric Power Syst. Res.}, vol. 210, Sep. 2022, {Art.} no. 108120.

\bibitem{le2021deep}
D.~V. Le, R.~Wang, Y.~Liu, R.~Tan, Y.-W. Wong, and Y.~Wen, ``Deep reinforcement learning for tropical air free-cooled data center control,'' \emph{ACM Trans. Sensor Netw.}, vol.~17, no.~3, pp. 1--28, 2021.

\bibitem{yu2023district}
P.~Yu, H.~Zhang, Y.~Song, H.~Hui, and G.~Chen, ``District cooling system control for providing operating reserve based on safe deep reinforcement learning,'' \emph{IEEE Trans. Power Syst.}, vol.~39, no.~1, pp. 40--52, Jan. 2024.

\bibitem{zhang2022safe}
C.~Zhang, S.~R. Kuppannagari, and V.~K. Prasanna, ``Safe building {HVAC} control via batch reinforcement learning,'' \emph{IEEE Trans. Sustain. Comput.}, vol.~7, no.~4, pp. 923--934, Oct.-Dec. 2022.

\bibitem{liang2021safe}
Z.~Liang, C.~Huang, W.~Su, N.~Duan, V.~Donde, B.~Wang, and X.~Zhao, ``Safe reinforcement learning-based resilient proactive scheduling for a commercial building considering correlated demand response,'' \emph{IEEE Open Access J. Power Energy}, vol.~8, pp. 85--96, 2021.

\bibitem{qiu2022safe}
D.~Qiu, Z.~Dong, X.~Zhang, Y.~Wang, and G.~Strbac, ``Safe reinforcement learning for real-time automatic control in a smart energy-hub,'' \emph{Appl. Energy}, vol. 309, Mar. 2022, {Art.} no. 118403.

\bibitem{sun2024energy}
Y.~Sun, S.~Zhang, M.~Liu, R.~Zheng, and S.~Dong, ``Energy management based on safe multi-agent reinforcement learning for smart buildings in distribution networks,'' \emph{Energy Build.}, vol. 318, Sep. 2024, {Art.} no. 114410.

\bibitem{wang2025safe}
X.~Wang, P.~Wang, R.~Huang, X.~Zhu, J.~Arroyo, and N.~Li, ``Safe deep reinforcement learning for building energy management,'' \emph{Appl. Energy}, vol. 377, Jan. 2025, {Art.} no. 124328.

\bibitem{wang2022green}
R.~Wang, Z.~Cao, X.~Zhou, Y.~Wen, and R.~Tan, ``Green data center cooling control via physics-guided safe reinforcement learning,'' \emph{ACM Trans. Cyber-Phys. Syst.}, 2022.

\bibitem{wan2023safecool}
J.~Wan, Y.~Duan, X.~Gui, C.~Liu, L.~Li, and Z.~Ma, ``{SafeCool}: safe and energy-efficient cooling management in data centers with model-based reinforcement learning,'' \emph{IEEE Trans. Emerg. Topics Comput. Intell.}, vol.~7, no.~6, pp. 1621--1635, Dec. 2023.

\bibitem{cao2023toward}
Z.~Cao, R.~Wang, X.~Zhou, and Y.~Wen, ``Toward model-assisted safe reinforcement learning for data center cooling control: A {Lyapunov}-based approach,'' in \emph{Proc. ACM Int. Conf. Future Energy Syst.}, 2023, pp. 333--346.

\bibitem{golpira2019multi}
H.~Golp{\^\i}ra and S.~A.~R. Khan, ``A multi-objective risk-based robust optimization approach to energy management in smart residential buildings under combined demand and supply uncertainty,'' \emph{Energy}, vol. 170, pp. 1113--1129, Mar. 2019.

\bibitem{liu2022safe}
H.-Y. Liu, B.~Balaji, S.~Gao, R.~Gupta, and D.~Hong, ``Safe {HVAC} control via batch reinforcement learning,'' in \emph{Proc. ACM/IEEE Int. Conf. Cyber- Phys. Syst.}, 2022, pp. 181--192.

\bibitem{zheng2021vulnerability}
Y.~Zheng, Z.~Yan, K.~Chen, J.~Sun, Y.~Xu, and Y.~Liu, ``Vulnerability assessment of deep reinforcement learning models for power system topology optimization,'' \emph{IEEE Trans. Smart Grid}, vol.~12, no.~4, pp. 3613--3623, Jul. 2021.

\bibitem{hao2024safe}
G.~Hao, Y.~Li, Y.~Li, K.~Guang, and Z.~Zeng, ``Safe reinforcement learning for active distribution networks reconfiguration considering uncertainty,'' \emph{IEEE Trans. Ind. Appl.}, vol.~61, no.~1, pp. 1757--1769, Jan.-Feb. 2025.

\bibitem{anderson2022power}
P.~M. Anderson, C.~F. Henville, R.~Rifaat, B.~Johnson, and S.~Meliopoulos, \emph{Power system protection}.\hskip 1em plus 0.5em minus 0.4em\relax John Wiley \& Sons, 2022.

\bibitem{ademola2020frequency}
A.~Ademola-Idowu and B.~Zhang, ``Frequency stability using {MPC}-based inverter power control in low-inertia power systems,'' \emph{IEEE Trans. Power Syst.}, vol.~36, no.~2, pp. 1628--1637, Mar. 2021.

\bibitem{yang2017linearized}
Z.~Yang, H.~Zhong, A.~Bose, T.~Zheng, Q.~Xia, and C.~Kang, ``A linearized {OPF} model with reactive power and voltage magnitude: A pathway to improve the {MW}-only {DC OPF},'' \emph{IEEE Trans. Power Syst.}, vol.~33, no.~2, pp. 1734--1745, Mar. 2018.

\bibitem{panteli2016power}
M.~Panteli, C.~Pickering, S.~Wilkinson, R.~Dawson, and P.~Mancarella, ``Power system resilience to extreme weather: Fragility modeling, probabilistic impact assessment, and adaptation measures,'' \emph{IEEE Trans. Power Syst.}, vol.~32, no.~5, pp. 3747--3757, Sep. 2017.

\bibitem{duchesne2020recent}
L.~Duchesne, E.~Karangelos, and L.~Wehenkel, ``Recent developments in machine learning for energy systems reliability management,'' \emph{Proc. IEEE}, vol. 108, no.~9, pp. 1656--1676, Sep. 2020.

\bibitem{hu2017recent}
L.~Hu, Z.~Wang, X.~Liu, A.~V. Vasilakos, and F.~E. Alsaadi, ``Recent advances on state estimation for power grids with unconventional measurements,'' \emph{IET Control Theory Appl.}, vol.~11, no.~18, pp. 3221--3232, Nov. 2017.

\bibitem{siemensGTautotuner}
\BIBentryALTinterwordspacing
{Siemens Energy}, ``{GT Auto Tuner},'' Accessed: Mar. 24, 2025. [Online]. Available: \url{https://www.siemens-energy.com/global/en/home/products-services/service/gt-autotuner.html}
\BIBentrySTDinterwordspacing

\bibitem{googleAICooling}
\BIBentryALTinterwordspacing
J.~Temple, ``{Google just gave control over data center cooling to an AI},'' Accessed: Mar. 24, 2025. [Online]. Available: \url{https://www.technologyreview.com/2018/08/17/140987/google-just-gave-control-over-data-center-cooling-to-an-ai/}
\BIBentrySTDinterwordspacing

\bibitem{deepmindCoolingAI}
\BIBentryALTinterwordspacing
DeepMind, ``{Safety-first AI for autonomous data centre cooling and industrial control},'' Accessed: Mar. 24, 2025. [Online]. Available: \url{https://deepmind.google/discover/blog/safety-first-ai-for-autonomous-data-centre-cooling-and-industrial-control/}
\BIBentrySTDinterwordspacing

\bibitem{luo2022controlling}
J.~Luo, C.~Paduraru, O.~Voicu, Y.~Chervonyi, S.~Munns, J.~Li, C.~Qian, P.~Dutta, J.~Q. Davis, N.~Wu \emph{et~al.}, ``Controlling commercial cooling systems using reinforcement learning,'' \emph{arXiv preprint arXiv:2211.07357}, 2022.

\bibitem{telusAIClimate}
\BIBentryALTinterwordspacing
{TELUS} and {Vector Institute}, ``{Using AI for good: TELUS and Vector Institute partner to reduce climate impacts from data centres with new Energy Optimization System},'' Accessed: Mar. 24, 2025. [Online]. Available: \url{https://www.telus.com/en/about/news-and-events/media-releases/using-ai-for-good-telus-and-vector-institute-partner-to-reduce-climate-impacts-from-data-centres}
\BIBentrySTDinterwordspacing

\bibitem{foobotAIHVAC}
\BIBentryALTinterwordspacing
A.~Galataud, ``{Nine months of AI-based control optimization on a modern office building HVAC},'' Accessed: Mar. 24, 2025. [Online]. Available: \url{https://techblog.foobot.io/hvac/control/ai/reinforcement_learning/sab_after_9.html}
\BIBentrySTDinterwordspacing

\bibitem{zhan2022deepthermal}
X.~Zhan, H.~Xu, Y.~Zhang, X.~Zhu, H.~Yin, and Y.~Zheng, ``Deepthermal: Combustion optimization for thermal power generating units using offline reinforcement learning,'' in \emph{Proc. AAAI Conf. Artif. Intell.}, vol.~36, no.~4, 2022, pp. 4680--4688.

\bibitem{ma2024efficient}
C.~Ma, A.~Li, Y.~Du, H.~Dong, and Y.~Yang, ``Efficient and scalable reinforcement learning for large-scale network control,'' \emph{Nat. Mach. Intell.}, vol.~6, no.~9, pp. 1006--1020, 2024.

\bibitem{vishwanath2024reinforcement}
A.~Vishwanath, L.~A. Dennis, and M.~Slavkovik, ``Reinforcement learning and machine ethics: A systematic review,'' \emph{arXiv preprint arXiv:2407.02425}, 2024.

\bibitem{fulton2018safe}
N.~Fulton and A.~Platzer, ``Safe reinforcement learning via formal methods: Toward safe control through proof and learning,'' in \emph{Proc. AAAI Conf. Artif. Intell.}, vol.~32, no.~1, 2018.

\bibitem{dulac2019challenges}
G.~Dulac-Arnold, D.~Mankowitz, and T.~Hester, ``Challenges of real-world reinforcement learning,'' in \emph{Proc. Int. Conf. Mach. Learn.}, 2019, pp. 1--14.

\bibitem{ingleson2005tracking}
J.~W. Ingleson and D.~M. Ellis, ``Tracking the eastern interconnection frequency governing characteristic,'' in \emph{Proc. IEEE Power Energy Soc. Gen. Meeting}, 2005, pp. 1461--1466.

\bibitem{lu2024overcoming}
C.~Lu, L.~Shi, Z.~Chen, C.~Wu, and A.~Wierman, ``Overcoming the curse of dimensionality in reinforcement learning through approximate factorization,'' \emph{arXiv preprint arXiv:2411.07591}, 2024.

\bibitem{hreinsson2021new}
K.~Hreinsson, A.~Scaglione, M.~Alizadeh, and Y.~Chen, ``New insights from the {Shapley-Folkman} lemma on dispatchable demand in energy markets,'' \emph{IEEE Trans. Power Syst.}, vol.~36, no.~5, pp. 4028--4041, Sep. 2021.

\bibitem{beck2023survey}
J.~Beck, R.~Vuorio, E.~Z. Liu, Z.~Xiong, L.~Zintgraf, C.~Finn, and S.~Whiteson, ``A survey of meta-reinforcement learning,'' \emph{arXiv preprint arXiv:2301.08028}, 2023.

\bibitem{da2019survey}
F.~L. Da~Silva and A.~H.~R. Costa, ``A survey on transfer learning for multiagent reinforcement learning systems,'' \emph{J. Artif. Intell. Res.}, vol.~64, pp. 645--703, Mar. 2019.

\bibitem{tang2022leveraging}
S.~Tang, M.~Makar, M.~Sjoding, F.~Doshi-Velez, and J.~Wiens, ``Leveraging factored action spaces for efficient offline reinforcement learning in healthcare,'' in \emph{Proc. Adv. Neural Inf. Process. Syst.}, vol.~35, 2022, pp. 34\,272--34\,286.

\bibitem{xiong2022hisarl}
Z.~Xiong, I.~Agarwal, and S.~Jagannathan, ``{HiSaRL}: A hierarchical framework for safe reinforcement learning.'' in \emph{Proc. AAAI SafeAI Workshop}, 2022.

\bibitem{xia2024hierarchical}
Y.~Xia, Y.~Xu, and X.~Feng, ``Hierarchical coordination of networked-microgrids towards decentralized operation: A safe deep reinforcement learning method,'' \emph{IEEE Trans. Sustain. Energy}, vol.~15, no.~3, pp. 1981--1993, Jul. 2024.

\bibitem{zhu2024parallel}
J.~Zhu, D.~Li, Y.~Chen, J.~Chen, and Y.~Luo, ``Parallel hybrid deep reinforcement learning for real-time energy management of microgrid,'' \emph{J. Modern Power Syst. Clean Energy}, vol.~13, no.~3, pp. 991--1002, May 2025.

\bibitem{yang2023safety}
Q.~Yang, T.~D. Sim{\~a}o, S.~H. Tindemans, and M.~T. Spaan, ``Safety-constrained reinforcement learning with a distributional safety critic,'' \emph{Mach. Learn.}, vol. 112, no.~3, pp. 859--887, 2023.

\bibitem{verbraeken2020survey}
J.~Verbraeken, M.~Wolting, J.~Katzy, J.~Kloppenburg, T.~Verbelen, and J.~S. Rellermeyer, ``A survey on distributed machine learning,'' \emph{{ACM} Comput. Surv.}, vol.~53, no.~2, pp. 1--33, Mar. 2020.

\bibitem{lu2021decentralized}
S.~Lu, K.~Zhang, T.~Chen, T.~Ba{\c{s}}ar, and L.~Horesh, ``Decentralized policy gradient descent ascent for safe multi-agent reinforcement learning,'' in \emph{Proc. AAAI Conf. Artif. Intell.}, vol.~35, no.~10, 2021, pp. 8767--8775.

\bibitem{zhang2021distributional}
P.~Zhang, X.~Chen, L.~Zhao, W.~Xiong, T.~Qin, and T.-Y. Liu, ``Distributional reinforcement learning for multi-dimensional reward functions,'' in \emph{Proc. Adv. Neural Inf. Process. Syst.}, vol.~34, 2021, pp. 1519--1529.

\bibitem{chung2020distributed}
H.-M. Chung, S.~Maharjan, Y.~Zhang, and F.~Eliassen, ``Distributed deep reinforcement learning for intelligent load scheduling in residential smart grids,'' \emph{IEEE Trans. Ind. Inform.}, vol.~17, no.~4, pp. 2752--2763, Apr. 2021.

\bibitem{lyu2021contrasting}
X.~Lyu, Y.~Xiao, B.~Daley, and C.~Amato, ``Contrasting centralized and decentralized critics in multi-agent reinforcement learning,'' in \emph{Proc. Int. Joint Conf. Auton. Agents Multiagent Syst.}, 2021, pp. 844--852.

\bibitem{chen2021communication}
T.~Chen, K.~Zhang, G.~B. Giannakis, and T.~Ba{\c{s}}ar, ``Communication-efficient policy gradient methods for distributed reinforcement learning,'' \emph{IEEE Trans. Control Netw. Syst.}, vol.~9, no.~2, pp. 917--929, Jun. 2022.

\bibitem{koursioumpas2024safe}
N.~Koursioumpas, L.~Magoula, N.~Petropouleas, A.-I. Thanopoulos, T.~Panagea, N.~Alonistioti, M.~A. Gutierrez-Estevez, and R.~Khalili, ``A safe deep reinforcement learning approach for energy efficient federated learning in wireless communication networks,'' \emph{IEEE Trans. Green Commun. Netw.}, vol.~8, no.~4, pp. 1862--1874, Dec. 2024.

\bibitem{chen2023distributed}
Y.~Chen, J.~Zhu, Y.~Liu, L.~Zhang, and J.~Zhou, ``Distributed hierarchical deep reinforcement learning for large-scale grid emergency control,'' \emph{IEEE Trans. Power Syst.}, vol.~39, no.~2, pp. 4446--4458, Mar. 2024.

\bibitem{fujimoto2024assessing}
T.~Fujimoto, J.~Suetterlein, S.~Chatterjee, and A.~Ganguly, ``Assessing the impact of distribution shift on reinforcement learning performance,'' \emph{arXiv preprint arXiv:2402.03590}, 2024.

\bibitem{laskin2020reinforcement}
M.~Laskin, K.~Lee, A.~Stooke, L.~Pinto, P.~Abbeel, and A.~Srinivas, ``Reinforcement learning with augmented data,'' in \emph{Proc. Adv. Neural Inf. Process. Syst.}, vol.~33, 2020, pp. 19\,884--19\,895.

\bibitem{du2024real}
S.~Du, T.~Ding, Y.~Xiao, J.~Wan, J.~Liu, and F.~Meng, ``Real-time scheduling of high-penetrated renewable power systems: An expert knowledge and reinforcement learning hybrid approach,'' \emph{IEEE Trans. Power Syst.}, vol.~40, no.~2, pp. 1545--1557, Mar. 2025.

\bibitem{huang2022constrained}
S.~Huang, A.~Abdolmaleki, G.~Vezzani, P.~Brakel, D.~J. Mankowitz, M.~Neunert, S.~Bohez, Y.~Tassa, N.~Heess, M.~Riedmiller \emph{et~al.}, ``A constrained multi-objective reinforcement learning framework,'' in \emph{Proc. Conf. Robot Learn.}, 2022, pp. 883--893.

\bibitem{kim2024trust}
D.~Kim, K.~Lee, and S.~Oh, ``Trust region-based safe distributional reinforcement learning for multiple constraints,'' in \emph{Proc. Adv. Neural Inform. Process. Syst.}, vol.~36, 2024.

\bibitem{zhou2024multi}
Z.~Zhou, J.~Booher, W.~Liu, A.~Petiushko, and A.~Garg, ``Multi-constraint safe {RL} with objective suppression for safety-critical applications,'' \emph{arXiv preprint arXiv:2402.15650}, 2024.

\bibitem{yao2024gradient}
Y.~Yao, Z.~Liu, Z.~Cen, P.~Huang, T.~Zhang, W.~Yu, and D.~Zhao, ``Gradient shaping for multi-constraint safe reinforcement learning,'' in \emph{Proc. Annu. Learn. Dyn. Control Conf.}, 2024, pp. 25--39.

\bibitem{roza2023safe}
F.~S. Roza, K.~Roscher, and S.~G{\"u}nnemann, ``Safe and efficient operation with constrained hierarchical reinforcement learning,'' in \emph{Proc. Eur. Workshop Reinforc. Learn.}, 2023.

\bibitem{vu2021safe}
T.~L. Vu, S.~Mukherjee, T.~Yin, R.~Huang, J.~Tan, and Q.~Huang, ``Safe reinforcement learning for emergency load shedding of power systems,'' in \emph{Proc. IEEE Power Energy Soc. General Meeting}, 2021, pp. 1--5.

\bibitem{calvo2023state}
M.~Calvo-Fullana, S.~Paternain, L.~F. Chamon, and A.~Ribeiro, ``State augmented constrained reinforcement learning: Overcoming the limitations of learning with rewards,'' \emph{IEEE Trans. Autom. Control}, vol.~69, no.~7, pp. 4275--4290, Jul. 2024.

\bibitem{cai2023safe}
M.~Cai, S.~Xiao, J.~Li, and Z.~Kan, ``Safe reinforcement learning under temporal logic with reward design and quantum action selection,'' \emph{Sci. Rep.}, vol.~13, no.~1, p. 1925, Feb. 2023.

\bibitem{nagabandi2018learning}
A.~Nagabandi, I.~Clavera, S.~Liu, R.~S. Fearing, P.~Abbeel, S.~Levine, and C.~Finn, ``Learning to adapt in dynamic, real-world environments through meta-reinforcement learning,'' in \emph{Proc. Int. Conf. Learn. Representations}, 2019, pp. 1--10.

\bibitem{retzlaff2024human}
C.~O. Retzlaff, S.~Das, C.~Wayllace, P.~Mousavi, M.~Afshari, T.~Yang, A.~Saranti, A.~Angerschmid, M.~E. Taylor, and A.~Holzinger, ``Human-in-the-loop reinforcement learning: A survey and position on requirements, challenges, and opportunities,'' \emph{J. Artif. Intell. Res.}, vol.~79, pp. 359--415, Jan. 2024.

\bibitem{zolfagharian2024smarla}
A.~Zolfagharian, M.~Abdellatif, L.~C. Briand, and S.~Ramesh, ``{SMARLA}: A safety monitoring approach for deep reinforcement learning agents,'' \emph{IEEE Trans. Softw. Eng.}, vol.~51, no.~1, pp. 82--105, Jan. 2025.

\bibitem{modares2023safe}
A.~Modares, N.~Sadati, B.~Esmaeili, F.~A. Yaghmaie, and H.~Modares, ``Safe reinforcement learning via a model-free safety certifier,'' \emph{IEEE Trans. Neural Netw. Learn. Syst.}, vol.~35, no.~3, pp. 3302--3311, Mar. 2024.

\bibitem{levine2020offline}
S.~Levine, A.~Kumar, G.~Tucker, and J.~Fu, ``Offline reinforcement learning: Tutorial, review, and perspectives on open problems,'' \emph{arXiv preprint arXiv:2005.01643}, 2020.

\bibitem{liu2023constrained}
Z.~Liu, Z.~Guo, Y.~Yao, Z.~Cen, W.~Yu, T.~Zhang, and D.~Zhao, ``Constrained decision transformer for offline safe reinforcement learning,'' in \emph{Proc. Int. Conf. Mach. Learn.}, 2023, pp. 21\,611--21\,630.

\bibitem{xue2024privacy}
L.~Xue, Y.~Zhang, J.~Wang, H.~Li, and F.~Li, ``Privacy-preserving multi-level co-regulation of {VPPs} via hierarchical safe deep reinforcement learning,'' \emph{Appl. Energy}, vol. 371, Oct. 2024, {Art.} no. 123654.

\bibitem{wang2019privacy}
B.~Wang and N.~Hegde, ``Privacy-preserving {Q}-learning with functional noise in continuous spaces,'' in \emph{Proc. Adv. Neural Inf. Process. Syst.}, vol.~32, 2019.

\bibitem{qiao2024offline}
D.~Qiao and Y.-X. Wang, ``Offline reinforcement learning with differential privacy,'' in \emph{Proc. Adv. Neural Inf. Process. Syst.}, vol.~36, 2024.

\bibitem{dvorkin2020differentially}
V.~Dvorkin, F.~Fioretto, P.~Van~Hentenryck, P.~Pinson, and J.~Kazempour, ``Differentially private optimal power flow for distribution grids,'' \emph{IEEE Trans. Power Syst.}, vol.~36, no.~3, pp. 2186--2196, May 2021.

\bibitem{qi2021federated}
J.~Qi, Q.~Zhou, L.~Lei, and K.~Zheng, ``Federated reinforcement learning: Techniques, applications, and open challenges,'' \emph{Intell. Robot.}, vol.~1, no.~1, pp. 18--57, 2021.

\bibitem{fan2021fault}
X.~Fan, Y.~Ma, Z.~Dai, W.~Jing, C.~Tan, and B.~K.~H. Low, ``Fault-tolerant federated reinforcement learning with theoretical guarantee,'' in \emph{Proc. Adv. Neural Inf. Process. Syst.}, vol.~34, 2021, pp. 1007--1021.

\bibitem{amos2017input}
B.~Amos, L.~Xu, and J.~Z. Kolter, ``Input convex neural networks,'' in \emph{Proc. Int. Conf. Mach. Learn.}, 2017, pp. 146--155.

\bibitem{chen2018optimal}
Y.~Chen, Y.~Shi, and B.~Zhang, ``Optimal control via neural networks: A convex approach,'' in \emph{Proc. Int. Conf. Learn. Representations}, 2019.

\bibitem{sayed2023optimal}
A.~R. Sayed, X.~Zhang, G.~Wang, C.~Wang, and J.~Qiu, ``Optimal operable power flow: Sample-efficient holomorphic embedding-based reinforcement learning,'' \emph{IEEE Trans. Power Syst.}, vol.~39, no.~1, pp. 1739--1751, Jan. 2024.

\bibitem{sun2023optimal}
X.~Sun, Z.~Xu, J.~Qiu, H.~Liu, H.~Wu, and Y.~Tao, ``Optimal {Volt/Var} control for unbalanced distribution networks with human-in-the-loop deep reinforcement learning,'' \emph{IEEE Trans. Smart Grid}, vol.~15, no.~3, pp. 2639--2651, May 2024.

\bibitem{yang2020optimal}
L.~Yang, Q.~Sun, N.~Zhang, and Z.~Liu, ``Optimal energy operation strategy for we-energy of energy internet based on hybrid reinforcement learning with human-in-the-loop,'' \emph{IEEE Trans. Syst., Man, Cybern.: Syst.}, vol.~52, no.~1, pp. 32--42, Jan. 2022.

\bibitem{cao2024survey}
Y.~Cao, H.~Zhao, Y.~Cheng, T.~Shu, Y.~Chen, G.~Liu, G.~Liang, J.~Zhao, J.~Yan, and Y.~Li, ``Survey on large language model-enhanced reinforcement learning: Concept, taxonomy, and methods,'' \emph{IEEE Trans. Neural Netw. Learn. Syst.}, vol.~36, no.~6, pp. 9737--9757, Jun. 2025.

\bibitem{Majumder2024}
S.~Majumder, L.~Dong, F.~Doudi, Y.~Cai, C.~Tian, D.~Kalathil, K.~Ding, A.~A. Thatte, N.~Li, and L.~Xie, ``Exploring the capabilities and limitations of large language models in the electric energy sector,'' \emph{Joule}, vol.~8, no.~6, pp. 1544--1549, 2024.

\end{thebibliography}

\end{document}